\documentclass[prd,twocolumn,nofootinbib,showpacs]{revtex4}
\usepackage[final]{graphicx}
\usepackage{mathrsfs}
\usepackage{amssymb}
\usepackage{amsmath}
\usepackage{calligra}
\usepackage{tensor}
\usepackage{color}

\newcommand{\e}{\varepsilon}
\newcommand{\A}[2]{\mathscr{A}_{#1}\coeff{#2}}
\newcommand{\B}[2]{\mathscr{B}_{#1}\coeff{#2}}
\newcommand{\C}[2]{\mathscr{C}_{#1}\coeff{#2}}
\newcommand{\D}[2]{\mathscr{D}_{#1}\coeff{#2}}
\newcommand{\E}[2]{\mathscr{E}_{#1}\coeff{#2}}
\newcommand{\F}[2]{\mathscr{F}_{#1}\coeff{#2}}
\newcommand{\G}[2]{\mathscr{G}_{#1}\coeff{#2}}
\renewcommand{\H}[2]{\mathscr{H}_{#1}\coeff{#2}}
\newcommand{\I}[2]{\mathscr{I}_{#1}\coeff{#2}}
\newcommand{\K}[2]{\mathscr{K}_{#1}\coeff{#2}}
\newcommand{\nhat}{\hat{n}}
\DeclareMathAlphabet{\mathcalligra}{T1}{calligra}{m}{n}
\DeclareFontShape{T1}{calligra}{m}{n}{<->s*[2.75]callig15}{}
\renewcommand{\r}{\ensuremath{\mathcalligra{r}}\hspace{0.45 mm}}
\newcommand{\rad}{\mathscr{R}}
\newcommand{\h}{\mathfrak{h}}
\renewcommand{\sb}{\bar\sigma}
\newcommand{\spb}{\bar\sigma'}
\newcommand{\spp}{\sigma''}
\newcommand{\av}[1]{\left\langle #1 \right\rangle}
\newcommand{\tail}{h^{{}^{{\!\scriptstyle\text{tail}}}}}
\newcommand{\order}[1]{O\!\left(#1\right)}
\newcommand{\etide}{\mathcal{E}}
\newcommand{\btide}{\mathcal{B}}
\newcommand{\U}{\mathfrak{U}}
\newcommand{\past}{I^{\;{}^\text{-}}}
\newcommand{\hmn}[2]{h^{^{\!\text{(#2)}}}_{#1}}
\newcommand{\hbarmn}[2]{\bar h^{^{\!\text{(#2)}}}_{#1}}
\newcommand{\del}[1]{\nabla_{\!\!#1}}

\DeclareMathOperator{\STF}{STF}
\newcommand{\trans}{\psi}
\newcommand{\map}{\varphi}
\newcommand{\man}{\mathcal{M}}
\newcommand{\exact}[1]{\mathsf{#1}}
\newcommand{\expand}{\Phi}

\newcommand{\gauge}[3]{L\coeff{#2}_{#1}\left[#3\right]}
\newcommand{\ddR}[3]{\delta^2 R\coeff{#2}_{#1}\!\Big[#3\Big]}
\newcommand{\coeff}[1]{^{\scriptscriptstyle{\text{(#1)}}}}
\newcommand{\an}[1]{a^{\scriptscriptstyle{(#1)}}}
\newcommand{\zn}[1]{z_{\scriptscriptstyle{(#1)}}}
\newcommand{\un}[1]{\dot{z}_{\scriptscriptstyle{(#1)}}}

\begin{document}
\title{The self-consistent gravitational self-force} 
\author{Adam Pound} 
\affiliation{Department of Physics, University of Guelph, Guelph, Ontario, N1G 2W1} 
\date{\today}

\begin{abstract}

I review the problem of motion for small bodies in General Relativity, with an emphasis on developing a self-consistent treatment of the gravitational self-force. An analysis of the various derivations extant in the literature leads me to formulate an asymptotic expansion in which the metric is expanded while a representative worldline is held fixed; I discuss the utility of this expansion for both exact point particles and asymptotically small bodies, contrasting it with a regular expansion in which both the metric and the worldline are expanded. Based on these preliminary analyses, I present a general method of deriving self-consistent equations of motion for arbitrarily structured (sufficiently compact) small bodies. My method utilizes two expansions: an inner expansion that keeps the size of the body fixed, and an outer expansion that lets the body shrink while holding its worldline fixed. By imposing the Lorenz gauge, I express the global solution to the Einstein equation in the outer expansion in terms of an integral over a worldtube of small radius surrounding the body. Appropriate boundary data on the tube are determined from a local-in-space expansion in a buffer region where both the inner and outer expansions are valid. This buffer-region expansion also results in an expression for the self-force in terms of irreducible pieces of the metric perturbation on the worldline. Based on the global solution, these pieces of the perturbation can be written in terms of a tail integral over the body's past history. This approach can be applied at any order to obtain a self-consistent approximation that is valid on long timescales, both near and far from the small body. I conclude by discussing possible extensions of my method and comparing it to alternative approaches.
\end{abstract}
\maketitle


\section{Introduction}
The problem of motion is of tremendous historical importance in General Relativity, both theoretically and experimentally. In conceiving of the theory, Einstein was fundamentally concerned with explaining the motion of bodies solely in terms of the geometric relationships between them. And much of the observational evidence for General Relativity---\emph{e.g.}, the deflection of light by massive objects, the post-Newtonian effects in solar-system dynamics, and the slow decay of binary pulsar orbits due to the emission of gravitational waves---is tied to analyses of motion.

Despite the historical importance of the problem, theoretical treatments of it have largely been confined to two limiting regimes: first, the Newtonian limit of weak gravity and slow motion, in which Newton's laws of motion and relativistic corrections to them can be derived for widely separated bodies; and second, the point particle limit, in which the geodesic equation and corrections to it can be derived for bodies of asymptotically small mass and size. Study of the Newtonian limit was pioneered by Einstein, Infeld, and Hoffmann \cite{Einstein, Einstein2} and is now fully developed in post-Newtonian theory (see the reviews \cite{Futamase_review,Blanchet_review} and references therein). Study of the point particle limit is less well developed, and it has typically focused on proving that at leading order, a small body behaves as a test particle, moving on a geodesic of some background spacetime (see, e.g., Refs.~\cite{Infeld, Kates_motion, DEath_paper, DEath, Geroch_particle1, Geroch_particle2}).

The advent of gravitational wave detectors such as LIGO~\cite{LIGO} and LISA~\cite{LISA} has rapidly broadened the scope of this research. Because these detectors have the potential to accurately measure the dynamics of bodies in regions of very strong gravity, there is now a pressing need to go beyond either the post-Newtonian or test-particle approximations. For example, in binary systems, which are potentially important sources of gravitational waves, the two bodies will emit gravitational radiation, thereby lose energy, and slowly spiral into one another (see Ref. \cite{binary_review} for an overview of such systems). Once the two bodies are very near one another, the post-Newtonian approximation breaks down. 

If the two bodies are of comparable mass, then their motion during these final stages must be determined via a numerical integration of Einstein's equation (see Refs.~\cite{numerical_review,Pretorius_review} for reviews of numerical approaches to this problem). However, if one of the bodies is much less massive than the other, then the entire inspiral can be treated analytically, rather than numerically, by utilizing the point particle limit. (See Refs.~\cite{Drasco_review,EMRI_review} for an overview of these systems, called extreme mass ratio inspirals (EMRIs).) In this case, an expansion in the point particle limit roughly corresponds to an expansion of the metric in powers of the mass ratio $m/M\sim\e$. At leading order in this expansion, the small body moves on a geodesic of the spacetime of the large body, while simultaneously emitting gravitational waves. Obviously this approximation breaks down after a brief time, since it implies that the body will travel forever on a geodesic even as it emits waves that carry off energy and angular momentum. Thus, one must proceed beyond the leading-order, geodesic approximation. At subleading order, the metric perturbations produced by the small body force it onto an accelerated worldline that slowly spirals into the large body. The acceleration of this worldline, caused by the body's interaction with its own gravitational field, is called the gravitational self-force. Along with all other corrections to the test-mass approximation, this force will be the principal subject of the present paper.

A formal expression for the gravitational self-force was first derived in 1996 by Mino, Sasaki, and Tanaka \cite{Mino_Sasaki_Tanaka} and Quinn and Wald \cite{Quinn_Wald}; the resulting equation of motion is now known as the MiSaTaQuWa equation. Since then, numerous other derivations have been offered (see Refs.~\cite{Mino_matching, Detweiler_Whiting, Eran_force, Galley_Hu, Galley_QFT, Fukumoto, Gal'tsov, Gralla_Wald} and the reviews \cite{Eric_review, Detweiler_review} for examples). However, none amongst this plethora of derivations has overcome a fundamental difficulty in defining a point particle limit: how can one accurately, self-consistently, and systematically incorporate corrections into a worldline?

At its most fundamental level, the self-force problem consists of finding a pair $(\gamma,h_{\mu\nu})$ representing the worldline $\gamma$ and metric perturbation $h_{\mu\nu}$ of an asymptotically small body. This problem is far from trivial. Unlike in other field theories such as electrodynamics, equations of motion are not independent of the field equation in General Relativity---in fact, they are integrability conditions for the Einstein field equation, following from the restriction imposed by the Bianchi identity \cite{Einstein,Detweiler_review}. This means that at each order in perturbation theory, the equation of motion, and hence the worldline, is fixed by the Bianchi identity; using any other worldline means that a given $n$th-order perturbation $\hmn{\mu\nu}{\emph{n}}$ is not a solution to the $n$th-order Einstein equation. But at each order, the worldline determined by the Bianchi identity differs from that at every other order. It seems clear that the higher-order equations of motion are corrections to the lower-order ones, but there is no obvious way to self-consistently and systematically incorporate these corrections.

For example, suppose that in an EMRI system, one expands the metric and the Einstein equation in powers of $\e$, and that at order $\e$ the stress-energy tensor of the small body can be approximated by that of a point mass. Then the linearized Einstein equation reads $\delta G_{\mu\nu}[h]=8\pi T_{\mu\nu}[\gamma]$, where $\delta G_{\mu\nu}[h]$ is linear in the perturbation $h_{\alpha\beta}$ and $T_{\mu\nu}[\gamma]$ is the stress-energy tensor of a point mass moving on a worldline $\gamma$ in the background spacetime. The linearized Bianchi identity implies that the point particle source must be conserved, which in turn implies that the particle must move on a geodesic of the background spacetime. Thus, at first order in the expansion, the body travels forever on a geodesic. In order to correctly derive the self-force, we must proceed to second order. But even if that is accomplished, it seems that the self-force cannot trivially be identified as the equation of motion of the worldline, since the worldline is fixed by the first-order Bianchi identity.

As in any problem involving a small parameter, two options present themselves: first, one can assume a regular Taylor series expansion of every function in the problem, which leads to a succession of equations that can be solved exactly, order-by-order; or second, one can be satisfied with the construction of an approximate solution to the exact equation, however the solution may be arrived at. If the first approach is adopted, then the linearized Bianchi identity fixes the worldline to be a geodesic, and the self-force can only be interpreted in a perturbative sense, as the equation of motion for a ``deviation vector" describing an infinitesimal correction to the geodesic. But this interpretation is meaningful only for a brief time: since the particle will eventually plunge into the large body, the ``small" correction to the geodesic will eventually grow large. At that point, the entire expansion in powers of $\e$ will have broken down. In other words, the straightforward expansion of the Einstein equation is a valid approximation to the actual Einstein equation only on short timescales.

However, in studies of the problem of motion in General Relativity, this first approach is atypical; instead, the second approach is typically adopted. In the self-force problem, this has been realized in the procedure of ``gauge relaxation," \cite{Quinn_Wald} in which the linearized Einstein equation is written in the Lorenz gauge, leading to a wave equation that can be solved for a point particle source moving on an arbitrary worldline---but then, in order to circumvent the conclusion that that worldline must be a geodesic, the solution to the linear equation is allowed to slightly violate the Lorenz gauge condition \cite{Mino_Sasaki_Tanaka, Quinn_Wald}. This procedure yields an approximate solution to the exact Einstein equation, as desired, and it leads to a single worldline obeying a self-consistent equation of motion. It is also similar to successful methods of post-Newtonian theory, in which, prior to any expansion, the exact Einstein equation is written in a ``relaxed" form that can be solved for an arbitrary source. However, the gauge-relaxation used in the self-force problem lacks the systematic nature of the post-Newtonian method, in the sense that the relaxed linear equation has not been shown to follow from a systematic expansion of the exact Einstein equation, and the solution to the relaxed linear problem has not been related to a solution to the exact problem.

One of the goals of this paper is to provide such a systematic justification of the gauge-relaxation procedure, by breaking the exact Einstein equation into a sequence of equations that can be solved for a fixed, $\e$-dependent worldline. Of course, trying to do so introduces another problem: a point particle is a sensible source only for the first-order, \emph{linearized} Einstein equation. In the full, nonlinear theory, point particles are not mathematically well-defined sources.\footnote{At least this is true within classical distribution theory \cite{linear_distributions}, since the Einstein tensor of a point particle would contain products of delta distributions and hence be too singular to be treated as a distribution. However, more general methods based on Colombeau algebras, which allow for multiplication of distributions, have been devised to overcome this problem \cite{nonlinear_distributions}.} Going beyond first order thus means that one must abandon the point particle approximation.

Hence, we can see that an accurate, self-consistent solution to the self-force problem should accomplish all of the following:
\begin{enumerate}
\item go beyond first order perturbation theory,
\item define a self-consistent, ``corrected" worldline,
\item treat the body as extended, but asymptotically small.
\end{enumerate}
Note that these criteria are interdependent, since defining a meaningful worldline---one that incorporates corrections---requires going beyond first order, and going beyond first order requires accounting for the extension of the body. Furthermore, note that these theoretical goals are closely related to experimental ones: in order to extract the parameters of an EMRI from a gravitational wave signal, one requires a template that relates the signal to the motion of the body over a large portion of the inspiral. Such a template must be based on an approximation scheme that is uniform on a domain of size $\sim 1/\e$; in other words, the errors in the approximation must remain small over a long span of time. This can be accomplished only in a scheme that self-consistently incorporates the corrected motion.

\subsection{Organization of this paper}
The paper contains two main parts. The first part, comprising Secs.~\ref{point-particle} and \ref{extended_body}, consists of a more thorough explication of the problem. This explication serves two purposes: first, to review the foundations of the problem and the various derivations in the literature. Much of this review overlaps with previous discussions by Mino \cite{Mino_expansion1, Mino_expansion2, Mino_Price}, Hinderer and Flanagan \cite{Hinderer_Flanagan}, and Gralla and Wald \cite{Gralla_Wald}. However, because of the wealth of derivations performed to date, a more comprehensive review is timely. In addition, my presentation differs significantly from those earlier discussions and serves to motivate and provide the necessary context for my own approach. The second purpose of the explication is to introduce the notion of a self-consistent expansion in which the metric perturbation is first written as a functional of a worldline and then expanded while holding the worldline fixed; in this expansion, the solution to the Einstein equation is consistent with what would result from an evolution in time that began with (1) some arbitrary initial data and (2) a system of evolution equations that involve only local values of the position, momentum, and metric perturbation of the small body at each value of time. Section~\ref{point-particle} presents this expansion in the context of a point particle; Section~\ref{extended_body} then generalizes it to asymptotically small bodies. After laying that groundwork, the first part of the paper concludes in Sec.~\ref{outline} with an outline of my approach. Readers who are uninterested in the preliminary material can skip directly to that section.

The second part of the paper, comprising Secs.~\ref{self-force calculation} and \ref{perturbation calculation}, presents my calculation of the first-order gravitational self-force and the metric perturbation created by an asymptotically small (sufficiently compact) massive object. Section~\ref{self-force calculation} presents the derivation of the first-order gravitational self-force; the result of this calculation is that the body moves on a geodesic of the spacetime $g+h^R$, where $h^R$ is a homogenous perturbation that is regular on the body's worldline. In Sec.~\ref{perturbation calculation}, I calculate the metric perturbation induced by the body, which determines $h^R$ in terms of tail integrals and recovers the usual MiSaTaQuWa equation. Section~\ref{Conclusion} concludes the paper with a comparison to other methods and a discussion of higher-order and globally accurate approximations.

Throughout this paper, Greek indices $\alpha,\beta,\gamma,...$ run from 0 to 3 and refer to a coordinate basis, uppercase Latin indices from the middle of the alphabet ($I,J,K,...$) run from 0 to 3 and refer to an orthonormal tetrad basis, lowercase Latin indices $i,j,k,...$ run from 1 to 3 and refer to the spatial part of both the coordinate basis and the tetrad basis, and uppercase Latin indices from the beginning of the alphabet ($A,B,C,...$) run from 1 to 2 and refer to angular coordinates. Sans-serif symbols $\exact{g}, \exact{R}, \exact{T},...$ denote exact quantities to be expanded, and ${}^{\exact{g}\!}\del{\nu}$ denotes the covariant derivative compatible with an exact metric $\exact{g}$; a semi-colon and $\nabla$ are used interchangeably to indicate a covariant derivative compatible with a background metric $g$. I work in geometrical units in which $G=c=1$. I will frequently omit indices for simplicity.

\section{Preliminary analysis I: the motion of a point particle}\label{point-particle}
Assume for the moment that the exact Einstein equation, $\exact{G}_{\mu\nu}=8\pi\exact{T}_{\mu\nu}$ , can be made sense of with the point particle source
\begin{equation}
\exact{T}^{\mu\nu}[\exact{g},\gamma]=\int_\gamma mu^\mu u^\nu\delta(x,z(\exact{t}))d\exact{t},
\end{equation}
where $\gamma$ is the worldline of the particle, $z^\alpha(\exact{t})$ are the coordinates on $\gamma$, $u^\mu$ is its four-velocity, $\exact{t}$ is proper time with respect to $\exact{g}$ on $\gamma$, and $\delta(x,x')=\delta^4(x^\mu-x'^\mu)/\sqrt{|\exact{g}|}$ is a covariant delta function in the spacetime of the exact solution $\exact{g}_{\mu\nu}$, with $|\exact{g}|$ denoting the absolute value of the determinant of $\exact{g}_{\mu\nu}$. The motion of the particle is constrained by the Bianchi identity ${}^{\exact{g}\!}\del{\nu}\exact{G}^{\mu\nu}=0$, which implies the conservation equation ${}^{\exact{g}\!}\del{\nu}\exact{T}^{\mu\nu}=0$. The conservation equation in turn implies $\exact{a}^\mu\equiv\left({}^{\exact{g}\!}\del{\nu}u^\mu\right)u^\nu=0$; that is, $\gamma$ must be a geodesic in the full spacetime described by $\exact{g}$.

\subsection{Non-systematic expansion}\label{nonsystematic}
Now suppose that $m$ is small compared to all other length scales of the system, denoted collectively by $\mathcal{R}$, and that we wish to construct an approximation to $\exact{g}_{\mu\nu}$ and $\gamma$ for $\e\equiv m/\inf{\mathcal{R}}\ll1$. In this limit, the metric can be expanded as $\exact{g}=g+h$, where $h\sim\e$. And the exact equation of motion $\exact{a}^\mu=0$ can then be expanded as \cite{Eric_review}
\begin{equation}\label{eq_motion}
a^\mu = -\tfrac{1}{2}(g^{\mu\nu}+u^\mu u^\nu)(2h_{\nu\rho;\sigma}-h_{\rho\sigma;\nu})u^\rho u^\sigma+\order{\e^2},
\end{equation}
where $a^\mu\equiv u^\mu{}_{;\nu}u^\nu$ is the acceleration in the background spacetime $g$. Since the metric perturbation of a point particle will diverge as $1/r$ near $\gamma$, this equation of motion is ill-behaved. Hence, some form of regularization is required.

The very first derivation of the self-force, given by Mino, Sasaki, and Tanaka \cite{Mino_Sasaki_Tanaka}, followed earlier derivations of the electromagnetic self-force \cite{Dirac, DeWitt_Brehme} by using conservation of energy-momentum, calculating the flux through a surface around the body and setting it equal to the change of energy-momentum within the tube. Unfortunately, the regularization in this method essentially consists of discarding various divergent integrals and assuming results for others \cite{Eric_review}.

Other derivations of the self-force \cite{Quinn_Wald, Detweiler_Whiting, Eric_review} began by assuming that Eq.~\eqref{eq_motion} is essentially valid, but that only a certain \emph{regular} part of $h$ actually contributes to the acceleration. This regular part was assumed to be either the angle-averaged field~\cite{Quinn_Wald} or the Detweiler-Whiting~\cite{Detweiler_Whiting} regular field; the two assumptions yield the same force. All of these derivations are axiomatic---in the sense that they simply assume a form for the force---with their axioms supported by various plausibility arguments. Gal'tsov \emph{et al.} \cite{Gal'tsov} later showed that Eq.~\eqref{eq_motion} can be regularized via a straightforward expansion along the worldline, without making any assumption about which part of the field contributes to the force.

These derivations are based on solving the linearized Einstein equation, substituting it into Eq.~\eqref{eq_motion}, and then regularizing the result in one way or another. Expanding the exact Einstein equation to linear order in $\e$, we find at zeroth order that $G^{\mu\nu}=0$, which tells us that the background metric $g$ is that of a vacuum. At linear order we find
\begin{equation}\label{linear_EFE}
\delta G^{\mu\nu}[h] = 8\pi T^{\mu\nu}[\gamma],
\end{equation}
where $\delta G^{\mu\nu}$ is the linearized Einstein tensor, and $T^{\mu\nu}[\gamma]$ is the stress-energy tensor of a point-mass moving on a worldline $\gamma$ in the background spacetime defined by $g$. A formal solution to this equation can easily be obtained by imposing the Lorenz gauge condition. In the Lorenz gauge, the first-order Einstein equation is split into a wave equation, which I will write as
\begin{equation}\label{linear_wave_eqn}
E_{\mu\nu}[\bar h] = -16\pi T_{\mu\nu}[\gamma],
\end{equation}
and the gauge condition $\nabla^\nu\bar h_{\mu\nu}^{\!\coeff{1}}=0$, which I will write as
\begin{equation}
L_{\mu}[h] = 0,
\end{equation}
where $E_{\mu\nu}$ and $L_\mu$ are linear operators defined by
\begin{align}
E_{\mu\nu}[h] & = \left(g^\rho_\mu g^\sigma_\nu\nabla^\gamma\del{\gamma} +2R\indices{_\mu^\rho_\nu^\sigma}\right)\!h_{\rho\sigma},\label{wave op def}\\
L_\mu[h] & =\left(g^\rho_\mu g^{\sigma\gamma}-\tfrac{1}{2}g^\gamma_\mu g^{\rho\sigma}\right)\!\del{\gamma}h_{\rho\sigma}\label{gauge op def}.
\end{align}
An overbar indicates trace-reversal with respect to $g$; \emph{e.g.}, $\bar h_{\mu\nu}=h_{\mu\nu}-\tfrac{1}{2}g_{\mu\nu}g^{\rho\sigma}h_{\rho\sigma}$.

The retarded solution to the wave equation is
\begin{align}
h_{\mu\nu} & = 4\int \bar G_{\mu\nu\mu'\nu'} T^{\mu'\nu'}[\gamma]dV'\nonumber\\
&= 4m\int\limits_\gamma \bar G_{\mu\nu\mu'\nu'}u^{\mu'}u^{\nu'}dt',
\end{align}
where $G_{\mu\nu\mu'\nu'}$ is the retarded Green's function for the operator $E_{\mu\nu}$ and $t'$ is the proper time with respect to $g$ on the worldline. (My conventions for Green's functions, along with useful identities, are given in Appendix \ref{Greens_functions}.) Near the worldline, this solution can be decomposed into a local term, which diverges as $1/r$, and a so-called tail term $\tail$, defined as
\begin{equation}
\tail_{\mu\nu} = 4m\int_{-\infty}^{t_{\text{ret}}^-} \bar G_{\mu\nu\mu'\nu'}u^{\mu'}u^{\nu'}dt',
\end{equation}
where the upper limit of the integral is cut off just prior to the retarded time $t_{\text{ret}}$ in order to avoid the divergence of the Green's function there. The regularized equation of motion is obtained by replacing the exact field $h$ with the regular field $\tail$ in Eq.~\eqref{eq_motion}. 

But in order for the solution to the wave equation to also be a solution to the first-order Einstein equation, it must satisfy the gauge condition, which now reads 
\begin{align}
0 &= L_\mu[h]\nonumber\\
  &= 4\int \nabla^\nu G_{\mu\nu\mu'\nu'} T^{\mu'\nu'}[\gamma]dV'\nonumber\\
  &= 4\int G_{\mu\mu'}\del{\nu'}T^{\mu'\nu'}[\gamma]dV'\nonumber\\
  &= 4m\int_\gamma G_{\mu\nu'}a^{\nu'}dt',
\end{align}
where I have used the identity $\nabla^{\nu}G_{\mu\nu\mu'\nu'}=-G_{\mu(\mu';\nu')}$ (Eq.~\eqref{Green1}) and integrated by parts in going from the second line to the third. Thus, we find that imposing the gauge condition is equivalent to imposing the conservation equation $T^{\mu\nu}{}_{;\nu}=0$, which is equivalent to imposing the first-order Bianchi identity $\delta G^{\mu\nu}{}_{;\nu} = 0$. (The same equivalences could also be found by taking the divergence of the wave equation.) The consequence of any of these conditions is that $\gamma$ must be a geodesic in the background spacetime.

This requirement obviously contradicts the equation of motion \eqref{eq_motion}, regularized or not. In the earliest derivations of the self-force \cite{Mino_Sasaki_Tanaka, Quinn_Wald}, the contradiction was overcome by allowing the Lorenz gauge to be slightly violated by the first-order perturbation, effectively sidestepping the requirement that $\gamma$ must be a geodesic. This would mean that the metric perturbation $h$ is a solution to the wave equation \eqref{linear_wave_eqn} but not a solution to the linearized Einstein equation. Besides the fact that this gauge-relaxation prevents $h$ from exactly solving the linearized field equation, it also calls into question the expression for the self-force. The force is presumably an integrability condition for the second-order Einstein equation, but the point particle source is meaningful only at linear order. By taking the accelerated motion into account, we implicitly assume the solubility of the second-order problem, even though a point particle source ceases to makes sense at second order.

As Mino, Sasaki, and Tanaka stated in their original derivation \cite{Mino_Sasaki_Tanaka}, neither of these problems are too severe. The failure to exactly solve the first-order Einstein equation is not devastating, since if the acceleration is of order $\e$, then the errors in the gauge condition are of order $\e^2$; presumably these errors would be cancelled at second order. And we might comfortably presume that if the extension of the body were somehow taken into account at second order, the equation of motion would remain consistent with the MiSaTaQuWa result, which is, after all, fairly well motivated. However, can we justify these reassurances more systematically? In the remainder of this section, I will examine methods of self-consistently incorporating the acceleration into the point particle solution. In the next section, I will begin to discuss how these methods can be applied to an asymptotically small extended body.

But first, allow me to introduce some nomenclature. In general, for any function $f(x,\e)$, there are two types of expansions to consider: a \emph{regular expansion}, of the form
\begin{equation}
f(x,\e)=\sum_{n=0}^N\e^nf\coeff{\emph{n}}(x)+\order{\e^{N+1}},
\end{equation}
where the coefficients $f\coeff{\emph{n}}(x)$ are independent of $\e$;\footnote{More generally, a regular expansion can be of the form $f(x,\e)=\sum_{n=0}^N\lambda_n(\e)f\coeff{\emph{n}}(x)+\order{\lambda_{N+1}}$ for any set of functions $\lambda_n$ satisfying $\lim_{\e\to0}\lambda_{n+1}(\e)/\lambda_n(\e)=0$. An analogous generalization holds for singular expansions.} and a \emph{singular expansion}, of the form
\begin{equation}
f(x,\e)=\sum_{n=0}^N\e^nf\coeff{\emph{n}}(x,\e) +\order{\e^{N+1}},
\end{equation}
where the coefficients $f\coeff{\emph{n}}(x,\e)$ depend on $\e$ but are of order 1, in the sense that there exist positive constants $k$ and $\e_0$ such that $|f\coeff{\emph{n}}(x,\e)|\le k$ for $0\le\e\le\e_0$, but $\displaystyle\lim_{\e\to0}f\coeff{\emph{n}}(x,\e) \not\equiv 0$ (unless $f\coeff{\emph{n}}(x,\e)$ is itself identically zero). Put simply, the goal of a singular expansion is to expand only \emph{part} of a function's $\e$-dependence, while holding fixed some specific $\e$-dependence that captures one or more of the function's essential features. Further details of such expansions, and their role in singular perturbation theory, can be found in numerous textbooks \cite{Holmes, Verhulst, Lagerstrom, Kevorkian_Cole}; see the text by Eckhaus \cite{Eckhaus} for a rigorous treatment of some aspects. See Refs.~\cite{Kates_structure, perturbation_techniques} for discussions of singular perturbation theory in the context of General Relativity.

\subsection{Regular expansion}\label{regular_expansion_point_particle}
Let us begin by considering a regular expansion in powers of $\e$. Although such an expansion might be at the backs of most researchers' minds when they derive an expression for the self-force, only one extant derivation of the MiSaTaQuWa equation \cite{Gralla_Wald} has explicitly sought to remain within the framework of such an expansion. We begin by expanding the metric as 
\begin{equation}
\exact{g}_{\mu\nu}(x,\e)=g(x)+\e\hmn{\mu\nu}{1}(x)+\e^2\hmn{\mu\nu}{2}(x)+\order{\e^3},
\end{equation}
and the Einstein tensor as 
\begin{align}
\exact{G}^{\mu\nu}[\exact{g}]&=G^{\mu\nu}+\e\delta G^{\mu\nu}[\hmn{}{1}]+\e^2\delta G^{\mu\nu}[\hmn{}{2}]\nonumber\\
&\quad+\e^2\delta^2G^{\mu\nu}[\hmn{}{1}] +\order{\e^3},
\end{align}
where $G$ is the Einstein tensor of the background metric $g$, $\delta G^{\mu\nu}[h]$ is linear in $h$ and its derivatives, and $\delta^2 G^{\mu\nu}[h]$ is quadratic in them. Similarly, by expanding the $\sqrt{|\exact{g}|}$ that appears in it, and converting from the proper time in $\exact{g}$ to the proper time in $g$, the stress-energy tensor can be expanded as
\begin{equation}
\exact{T}^{\mu\nu}[\exact{g},\gamma]=\e T^{\mu\nu}[\gamma]+\e^2\delta T^{\mu\nu}[\hmn{}{1},\gamma]+\order{\e^3},
\end{equation}
where the factor of $\e$ is pulled out of $T$ for clarity. However, given that $\gamma$ satisfies the geodesic equation $\exact{a}^\mu=0$ in the full spacetime, it should be obvious that it will generically depend on $\e$, so the above expansion is not yet regular. To make it regular, we must expand the worldline as $\gamma=\gamma\coeff{0}+\e\gamma\coeff{1}+\order{\e^2}$. With the coordinates of the worldline defined by $z^\alpha(t,\e)$, this expansion takes the form
\begin{equation}
z^\alpha(t,\e)=\zn{0}^\alpha(t)+\e\zn{1}^{\alpha}(t)+\order{\e^2},
\end{equation}
where for the remainder of this subsection, $t$ will indicate proper time on the leading-order worldline $\gamma\coeff{0}$. To make this expansion most meaningful, we can insist that at some time $t=t_0$ the exact curve $z^\alpha$ is tangential to the leading-order curve $\zn{0}^\alpha$; the corrections $\zn{n}^\alpha$, $n\geq1$, then determine the deviation of the exact curve from the geodesic as time progresses away from $t=t_0$. Since the different terms in the expansion cannot map to different points in a curved spacetime, the ``corrections" are in fact vectors defined on the leading-order worldline; they ``connect" $\zn{0}^\alpha(t)$ to $z^\alpha(t)$, in the same sense as a geodesic deviation vector connects two neighbouring geodesics. Hence, in this expansion, one does not arrive at an equation for the acceleration of a worldline. Instead, one arrives at an equation for the acceleration of a deviation vector. This acceleration will naturally include a term identical to that of the geodesic deviation equation \cite{Gralla_Wald}, due to the drift of the the true worldline $\gamma$ away from the reference worldline $\gamma\coeff{0}$.

By using this expansion of the worldline, we can construct a regular expansion of the stress-energy tensor,
\begin{align}
\exact{T}^{\mu\nu}(\gamma)&=\e T^{\mu\nu}[\gamma\coeff{0}]+\e^2\delta T^{\mu\nu}[\hmn{}{1},\gamma\coeff{0}]\nonumber\\
&\quad+\e^2\tilde{\delta}T_{\mu\nu}[\gamma\coeff{0},\gamma\coeff{1}]+\order{\e^3},
\end{align}
where $\delta T^{\mu\nu}$ is linear in $\hmn{}{1}$, and $\tilde\delta T^{\mu\nu}$ is linear in $\zn{1}$. Substituting this expansion into the Einstein equation, we arrive at a sequence of field equations, written schematically as
\begin{align}
G^{\mu\nu} &= 0,\\
\delta G^{\mu\nu}[\hmn{}{1}] & = 8\pi T^{\mu\nu}[\gamma\coeff{0}], \\
\delta G^{\mu\nu}[\hmn{}{2}] & = 8\pi\delta T^{\mu\nu}[\hmn{}{1},\gamma\coeff{0}]+8\pi\tilde\delta T^{\mu\nu}[\gamma\coeff{0},\gamma\coeff{1}]\nonumber\\
&\quad-\delta^2 G^{\mu\nu}[\hmn{}{1}]\\
&\ \ \vdots\nonumber
\end{align}
These equations can be solved order-by-order for the background metric $g$, the perturbations $\hmn{\mu\nu}{{\it n}}$, and the curves $\zn{n}$.

At linear order, repeating the analysis above, we can split the Einstein equation into a wave equation and a gauge condition, and the first-order perturbation $\hmn{}{1}$ can be written as
\begin{equation}\label{1st order regular}
\hmn{\mu\nu}{1}=4\int_{\gamma\coeff{0}} \bar G_{\mu\nu\mu'\nu'}\un{0}^{\mu'}\un{0}^{\nu'}dt',
\end{equation}
where $\un{0}^\mu\equiv\displaystyle\frac{d\zn{0}^\mu}{dt}$. Imposition of either the gauge condition, the Bianchi identity, or the conservation of the source implies that $\gamma\coeff{0}$ must be a geodesic in the background spacetime.

But $\gamma\coeff{0}$ does not describe the true worldline of the particle: the effect of radiation at first order is incorporated into the correction $\zn{1}^\mu$, which is again determined by the Bianchi identity (or, equivalently, the conservation of the source $8\pi\delta T^{\mu\nu}+8\pi\tilde\delta T^{\mu\nu}-\delta^2 G^{\mu\nu}$). Explicitly, after splitting the second-order field equation into a wave equation and a gauge condition, the solution to the wave equation is given by
\begin{align}
\hmn{\mu\nu}{2} &= 2\int_{\gamma\coeff{0}} \bar G_{\mu\nu\mu'\nu'}\un{0}^{\mu'}\un{0}^{\nu'}(\un{0}^{\rho'}\un{0}^{\sigma'} -g^{\rho'\sigma'})\hmn{\rho'\sigma'}{1}dt' \nonumber\\
&\quad +4\int_{\gamma\coeff{0}}\bar G_{\mu\nu\mu'\nu'}\left(2\un{0}^{\mu'}\un{1}^{\nu'} +\un{0}^{\mu'}\un{0}^{\nu'}g_{\gamma'\delta'}\un{0}^{\delta'}\un{1}^{\gamma'}\right)dt' \nonumber\\
&\quad +4\int_{\gamma\coeff{0}}\bar G_{\mu\nu\mu'\nu';\rho'}\un{0}^{\mu'}\un{0}^{\nu'}\zn{1}^{\rho'}dt'\nonumber\\
&\quad-\frac{1}{2\pi}\int \bar G_{\mu\nu\mu'\nu'}\delta^2 G^{\mu'\nu'}dV',
\end{align}
where $\un{1}^\mu\equiv\del{\un{0}}\zn{1}^\mu$. The first line in this solution arises from $\delta T$, while the second and third arise from $\tilde\delta T$.

Imposing the gauge condition $L_\mu[\hmn{}{2}]=0$, making use of Eq.~\eqref{Green1}, integrating by parts, and then making use of the Ricci identity and the second-order Bianchi identity (given by $\del{\nu}\delta^2 G^{\mu\nu}=-\delta\Gamma^\mu_{\beta\gamma}\delta G^{\beta\gamma}-\delta\Gamma^\beta_{\beta\gamma}\delta G^{\mu\gamma}$, where $\delta\Gamma$ is the linear correction to the background Christoffel symbol) we arrive at
\begin{align}\label{force regular}
\ddot{z}_{\scriptscriptstyle{(1)}}^\mu &= R^\mu{}_{\nu\rho\sigma}\un{0}^\nu\un{0}^\rho\zn{1}^\sigma\nonumber\\
&\quad-\tfrac{1}{2}(g^{\mu\nu}+\un{0}^\mu \un{0}^\nu)(2\hmn{\nu\rho;\sigma}{1} -\hmn{\rho\sigma;\nu}{1})\un{0}^\rho\un{0}^\sigma,
\end{align}
where $\ddot{z}_{\scriptscriptstyle{(1)}}^\mu\equiv \del{\un{0}}\del{\un{0}}\zn{1}^\mu$. This equation describes the spatial deviation of the true worldline away from the reference geodesic $\zn{0}^\mu$;\footnote{I have assumed for simplicity that $\zn{1}^\mu$ is perpendicular to $\un{0}^\mu$. In the general case, the left-hand side of the equation reads $(g^{\mu\nu}+\un{0}^\mu\un{0}^\nu)g_{\nu\rho} \ddot{z}_{\scriptscriptstyle{(1)}}^\rho$, such that the result still describes only the spatial part of the acceleration.} because it is an equation for a deviation vector, it includes a term proportional to the Riemann tensor.

While the basic idea of this approach is valid and rigorous, it is unsatisfactory because of its limited realm of validity. For example, in a typical EMRI orbit, the radial coordinate $\zn{0}^r$ on the particle's leading-order worldline will be of order $\e^0$ for all time, while the deviation vector $\zn{1}^r$ will grow as $\an{1}\cdot(t-t_0)$. This means that the expansion of $\gamma$ is valid only on timescales $t\sim\mathcal{R}$: after a radiation-reaction time $t\sim\mathcal{R}/\e$, the ``correction" $\e\zn{1}$ will be of the same order as the leading-order term $\zn{0}$. In other words, the expansion is not uniform in time. And once we commit ourselves to a nonuniform expansion, we must restrict the entire problem to a bounded time-interval $[t_i,t_f]$. Within this fixed interval, the expansion is valid in the sense that we can guarantee that our approximation will be accurate to any given numerical value by making $\e$ sufficiently small; on an unbounded, or a generically $\e$-dependent interval, this statement would not hold true. The restriction to this bounded region has several important consequences. Most obviously, as previously stated, we are specifically interested in large changes that occur on the time-interval $\sim \mathcal{R}/\e$---such as the particle's slow inspiral in an EMRI. Thus, the entire expansion scheme fails on the timescale of interest.

The restriction to a bounded time-interval also restricts the formalism in an important way: since the expansion of the Einstein equation is valid only on a bounded region, the solution to it cannot (necessarily) be expressed in terms of an unbounded past history. In other words, the boundedness of the domain effectively forces us to cast the problem in an initial value formulation from the beginning. This means that we cannot express the force purely in terms of the usual tail integral; as soon as one writes down the solution as an integral over the entire past history, one assumes that one's expansion is globally valid, rather than just locally valid.\footnote{This point seems to have been missed in Ref.~\cite{Gralla_Wald}.} We can easily see this from the following argument: The correction terms $\zn{n}$ grow large not only for times far in the future of $t_0$, but also for times far in the past of $t_0$. Hence, at any time $t$, the difference between the tail as calculated on $\gamma\coeff{0}$ and the tail as calculated on $\gamma\coeff{0}+\e\gamma\coeff{1}$ will differ by a significant amount, given by $\left|\tail[\gamma\coeff{0}] -\tail[\gamma\coeff{0} +\e\gamma\coeff{1}]\right|\sim \e^2\int_{-\infty}^t \zn{1}^{\alpha'}(t')\partial_{\alpha'} G(x,\zn{0}(t'))dt'$; since $\zn{1}$ grows with $|t-t_0|$, the difference between the two tails appears to be potentially infinite. It is quite likely that the decay of the retarded Green's function would ameliorate this divergence in any case of interest. But there is no obvious reason for extending the domain of the solution beyond the domain of validity of the expansion.

Hence, at each order, the integral over the source must be cut off at the initial time $t=t_i$, and the remainder of the tail must be replaced by Cauchy data on that initial timeslice. A consequence of this is that the self-force is not naturally expressed in terms of a tail integral over an infinite past history; instead, it is more naturally expressed in terms of a purely local regular field, defined as the retarded field minus a certain local, singular part, in the manner of Detweiler and Whiting. Besides making the solution valid, this also has the advantage of expressing the force in terms of local quantities, with no reference to the past history of the particle; this is useful for a numerical integration in the time domain---and for developing a two-timescale method, as I will discuss presently.

\subsection{Singular expansion}\label{singular_expansion_point_particle}
Given the limitations of a regular expansion, let us now consider a singular expansion. In effect, this expansion will formalize the non-systematic procedure presented in Sec.~\ref{nonsystematic} and provide a systematic justification of the relaxation of the Lorenz gauge in the first-order problem.

Recall that our basic goal is to find a pair $(\gamma,h)$ satisfying Einstein's equation. In a regular expansion, both the worldline and the metric perturbation are expanded in the limit of small $\e$. In the singular expansion we shall now consider, the worldline is held fixed. To find each term in this expansion, I seek an expansion of both the exact Einstein equation and the exact equation of motion that can be solved with this fixed worldline. Hence, I decompose the metric as
\begin{equation}\label{worldline_decomposition}
\exact{g}_{\mu\nu}(x,\e)=g(x)+h_{\mu\nu}(x,\e;\gamma),
\end{equation}
where the semicolon is used to separate ordinary coordinate-dependence from functional dependence; when leaving the dependence on coordinates implicit, I will write, e.g., $h_{\mu\nu}[\gamma]$. I assume that the perturbation can be expanded while holding this functional dependence fixed:
\begin{equation}\label{worldline_expansion}
h_{\mu\nu}(x;\gamma)=\sum_{n=1}^N\e^n\hmn{\mu\nu}{{\it n}}(x;\gamma)+\order{\e^{N+1}},
\end{equation}
where each term $\hmn{\mu\nu}{{\it n}}$ is a functional of the true worldline $\gamma$ but is nevertheless of order unity. Note that the approximation scheme fails---becoming both inaccurate and internally inconsistent---if any of these coefficients are found to grow larger than order unity. Substituting Eq.~\eqref{worldline_decomposition} into the Einstein equation, we arrive at
\begin{align}\label{not wave-like}
\delta G^{\mu\nu}\big[h\big] &= -G^{\mu\nu}+8\pi\e T^{\mu\nu}[\gamma]+8\pi\e\delta T^{\mu\nu}[h,\gamma]\nonumber\\
&\quad-\delta^2G^{\mu\nu}\big[h\big]+...
\end{align}

Now, one might seek to expand and solve this equation at fixed $\gamma$. Unfortunately, if one substitutes the expansion \eqref{worldline_expansion} into Eq.~\eqref{not wave-like}, then one arrives at the linear equation $\delta G^{\mu\nu}\big[\hmn{}{1}\big]=8\pi T^{\mu\nu}[\gamma]$. And since the linearized Bianchi identity holds for any $\hmn{}{1}$, this equation implies that $\del{\nu}T^{\mu\nu}[\gamma]=0$---and hence that $\gamma$ must be a geodesic. Thus, this method immediately fails. To avoid such a problem, I recast the full equation in a more useful form by imposing the Lorenz gauge condition on the entire perturbation $h$ (rather than on any individual term in its expansion):\footnote{I assume that this condition can be imposed up to any desired order of accuracy. At second order, it can be imposed via a gauge transformation generated by a gauge vector $\xi$ satisfying $\Box\xi^\alpha=L^{\alpha}[h]+S^\alpha[h,\xi]+\order{\e^3}$, where $S^\alpha$ is quadratic in $h$ and $\xi$. At higher order, higher-order terms would appear on the right-hand side. Beginning with $h$ in an arbitrary gauge, this weakly nonlinear wave equation can be solved iteratively for $\xi$, using the same methods as those used to solve for $h$. Note that in the singular expansion presented here, gauge transformations have the coordinate form $x^\alpha\to x^\alpha-\e\xi^\alpha(x,\gamma)+\order{\e^2}$.}
\begin{equation}
L_{\mu}\big[h\big]=0.
\end{equation}
This transforms the Einstein equation into the weakly nonlinear wave equation
\begin{align}\label{wave-like}
E^{\mu\nu}\big[\bar h\big] &= 2G^{\mu\nu}-16\pi\e T^{\mu\nu}[\gamma]-16\pi\e\delta T^{\mu\nu}[h,\gamma]\nonumber\\
&\quad+2\delta^2G^{\mu\nu}\big[h\big]+...
\end{align}
Unlike Eq.~\eqref{not wave-like}, this equation can be solved for an arbitrary worldline. Its form is essentially identical to the relaxed Einstein equation, which forms the basis of most post-Minkowski expansions \cite{relaxed_EFE1,relaxed_EFE2}. Both equations are ``relaxed" in the sense that they can be solved without specifiying the motion of the source; the motion of the source is determined by the gauge condition. Also, in both cases, nonlinearities are treated as source terms for a hyperbolic wave operator; this means that corrections to the null cones are incorporated into the perturbations, rather than into the characteristics of the wave equation.

Substituting the expansion \eqref{worldline_expansion} into the wave-like equation \eqref{wave-like} and solving at fixed $\gamma$ now yields the sequence of equations
\begin{align}
G^{\mu\nu} &= 0,\\
E^{\mu\nu}\big[\hbarmn{}{1}\big] & = -16\pi T^{\mu\nu}[\gamma], \label{wave_first}\\
E^{\mu\nu}\big[\hbarmn{}{2}\big] & = -16\pi\delta T^{\mu\nu}\big[\hmn{}{1},\gamma\big]+2\delta^2 G^{\mu\nu}\big[\hmn{}{1}\big] \label{wave_second}\\
&\ \ \vdots\nonumber
\end{align}

Along with this expansion of the Einstein equation, I seek an analogous expansion of the matter equation of motion $\exact{a}_\mu=0$:
\begin{equation}
a_\mu(t,\e) = \an{0}_\mu(t)+\e\an{1}_\mu(t;\gamma)+\order{\e^2}.
\end{equation}
This is an expansion of a function of time along the fixed worldline. Of course, it must also follow from the Bianchi identity, and it must therefore also follow from the Lorenz gauge condition. Hence, I seek an expansion of the gauge condition that is equivalent to the expansion of the acceleration. Substituting the expansions of the perturbation and the acceleration into the exact gauge condition $L_\mu[h]=0$, and solving at fixed $\gamma$, yields the sequence of equations
\begin{align}
L\coeff{0}_\mu\big[\hmn{}{1}\big] &=0, \label{gauge_expansion 1}\\
L\coeff{1}_\mu\big[\hmn{}{1}\big] &= -L\coeff{0}_\mu\big[\hmn{}{2}\big],
 \label{gauge_expansion 2}\\
&\ \ \vdots\nonumber
\end{align}
where $L\coeff{0}[f]\equiv L[f]\big|_{a=\an{0}}$, $L\coeff{1}[f]$ is linear in $\an{1}$, $L\coeff{2}[f]$ is linear in $\an{2}$ and quadratic in $\an{1}$, and so on. Note the meaning of the notation here: $\hmn{}{\emph{n}}[\gamma]$ is the coefficient of $\e^n$ at fixed $\gamma$---including, in particular, at fixed values of the acceleration on $\gamma$---while for $f\sim1$, $L\coeff{\emph{n}}[f]$ is the coefficient of $\e^n$ in an expansion that incorporates an expansion of $a$ on $\gamma$.

This expansion of the gauge condition serves to determine increasingly accurate equations of motion for the single fixed worldline $\gamma$. Alternatively, it might be thought of as an iterative improvement of the choice of worldline. In either case, in any concrete calculation, the metric perturbations would be treated as functionals of the worldline described by the highest-order equation of motion available.

The solution to the first-order wave equation, \eqref{wave_first}, is given by
\begin{equation}
\hmn{\mu\nu}{1}[\gamma] = 4m\int_\gamma \bar G_{\mu\nu\mu'\nu'} u^{\mu'}u^{\nu'}dt',\label{1st_funct}
\end{equation}
where $t'$ and $u^{\mu'}$ are, respectively, the proper time and four-velocity on $\gamma$. Note that this is the ``usual" solution obtained by solving the linearized wave equation, as in Sec.~\ref{nonsystematic}. The acceleration of the true worldline is determined from Eq.~\eqref{gauge_expansion 1}:
\begin{align}
0 &= L\coeff{0}_\mu\big[\hmn{}{1}\big]\nonumber\\
 & = 4m\int_\gamma G_{\mu}{}^{\mu'}\an{0}_{\mu'}dt,
\end{align}
which implies that $\an{0}_\mu=0$. It does not imply that $a_\mu=0$, however.

Proceeding to second order, the solution to Eq.~\eqref{wave_second} is
\begin{align}
\hmn{\mu\nu}{2}[\gamma] &= 2\int_{\gamma} \bar G_{\mu\nu\mu'\nu'}u^{\mu'}u^{\nu'}(u^{\rho'}u^{\sigma'} -g^{\rho'\sigma'})\hmn{\rho'\sigma'}{1}dt' \nonumber\\
&\quad-\frac{1}{2\pi}\int \bar G_{\mu\nu\mu'\nu'}\delta^2 G^{\mu'\nu'}dV'.\label{2nd_funct}
\end{align}
Imposing the gauge condition \eqref{gauge_expansion 2}, making use of Eq.~\eqref{Green1} and the second-order Bianchi identity, and integrating by parts determines the acceleration to order $\e$:
\begin{equation}
\an{1}_\mu = -\tfrac{1}{2}\!\!\left(g_{\alpha}{}^{\beta}\!+\!u_\alpha u^\beta\right)\!\!\left(2\hmn{\beta\gamma;\delta}{1} -\hmn{\delta\gamma;\beta}{1}\right)\!u^\gamma u^\delta\Big|_{a=0}.
\end{equation}
Note that the right-hand side of this equation is evaluated on the worldline, and once evaluated, it contains a term proportional to $-\dot{a}_\mu$, corresponding to the antidamping phenomenon discovered by Havas \cite{damping} (as corrected by Havas and Goldberg \cite{damping2}). However, my assumed expansion of the acceleration has forced the right-hand side to be evaluated for $a=\an{0}=0$, which serves to automatically yield an ``order-reduced" equation with no higher-order derivatives (assuming, of course, that a time-derivative does not change the order of a term in the expansion).

From this calculation, we see that this method yields an equation of motion that agrees with the expansion given in Eq.~\eqref{eq_motion}. In both cases, the equation of motion applies to the actual worldline $\gamma$, not to a correction to a reference geodesic. Combined with the first-order perturbation given in Eq.~\eqref{1st_funct}, the equation of motion defines a self-consistent solution to the Einstein equation, up to errors of order $\e^2$ on a timescale $\mathcal{R}/\e$; combined with the sum of the first- and second-order perturbations, it defines a solution accurate up to errors of order $\e^3$ on a timescale of order $\mathcal{R}$.

One should note two important facts about the results just derived. First, from these results, one can easily derive those of the regular expansion, given in Eqs.~\eqref{force regular} and \eqref{1st order regular}, by expanding the worldline and following the usual steps involved in deriving the geodesic deviation equation. Second, while the Lorenz gauge is especially useful for finding the metric perturbation in the singular expansion, it is not essential for finding the equation of motion, which could have equivalently been found from the conservation of the source $8\pi\e(T^{\mu\nu}+\delta T^{\mu\nu}) -\delta^2G^{\mu\nu}$.

Beyond these specifics, one should also note the broad similarity between this singular expansion and a post-Minkowksian expansion (in particular, the fast-motion approximation \cite{relaxed_EFE1}): the split of the Einstein equation into a wave equation and a gauge condition, the iterative solution to the wave equation in terms of an arbitrary worldline, and use of the gauge condition to fix the worldline. Given these commonalities and the many successes of the post-Minkowskian expansion, one might hope that the singular expansion suggested here will be equally successful in more general contexts. Note, however, that the character of the solutions given in Eqs.~\eqref{1st_funct} and \eqref{2nd_funct} is significantly different in a curved background than in a flat one, since curvature creates caustics in null cones and allows gravitational perturbations to propagate within, not just on, those cones. These complications suggest that the integrals in Eqs.~\eqref{1st_funct} and \eqref{2nd_funct} might much more easily display secular growth in a curved spacetime. Since the expansion is consistent only in the absence of such secular behaviour, it may be valid only in certain spacetimes and with certain initial conditions.

Perhaps a more significant difference between the above expansion and a post-Minkowskian one is the choice of gauge. The harmonic gauge used in post-Minkowksi expansions can be imposed as an exact coordinate condition ${}^{\exact{g}}\Box x^\alpha=0$ on the manifold of the exact solution; as long as the exact solution admits these coordinates, the gauge condition $\partial_\mu h^{\mu\nu}=0$ is automatically imposed on the entire metric perturbation, rather than on any particular order in its expansion. The Lorenz gauge used here, on the other hand, can be imposed only after decomposing the metric into a background plus perturbation, and it is typically formulated only in terms of a first-order perturbation; there is no proof that it can be imposed on the total perturbation to all orders (although it can seemingly be imposed to any desired order by solving a weakly nonlinear wave equation). Despite these caveats, and independent of the analogy with post-Minkowskian theory, the expansion discussed here has the concrete advantage of offering a systematic justification of the self-consistent solution $\eqref{1st_funct}$ and higher order corrections to it.

Other arguments have been made in favor of using the self-consistent solution \eqref{1st_funct} rather than the regular solution \eqref{1st order regular}. The simplest argument is one based on  ``adiabaticity": because the acceleration is very small, the true worldline deviates only very slowly from a geodesic, so the self-consistent solution can be ``patched together" from a collection of regular solutions. This argument has been made frequently in the past, most recently by Gralla and Wald~\cite{Gralla_Wald}. While it is intuitively reasonable, one must keep in mind its most basic assumption, which is that the (covariant derivative of the) tail integral as calculated over a geodesic $\gamma\coeff{0}$ is nearly identical to the tail integral as calculated over the true worldline $\gamma$. This assumption is a very strong one, since the tail integral potentially contains highly nonlocal contributions \cite{quasilocal, quasilocal2, quasilocal3}. As such, it is probably true only in a very particular set of situations, such as, for example, an EMRI system in the adiabatic limit \cite{Hughes_adiabatic}, in which the geodesic motion is periodic and the particle executes a large number of orbits before deviating noticeably from the geodesic.\footnote{In fact, the typical derivation of the MiSaTaQuWa equation, which begins with a source moving on a geodesic but ends with a self-force, has sometimes been called an adiabatic approximation \cite{Mino_expansion2}. However, others \cite{quasilocal} have suggested that the substitution $\tail[\gamma]\to\tail[\gamma\coeff{0}]$ in the equation of motion is more generally valid.} Obviously, we would like the self-force to be valid in more general regimes---for example, in the final moments of plunge in the EMRI orbit.

A more systematic method of ``patching together'' regular expansions has been devised by Hinderer and Flanagan \cite{Hinderer_Flanagan}. They perform a two-timescale expansion of the Einstein equation and the equation of motion, tailoring their approach to EMRI systems. This expansion captures both the fast dynamics of orbital motion and the slow dynamics of the particle's inspiral and the gravitational backreaction on the background spacetime. At each value of the ``slow time," one can perform a regular expansion, from which the self-force can be derived as discussed above, using the actual field and the position and momentum of the particle as initial data---this is one reason why the force as derived in a regular expansion should be expressed in terms of the actual field, rather than the tail integral over the entire past history of a geodesic. The slow dynamics are assumed to be irrelevant over the timescale of the regular expansion, such that the slow time variable appears as a fixed parameter during the expansion. By letting the slow time evolve continuously, a series of regular expansions are automatically patched together to arrive at a self-consistent evolution.

Another expansion has been devised by Mino~\cite{Mino_expansion1, Mino_expansion2}. He begins with an expansion similar to the one presented here, but he then performs a second expansion of each $\hmn{}{n}$, in such a way that each term in the expansion of $\hmn{}{n}$ depends only on information from the instantaneously tangential worldline governed by the $(n-1)$th-order self-force; that is, each term in the expansion of the leading-order perturbation $\hmn{}{1}$ depends only on the geodesic instantaneously tangential to the true worldline, $\hmn{}{2}$ depends only on the worldline governed by the first-order self-force, and so on.

The methods developed in this paper are intended to complement the above approaches. It is hoped that they will be valid in more general contexts, though more detailed studies would be required to bear out that hope.

\section{Preliminary analysis II: the motion of an extended body}\label{extended_body}
Since the singular expansion presented in the previous section is based on an exact point particle source, it is ill-behaved beyond first order. As such, we must now consider methods of accounting for the extension of an asymptotically small body. Specifically, we must consider how to formulate an asymptotic expansion in which a representative worldline for the small body is held fixed.

Perhaps the most obvious approach is to work with a body of arbitrary size and then take the limit as that size becomes small. Such a method has been used by Harte \cite{Harte} in deriving self-force expressions, following the earlier work of Dixon \cite{Dixon}. However, in this article I will be interested only in approaches that treat the body as asymptotically small from the start. The simplest means of doing so is to treat the body as an \emph{effective} point particle at leading order, with finite size effects introduced as higher-order effective fields, as done by Galley and Hu \cite{Galley_Hu}. However, while this approach is computationally efficient, allowing one to perform high-order calculations with (relative) ease, it requires one to introduce methods such as dimensional regularization and mass renormalization in order to arrive at meaningful results. Because of these undesirable requirements, I will not consider such a method here.

\subsection{Point particle limits}\label{point-particle limits}
In order to move from an exactly pointlike body to an asymptotically small one, we must consider a family of metrics $g(\e)$ containing a body whose mass scales as $\e$ in the limit $\e\to0$. (That is, $m\sim\e\mathcal{R}$.) If each member of the family is to contain a body of the same type, then the size of the body must also approach zero with $\e$. The precise scaling of the size with $\e$ is determined by the type of body, but this precise scaling is not generally relevant.\footnote{However, the calculations in this paper require the existence of a vacuum region of radius $m\ll r\ll\mathcal{R}$ around the body. If the body is not sufficiently compact, then this region will not exist and my calculation will not apply. Likewise, my calculation fails when a body becomes tidally disrupted.} What \emph{is} relevant is the ``gravitational size"---the length scale relevant to the metric outside the body---and this size always scales linearly with the mass. If the body is compact, as is a neutron star or a black hole, then its gravitational size is also its actual linear size.

Point particle limits such as this have been used to derive equations of motion many times in the past, including in derivations of geodesic motion at leading order \cite{Infeld,Geroch_particle1,Geroch_particle2} and in constructing post-Newtonian limits \cite{Futamase_particle1, Futamase_particle2, Futamase_review}. In general, deriving corrections to geodesic motion requires considering two types of point particle limits: an \emph{outer limit}, in which $\e\to0$ at fixed coordinate values; and an \emph{inner limit}, in which $\e\to0$ at fixed values of $\tilde r\equiv r/\e$, where $r$ is a radial coordinate centered on the body. In the outer limit, the body shrinks toward zero size as all other distances remain roughly constant; in the inner limit, the small body remains a constant size while all other distances blow up toward infinity. Thus, the inner limit serves to ``zoom in" on a small region around the body. The outer limit can be expected to be valid in regions where $r\sim \mathcal{R}$, while the inner limit can be expected to be valid in regions where $\tilde r\sim \mathcal{R}$, though both of these regions can be extended into larger domains.

These two limits can be utilized in multiple ways. For example, the outer limit can be used to examine the effect of the small body on the external spacetime, while the inner limit can be used to study the effect of the external spacetime on the metric of the small body (as studied in Refs.~\cite{Manasse, Thorne_Hartle, Eric_tidal}, for example). What is of interest for this paper is how the two limits mesh in a \emph{buffer region}---a region in which $m\ll r\ll \mathcal{R}$. The metric in this region will determine the motion. To understand this, note that the buffer region, at fixed time, is approximately flat, since it is simultaneously in the asymptotic far zone of the body (because $r\gg m$) and in a small local patch in the external spacetime (because $r\ll\mathcal{R}$). Thus, in the buffer region, the linear momentum of the small body, or some other measure of motion, can be defined. Speaking roughly, an equation for the derivative of this linear momentum will then provide an equation of motion for the body.

We can consider two basic methods of deriving equations of motion in the buffer region. The first method is that of matched asymptotic expansions, in which one approximation is constructed using the inner limit, another using the outer limit, and then any undetermined functions are fixed by insisting that the two approximations are equal to one another in the buffer region. The second method foregoes an explicit calculation of an approximation in either the inner or outer limit (or both), instead working entirely in the buffer region and using some local definition of the motion of the body. Although both make use of inner and outer expansions, the two methods are logically distinct. However, both methods have sometimes been referred to as the method of matched asymptotic expansions (\emph{e.g.}, in Ref.~\cite{Kates_motion}).

D'Eath was the first to apply these methods to the problem of motion in General Relativity. He used matched asymptotic expansions to show that at leading order, a rotating black hole moves on a geodesic of the external spacetime \cite{DEath, DEath_paper}. Since D'Eath's pioneering work, these methods have been used in many contexts: to show that an arbitrarily structured body follows a geodesic \cite{Kates_motion}; to show that a charged body follows a worldline governed by the Lorentz force law \cite{Kates_Lorenz_force}; to derive post-Newtonian equations of motion \cite{Kates_PN, Futamase_particle1, Futamase_particle2, Futamase_review, PN_matching}; to derive general laws of motion due to the coupling of the body's multipoles with those of the external spacetime \cite{Thorne_Hartle}; and most pertinently, to derive the gravitational self-force \cite{Mino_Sasaki_Tanaka, Eric_review, Detweiler_review, Fukumoto, Gralla_Wald}. These derivations of the gravitational self-force will occupy the remainder of this section.

Let us first consider the earliest such derivation, performed by Mino, Sasaki, and Tanaka \cite{Mino_Sasaki_Tanaka}, and in slightly different manners by Poisson \cite{Eric_review} and Detweiler \cite{Detweiler_review}. These derivations take the small body to be a Schwarzschild black hole (with the hope that more general bodies would obey the same equation of motion), such that in the inner limit the exact metric $\exact{g}$ can be approximated by $\exact{g}=g_B(\tilde r)+H(\tilde r)+\order{\e^2}$, where the internal background metric $g_B$ is the metric of the isolated black hole, and $H(\tilde r)$ consists of tidal perturbations. In the outer limit, the metric is written as $\exact{g}=g+h[\gamma]+\order{\e^2}$, where $g$ is an arbitrary vacuum metric and $h[\gamma]$ is the perturbation due to a point particle traveling on a worldline $\gamma$. Expanding the external metric in normal coordinates centered on the worldline, expanding the internal metric for $r\gg m$, and insisting that the results of these expansions are identical, then determines an equation of motion for $\gamma$.

Mino, Sasaki, and Tanaka \cite{Mino_matching} later used a similar method to determine an equation of motion for a small Kerr black hole; they followed Thorne and Hartle's \cite{Thorne_Hartle} approach of defining the spin and angular momentum of the body as an integral over a closed spatial surface in the buffer region, and they derived an equation of motion by combining these definitions with the assumed point particle perturbation in the external spacetime.

Allow me to more precisely state the underlying logic of these derivations, which is as follows: Suppose there exists a metric $\exact{g}$ such that (1) in a region $\mathcal{R}_1$, $\exact{g}$ is well approximated by the metric of a tidally perturbed black hole, (2) in a region $\mathcal{R}_2$, $\exact{g}$ is well approximated by the metric of some vacuum spacetime as perturbed by a point-like source moving on a worldline $\gamma \subset \mathcal{R}_1$, and (3) the regions $\mathcal{R}_1$ and $\mathcal{R}_2$ overlap. Then the worldline $\gamma$ is governed by the MiSaTaQuWa equation. Alternatively, a weaker formulation might be stated as follows: the approximate solutions to the Einstein equation given by $g+h[\gamma]$ and $g_B+H$, as defined above, can be combined to form a global approximate solution if and only if $\gamma$ is governed by the MiSaTaQuWa equation.\footnote{In actuality, the method of matched asymptotic expansions, as it has been used in derivations of the self-force, provides significantly weaker results than this, because at first order in $\e$ the relationship between points in $\mathcal{R}_1$ and points in $\mathcal{R}_2$ is not unique, which leads to a non-unique result for the acceleration~\cite{perturbation_techniques}.}

There are several problems with this approach. First, it suffers from the same problem described in the previous section: since the point particle solution solves the linearized Einstein equation only if the point particle travels on a geodesic, a non-systematic gauge relaxation must be invoked. Second, it does not offer any way to go beyond first order, since it provides no means of determining the external perturbations (though see Refs.~\cite{Eran_field, Eran_force} for an extension to second order). Third, it is a somewhat weak result, with many \emph{if}s in its construction. Of course, it does provide an improvement over the earliest point particle derivations, since it derives the self-force from the consistency of the field equation and makes no questionable assumptions about the behavior of singular quantities.

More recent derivations have been performed by Fukumoto, Futamase, and Itoh \cite{Fukumoto} and Gralla and Wald \cite{Gralla_Wald}. These derivations work entirely in the buffer region, rather than using matching; they do not assume that the external perturbation is that of a point particle at leading order; and they do not restrict the small body to be a Schwarzschild black hole. Fukumoto \emph{et al.}, following the work of Futamase \cite{Futamase_particle1, Futamase_particle2, Futamase_review} and Thorne and Hartle \cite{Thorne_Hartle}, defined the linear momentum of the body as an integral in the buffer region and derived the acceleration by simply differentiating this linear momentum. While this derivation is quite simple relative to most others, it contains at least one questionable aspect: it relies on an assumed relationship between the body's linear momentum, as defined in the buffer region, and the four-velocity of the body's worldline (justified by an analogy with post-Newtonian results).

Gralla and Wald explicitly restricted themselves to a regular expansion, defining the acceleration of the body via a regular expansion of the body's worldline, roughly as described in the previous section. The only questionable aspect of their derivation is that it writes the solution to the first-order Einstein equation as an integral over the past history of the leading-order, geodesic worldline $\gamma\coeff{0}$, and it expresses the force in terms of the tail integral $\tail[\gamma\coeff{0}]$. As discussed above, this is not obviously justified: in order to remain consistently within the domain of validity of a regular expansion, the tail should be cut off at some finite past time and complemented with an integral over an initial data surface. The practical drawback of this derivation is that it is obviously limited to short timescales, as discussed above. While a regular expansion such as this can be used to derive the self-force and then incorporated into a two-timescale expansion, my goal here is to provide a self-consistent approach in which the worldline is never treated as a geodesic.

\subsection{Definitions of the worldline}
While the derivations described above are increasingly satisfactory, none of them have satisfactorily defined the worldline of the asymptotically small body. To see this, we must examine the various definitions of this worldline.

Most of these definitions are in terms of the outer limit \cite{Infeld, DEath, DEath_paper, Geroch_particle1, Geroch_particle2, Gralla_Wald}. At each value of $\e$, the body can be surrounded by a worldtube $\Gamma_\e$ that has a radius which vanishes in the limit $\e\to0$; the worldline of the body is then defined as the limit $\Gamma_{\e=0}$ of these worldtubes, which defines a curve in the limiting spacetime $\exact{g}(\e=0)$. Kates generalized this to the case when the limit $\e\to0$ is singular \cite{Kates_motion}.

These definitions are extremely problematic, because the worldline seems to naturally emerge only when the body is exactly point-like, which is the case only in the limiting spacetime defined by $\e=0$. But in this spacetime, since $\e=0$, the worldline will naturally be $\e$-independent; any $\e$-dependence would automatically be ``pure gauge," such that the self-force could be set to zero over the entire domain of the regular expansion. How, then, can the worldline be defined such that it can accurately and meaningfully reflect the motion of the body for $\e>0$? If we reject the use of small ``corrections" to a worldline, represented by a deviation vector, how can we find the worldline that the deviation vector ``points" to?

One possible definition has been suggested by Futamase \cite{Futamase_particle1, Futamase_particle2, Futamase_review, Fukumoto}. Rather than defining the worldline as the limit of a family of worldtubes of radius $\sim\e$, he defines the worldline as the curve that remains within every such tube as $\e\to0$. One can easily show that if such a curve exists, then it is unique. However, one can just as easily show that, in general, such a curve exists only for a short time: for any two values of $\e$, say $\e_1$ and $\e_2$, the interiors of the worldtubes $\Gamma_{\e_1}$ and $\Gamma_{\e_2}$ will intersect only for a brief time period $t\lesssim\mathcal{R}$. Hence, this definition does not improve upon the previous one. It also has the disadvantage that it cannot apply to small black holes.

Yet another approach is to define the worldline implicitly. Consider how this is realized in matched asymptotic expansions (see Ref.~\cite{PN_matching} for the clearest example), in which the worldline is defined roughly as follows: if the perturbed metric of the external spacetime near a worldline $\gamma$ is equivalent (up to diffeomorphism) to the perturbed metric of the small body, then $\gamma$ is said to be the worldline of the body. In practice, this means that the worldline is defined operationally by the fact that, in the buffer region, the external metric in some coordinates centered on $\gamma$ is identical to the internal metric in some mass-centered coordinates (\emph{i.e.}, coordinates in which the internal metric has no mass dipole---see Ref.~\cite{Thorne_Hartle} for a discussion of mass-centered coordinates defined in the buffer region). More precisely, suppose we are given an inner expansion $g_I(\e)=g_B+H$ in some ``local" coordinates $X^\mu$ on a manifold $\man_I$, and an outer expansion $g_E(\e)=g+h$ in some ``global" coordinates $x^\mu$ on a manifold $\man_E$.\footnote{The subscript $E$ is derived from the title ``external spacetime'' given to the background metric $g$ on $\man_E$. Note that two manifolds are generically required. Consider, for example, the case of a small black hole orbiting a large black hole. The manifold $\man_I$ possesses a singularity at the ``position" of the small black hole but is otherwise smooth, while the manifold $\man_E$ possesses a singularity at the ``position" of the large black hole but possesses a smooth worldline where the small black hole should be.} For example, in an EMRI consisting of a small Schwarzschild black hole orbiting a large Kerr black hole, the coordinates $x^\mu$ might be the Boyer-Lindquist coordinates of the large black hole, and the coordinates $X^\mu$ might be the Schwarzschild coordinates of the small black hole. We can always transform $g_E$ into a coordinate system (\emph{e.g.}, Fermi coordinates) centered on a worldline $\gamma\subset\man_E$, via a map $\trans:\man_E\to\man_E$. The worldline $\gamma$ is then defined to be that of the small body if there exists a \emph{unique} map $\map:\man_E\to\man_I$ in a buffer region $\mathcal{R}\subset\man_E$ such that $(\map\circ\trans)_*(x^\mu)=X^\mu$ and $(\map\circ\trans)_*(g_E)=g_I$, where $f_*$ denotes the push-forward of $f$ and an equal sign indicates equality at the lowest common order.

So long as one restricts one's attention to approximate solutions of the Einstein equations, this operational definition seems to be valid. It will undoubtedly result in a metric that solves the Einstein equation to some specified order in some large region of spacetime. However, at first glance it might seem unlikely that the resulting approximate solution actually approximates any ``true" metric (that is, any exact solution to the Einstein equation), since it is not apparent how the $\e$-dependent worldline could arise from expanding an exact solution. If we wish the approximate metric to be an approximation to a true metric, rather than just an approximate solution, we must at least show that the approximation could plausibly be constructed from an exact solution. And this in turn means that we must relate our operational definition of the worldline to the asymptotic behavior of a family of exact solutions to the Einstein equation.

In the analysis of the point particle field equation presented in Sec.~\ref{point-particle}, I argued in favor of a singular expansion that holds the worldline of the particle fixed. In the case of a point particle, one can easily imagine performing such an expansion of an exact metric, since one can easily imagine that the exact solution could be written as a functional of the particle's worldline. But the situation seems somewhat obscure in the case of an asymptotically small body, for which a worldline seems to appear only when the size of the body becomes precisely zero. However, the matching calculation provides some insight into the problem. First, although the worldline is a curve in the external manifold $\man_E$, it need not be a curve in the manifold $\man$ on which the exact metric $\exact{g}$ lives; in fact, if the small body is a black hole, then there is obviously no such curve. ($\man$ might be thought of as the result of cutting out a portion of $\man_E$ around the worldline $\gamma$, then stitching part of the manifold $\man_I$ into the excised region.) Second, note that \emph{if} the outer expansion $g+h[\gamma]$, written in the coordinates $x^\mu$, and the inner expansion $g_B+H$, written in the coordinates $X^\mu$, are both approximations of the same exact solution, then they induce coordinates on the manifold $\man$. This means that the coordinate transformation $\map\circ\trans$ between $x^\mu$ and $X^\mu$ induces a transformation between the coordinates on the exact spacetime. And since $\trans$ is defined by a curve $\gamma$, this coordinate transformation is parametrized by that curve, even though $\gamma$ is not an actual worldline in $\man$. Thus, the metric in the external coordinates $x^\mu$ can naturally be written as a functional of $\gamma$. In this sense, the ``worldline" is simply a function that parametrizes the metric (\emph{c.f.} the discussion in Ref.~\cite{Racine_Flanagan}).

We can also consider this from a different angle. As shown by Sciama \emph{et al.}~\cite{Sciama}, any exact solution of the Einstein equation can be written in an integral formulation. Consider a bounded vacuum region $\Omega\subset\man$ with a boundary $\partial\Omega$. At any point $x$ in the interior of $\Omega$, the metric will satisfy
\begin{equation}\label{Sciama formal solution}
\exact{g}^{\alpha\beta}(x) = \int\limits_{\partial\Omega_\e}{}^\exact{g}\del{\sigma'} G^{\alpha\beta\nu'}{}_{\nu'}(x,x')dS^{\sigma'},
\end{equation}
where $G^{\alpha\beta\nu'}{}_{\nu'}(x,x')$ is a Green's function for the operator 
\begin{equation}
\exact{D}_{\mu\nu\rho\sigma} = \tfrac{1}{2}\exact{g}^{\alpha\beta}\exact{g}_{\mu(\rho}\exact{g}_{\sigma)\nu} {}^\exact{g}\del{\alpha}{}^\exact{g}\del{\beta} +\exact{R}_{\mu(\rho\sigma)\nu}.
\end{equation}
(The proof of the integral identity in Ref.~\cite{Sciama} is restricted to a convex normal neighbourhood of $x$, but for the sake of argument, assume that it is valid even if $\Omega$ extends beyond the normal neighbourhood.) Assume that in some region $\mathcal{U}$ around the body, a scalar field $r$ provides a measure of distance from the body. The region $\mathcal{U}$ need not include the body itself, but should have the topology $S^2\times[r_1,r_2]\times[t_1,t_2]$, where $r_2>r_1$, $t_2>t_1$, and $S^2$ is a spatial 2-sphere around the body. Now suppose that the ``inner" boundary of $\Omega$ is a timelike worldtube $\Gamma_\e\subset\mathcal{U}$ of fixed radius $r\equiv\rad\sim\e^p$, $0<p<1$, around the body. The surface $\Gamma_\e$ is parametrized by two angles $\theta^A$ ($A=1,2$) and by a time $t$. Thus, $\Gamma_\e$ is generated by a collection of timelike curves $\gamma_{(\rad,\theta)}:t\mapsto x^\alpha(t,\rad,\theta^A)$. Note that since $\rad$ is arbitrary within some interval, we can use it interchangeably with $r$, implying that $(t,\rad,\theta^A)$ defines a local coordinate system $X^\mu$ near the small body. The collection of maps $\gamma_{(\rad,\theta)}$ thus defines the coordinate transformation $\trans$ between these local coordinates and the global coordinates $x^\mu$.

The metric can be written as 
\begin{align}
\exact{g}^{\alpha\beta} &= \int\limits_{S^2}\int\limits_{\gamma_{(\rad,\theta)}}{}^\exact{g}\del{\sigma'} G^{\alpha\beta\nu'}{}_{\nu'}\sqrt{|\exact{g}'|}n^{\sigma'}dtd\theta^1d\theta^2 \nonumber\\
&\quad +\int\limits_{\partial\Omega -\Gamma_\e}{}^\exact{g}\del{\sigma'}G^{\alpha\beta\nu'}{}_{\nu'}dS^{\sigma'}.
\end{align}
For small values of $\e$, the radius $\rad$ of the tube is also small, so each of the curves $\gamma_\theta$ can be expanded about $\rad=0$. If $\gamma_{(\rad,\theta)}$ is sufficiently well behaved, this expansion is valid even if $x^\alpha(t,\rad=0)$ does not describe a timelike curve (or any curve) in $\man$. However, note that the integrand in the above integral will generically diverge at $\rad=0$, since the small body's contributution to the metric will contain terms diverging as $\rad^{-n}$; thus, the integrand itself cannot be naively expanded in powers of $\rad$ without carefully expanding it in powers of $\e$ at the same time. (Alternatively, one could perform such an expansion and introduce some regularization method afterward.) Nevertheless, in the limit of small $\e$, the metric outside of the tube will naturally be expressed as a functional of a single worldline $\gamma: t\mapsto x^\alpha(t,\rad=0)$. This curve can be made unique by demanding that the mass dipole of the body vanishes, up to some desired order, when calculated in the local coordinates $(t,\rad,\theta^A)$.

Based on these plausibility arguments, we can reasonably believe that an exact metric for a small body could naturally be expressed as a functional of an $\e$-dependent curve that represents the motion of the body. Hence, we can reasonably believe that a singular expansion in which the metric perturbations are treated as functionals of a fixed, $\e$-dependent worldline, can approximate an exact metric. Actually proving that the expansion in this paper approximates an exact solution would presumably require a monumental effort. However, as discussed above, the worldline of the body is uniquely defined in an operational sense, and the metric that depends on it will provide a (hopefully uniform) approximate solution to the Einstein equation, whether or not it provides an approximation to an exact solution.

\begin{figure}[tb]
\begin{center}
\includegraphics{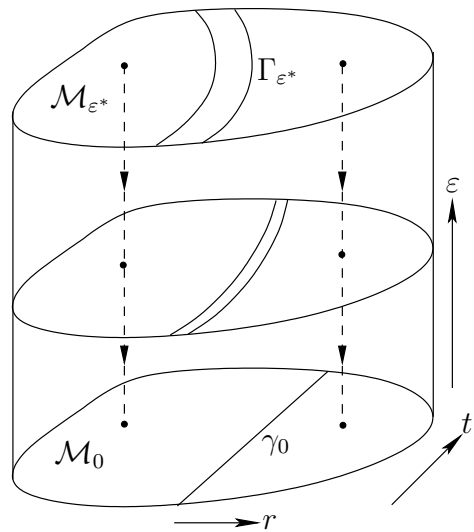}
\end{center}
\caption{Regular limit of a family of spacetimes. The dashed lines indicate a limit process in which $\e\to0$ at fixed coordinate values. As $\e\to0$, the worldtube of the body shrinks to zero size, leaving a geodesic remnant curve. The remnant curve lies in the manifold $\man_0$ defined by $\e=0$.} 
\label{regular_limit}
\end{figure}

\begin{figure}[tb]
\begin{center}
\includegraphics{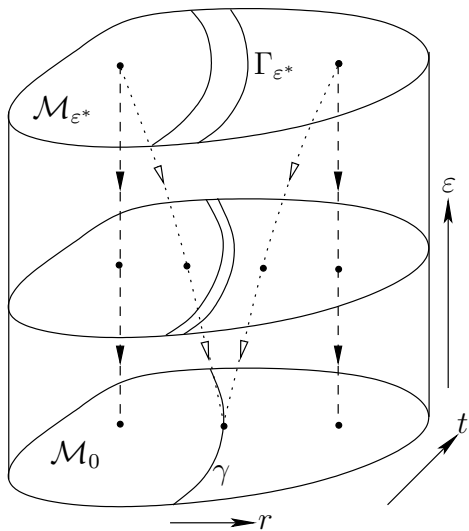}
\end{center}
\caption{Singular limits of a family of spacetimes. The dashed lines correspond to the external limit, which lets the body shrink to zero size but keeps its motion fixed. The dotted lines correspond to the internal limit, which keeps the size of the body fixed. The worldline lies in the manifold $\man_0$, but it does not correspond to the remnant curve defined by the regular limit; instead, it is allowed to have $\e$-dependence, and it is determined by the particular value of $\e$ (in this case, $\e=\e^*$) at which an approximate solution is sought.} 
\label{singular_limit}
\end{figure} 

Thus, the metric in the outer limit can be taken to be
\begin{equation}
\exact{g}(x,\e) = g(x)+h(x,\e;\gamma).
\end{equation}
In this expansion, the perturbations produced by the body are constructed about a fixed worldline determined by the particular value of $\e$ at which one seeks an approximation. Note that because the external background metric $g$ does not depend on $\gamma$, it is identical to the metric $\exact{g}(\e=0)$ in the regular limit, and the manifold $\man_E$ on which it lives is identical to the manifold $\man_0$.\footnote{Note that if $g$ is smooth, then it does not depend on $\gamma$. One can see this by referring to the formal solution \eqref{Sciama formal solution}. Taking part of the boundary to be a spatial surface $\Sigma$ that intersects the timelike worldtube $\Gamma$ and assuming that at leading order the interior of the worldtube is smooth, it follows that for the leading-order solution outside the tube, the integral over the tube can be replaced by an integral over a spacelike ``cap" that joins smoothly with $\Sigma$. Of course, since the regular limit $\e\to0$ might not exist at the ``position" $r=0$ of the small body (for example, if the body is a black hole), the manifold $\man_0$ may have the ``remnant" worldline $\gamma(\e=0)$ removed from it; but this discontinuity is obviously removable.} However, even if one writes the perturbation $h$ as an asymptotic series $\e\hmn{}{1}[\gamma] +\e^2\hmn{}{2}[\gamma]+...$, then the terms in the series will \emph{not} be equal to derivatives of $\exact{g}(\e)$ with respect to $\e$ at $\e=0$, since they depend on an $\e$-dependent worldline. Refer to Figs.~\ref{regular_limit} and \ref{singular_limit} for a schematic comparison between the approximations constructed with and without a fixed worldline.

Since the the inner limit is constructed in a local coordinate system around the body, which is effectively already centered on the worldline, it need not hold the worldline fixed. Instead, the worldline in the outer limit serves to fix the location at which the inner expansion is constructed. That is, the fixed worldline ensures a fixed relationship between the inner and outer expansions.

\subsection{Outline of the construction of a uniform and self-consistent approximation scheme}\label{outline}
The foregoing discussions have made two points clear: First, every derivation of the gravitational self-force has at least one questionable aspect. Some of these questionable aspects are fundamental---\emph{e.g.}, a reliance on an exact point particle source or an assumed form for the force---while others are relatively innocuous. However, at this point in time, nearly a dozen derivations have arrived at the same expression for the force---up to the ambiguity of whether the tail integral is to be evaluated over a geodesic or over the true worldline, which is not always clearly stated in the derivations. Thus, there can be little doubt, if there ever was, that the equation for the self-force is essentially correct.

Second, the discussions above have clearly shown that the heart of the problem lies in singular perturbation theory, which is signalled by the failure of a regular expansion due to the appearance of multiple length scales (the size $\sim \e$ of the body, the radius of curvature $\sim \e^0$ of the external spacetime, and the radiation-reaction time $\sim 1/\e$ over which the worldline deviates from geodesic motion). This is closely related to the ambiguity mentioned above: whether or not the tail integral is to be evaluated over the true past history of the body, or over a fictitious geodesic past. Straightforward analysis suggests that the integral \emph{must} be over a geodesic, even if, contradictorily, the motion is accelerated---but that is a faulty conclusion based on regular perturbation theory, which is valid only for short times, and which hence should never have included an integral over the entire past. Given these facts, the most obvious way to arrive at a self-consistent solution is to make use of singular perturbation techniques.

The derivation presented in the remainder of this paper is not intended to remedy the first condition mentioned above: it is not without questionable aspects of its own (though these are relatively few). What the derivation \emph{is} intended to do is utilize singular perturbation theory to construct a self-consistent approximation scheme. Because of its self-consistency, my scheme potentially provides a uniform approximation valid over times $t\lesssim\mathcal{R}/\e$. Its essential feature, which distinguishes it from previous methods, is that it uses expansions in which the worldline of the object is held fixed; while this idea was taken for granted in some earlier derivations, it has not been considered explicitly before now. Within the context of this overall approach, I consider two specific expansions: an inner expansion accurate at distances $r\sim\e$ from the body, and an outer expansion accurate at distances $r\sim 1$.

I will now outline the structure of my approximation scheme. I consider a family of metrics $\exact{g}(\e)$, where $\e>0$, in a large vacuum region $\Omega$ outside the body. Eventually, the parameter $\e$ will be identified with the mass $m_0$ of the body at some initial time. I will ensure, by construction, that the expansion is an asymptotic solution to the Einstein equation; I will hope, based on the plausibility arguments offered in the previous section, that the expansion is also an asymptotic approximation to an exact solution. As a technicality, I assume that all quantities have been rescaled by the infimum of the external length scales in $\Omega$, such that we can meaningfully speak of the mass of the body, or a radial coordinate near the body, being small or large relative to unity.

I choose $\Omega$ to lie outside a worldtube $\Gamma$ surrounding the body. The tube's radius $\rad$ is chosen to satisfy $\e\ll\rad\ll 1$; in other words, $\Gamma$ is chosen to be embedded in the buffer region where both inner and outer expansions are valid. Hence, from the point of view of the outer expansion, the radius of the worldtube is asymptotically small ($\rad\ll 1$), and its interior forms part of a smooth manifold $\man_E$, on which the external background metric is defined. The worldline lies in $\man_E$, at the center of this smooth interior. But from the point of view of the inner expansion, the radius of the tube is asymptotically large ($\rad\gg \e$), and its interior is a subset of a manifold $\man_I$, on which the internal background metric is defined, and in which there is potentially a black hole and no meaningful worldline. The worldline $\gamma$ at the center of the worldtube's interior---in $\man_E$---is defined to be the body's worldline if the body also lies at the center of the worldtube's interior---in $\man_I$---in the sense that its mass dipole vanishes on the worldtube. Using the worldtube to divide the spacetime into an inner region and an outer region in this way serves to ``cut out" the singularities that would appear in the metric perturbation in the outer limit, were it extended into the interior of the worldtube.

Although I am interested in the solution outside the tube, I will require some information from the metric in the inner limit. I assume the existence of some local polar coordinates $X^\alpha=(T,R,\Theta^A)$, such that the metric can be expanded for $\e\to 0$ while holding $\tilde R\equiv R/\e$, $\Theta^A$ and $T$ fixed. This leads to the ansatz
\begin{align}\label{internal ansatz}
\exact{g}(X,\e) &= g_B(T,\tilde R,\Theta^A)+H(T,\tilde R,\Theta^A,\e),
\end{align}
where $H$ at fixed $(T,\tilde R,\Theta^A)$ is a perturbation beginning at order $\e$. The leading-order term $g_B(T,\tilde R,\Theta^A)$ at fixed $T$ is the metric of the small body if it were isolated. For example, if the body is a small Schwarzschild black hole of ADM mass $\tilde m(T)$, then in Schwarzschild coordinates $g_B(T,\tilde R,\Theta^A)$ is given by
\begin{align}
ds^2 &= -\left(1-2m(T)/\tilde R\right)dT^2 +\left(1-2m(T)/\tilde R\right)^{-1}\e^2d\tilde{R}^2 \nonumber\\
&\quad + \e^2\tilde{R}^2\left(d\Theta^2+\sin^2\Theta d\Phi^2\right),
\end{align}
where $m(T)\equiv\tilde m(T)/\e$. Since the metric becomes one-dimensional at $\e=0$, the limit $\e\to0$ is singular. As discussed by D'Eath \cite{DEath_paper,DEath} (see also Ref.~\cite{Gralla_Wald}), the limit can be made regular by rescaling time as well, such that $\tilde T=(T-T_0)/\e$, and then rescaling the entire metric by a conformal factor $1/\e^2$. This is equivalent to using the above singular expansion and assuming that the metric $g_B$ and its perturbations are quasistatic (evolving only on timescales $\sim 1$). Both are equivalent to assuming that the exact metric contains no high-frequency oscillations occuring on the body's natural timescale $\sim\e$. If one were interested in the effect of the external spacetime on the metric near the small body, one could determine the perturbation $H$ by expanding it in powers of $\e$, solving the perturbative Einstein equation, and then matching the result to the external solution.

However, in this paper, the inner expansion will be used only to provide data for the outer expansion. In the outer limit, I expand for $\e\to 0$ while holding fixed some global coordinates $x^\alpha$ as well as the worldline $\gamma$. This leads to the ansatz
\begin{equation}\label{external ansatz}
\exact{g}(x,\e) = g(x)+h(x,\e;\gamma),
\end{equation}
where
\begin{equation}
h(x,\e;\gamma)=\sum_{n=1}^{N_E}\e^n\hmn{E}{\emph{n}}(x;\gamma) +\order{\e^{N_E+1}}.
\end{equation}

In order to solve the Einstein equation with a fixed worldline, I assume that the Lorenz gauge can be imposed everywhere in $\Omega$ on the entirey of $h$, such that $L_\mu[h]=0$.\footnote{Note that this is a stronger assumption than in the point particle case, because if the metric is given in some other gauge, the gauge vector(s) transforming to the Lorenz gauge must satisfy not only some weakly nonlinear wave equation, but also some suitable boundary conditions on the worldtube $\Gamma$. However, in practice I will be satisfied by the existence of an approximate solution to the Einstein equation that approximately satisfies the gauge condition up to errors of order $\e^3$.} With this gauge condition, the vacuum Einstein equation $\exact{R}_{\mu\nu}=0$ is reduced to a weakly nonlinear wave equation that can be expanded and solved at fixed $\gamma$, leading to the sequence of wave equations
\begin{align}
E_{\mu\nu}[\hmn{E}{1}] &=0, \label{h_E1 eqn}\\
E_{\mu\nu}[\hmn{E}{2}] &=2\delta^2 R_{\mu\nu}[\hmn{E}{1}], \label{h_E2 eqn}
\end{align}
and so on, where $E_{\mu\nu}$ is the wave operator defined in Eq.~\eqref{wave op def}. I discuss the formal solution to these equations in Sec.~\ref{perturbation calculation}.

For simplicity, I assume that each term in the expansion of the metric perturbation minimally violates the Lorenz gauge, in the sense that if a solution truncated at some finite order violates the Lorenz gauge, then that violation is solely due to the acceleration. Again solving at fixed $\gamma$, this assumption leads to the equations
\begin{align}
L\coeff{0}_\mu\big[\hmn{E}{1}\big] &=0,\label{h_E1 gauge}\\
L\coeff{1}_\mu\big[\hmn{E}{1}\big] &= -L\coeff{0}_\mu\big[\hmn{E}{2}\big], \label{h_E2 gauge}
\end{align}
which follow from an assumed expansion of the acceleration:
\begin{equation}
a_i(t,\e) = \an{0}_i(t)+\e\an{1}_i(t;\gamma)+\order{\e^2}.\label{a expansion}
\end{equation}
I remind the reader that $L_\mu$ is the gauge operator defined in Eq.~\eqref{gauge op def}, $L\coeff{0}[f]\equiv L[f]\big|_{a=\an{0}}$, and $L\coeff{1}[f]$ consists of the terms in $L[f]$ that are linear in $\an{1}$.

In Sec.~\ref{force calculation}, following the approach of Kates \cite{Kates_motion} and Gralla and Wald \cite{Gralla_Wald}, I determine the acceleration by solving the Einstein equation in the buffer region around the body. In this region, I work in Fermi normal coordinates centered on the worldline, and I expand the metric for small $\e$ and $r$, allowing each coefficient to have a functional dependence on $\gamma$. I never make use of the global coordinates $x^\mu$ or the local coordinates $X^\mu$, assuming only that they can be transformed into Fermi coordinates in the buffer region---which is necessarily true if both the inner and outer expansions are approximations to the same exact solution. Since the worldline is fixed, its acceleration will appear explicitly in the external background metric in Fermi coordinates. Hence, by solving the Einstein equation in these coordinates, the acceleration will naturally be determined. Although I perform this calculation in the Lorenz gauge, the choice of gauge should be of little significance.

The result of this calculation is that the external metric perturbation in the buffer region is expressed as the sum of two solutions: one solution that diverges at $r=0$ and which is entirely determined from a combination of (i) the multipole moments of the internal background metric $g_B$, (ii) the Riemann tensor of the external background $g$, and (iii) the acceleration of the worldline $\gamma$; and a second solution that is regular at $r=0$ and must be determined from the global past history of the body. At leading order, these two solutions are identified as the Detweiler-Whiting singular and regular fields $h^S$ and $h^R$ \cite{Detweiler_Whiting}, and the self-force is determined entirely by $h^R$. Along with the self-force, the acceleration of the worldline includes the Papapetrou spin-force. This leaves us with the self-force in terms of the metric perturbation induced by the body.

In Sec.~\ref{perturbation calculation}, I proceed to obtain a global, formal solution for the metric perturbation in the Lorenz gauge. Following the method of D'Eath \cite{DEath_paper, DEath}, I write the formal solution to the wave equation in an integral representation, whereby the value of the metric perturbation at any point in the exterior region is related to an integral over the worldtube around the body. Since the tube is chosen to lie in the buffer region, the previously obtained expansion in that region then serves to provide the boundary data on the tube. This approach allows me to determine $h^R$ in the buffer-region expansion by appealing to the consistency of the integral representation of the wave equation. Given the results of the buffer-region expansion as boundary values, evaluating the integral representation at a point just outside the worldtube must return the general solution in the buffer region. This consistency condition determines the unknown functions in terms of a tail integral. With the solution in the buffer region determined, the worldline is also determined; at the same time, since the boundary values are determined, the solution in the external spacetime is also determined.

\section{The self-force in terms of the metric perturbation}\label{self-force calculation}\label{force calculation}

\subsection{Expansion of the metric in the buffer region} \label{buffer_expansion}
In the buffer region, I adopt Fermi normal coordinates $(t,x^a)$ centered on the worldline $\gamma$. This coordinate system is constructed from a tetrad $e_I^\alpha$, where on the worldline $e_0^\alpha=u^\alpha$ and $e_a^\alpha$ is a spatial triad, and off the worldline the tetrad is defined by parallel propagation along a spatial geodesic perpendicular to $u^\alpha$. Refer to Ref.~\cite{Eric_review} for a detailed description.

Although the solution to the wave equation is more naturally expressed in terms of retarded coordinates \cite{Eric_review}, in the calculations here Fermi coordinates are more advantageous; for example, the solution to the wave equation with a point particle source is expressed as an integral over the worldline up to a retarded time $t_{\text{ret}}$, but in my calculation, the solution to the wave equation will be expressed as an integral over a worldtube, which will be evaluated just as easily in Fermi coordinates as in retarded coordinates. Thus, in this calculation, the simpler form of the background metric in Fermi coordinates outweighs the advantages of retarded coordinates.

I will be interested only in components in the Cartesian coordinates $(t,x^a)$, but I will express these components in terms of the geodesic distance $r\equiv\sqrt{\delta_{ij}x^ix^j}$ and the angles $\theta^A$, which are defined in the usual way in terms of $x^a$. I also introduce the unit one-form $n_\alpha\equiv \partial_\alpha r$, which has components $n_\alpha = (0,n_a)$ in Fermi coordinates, where $n_a\equiv \delta_{ab}x^b/r$. Note that this one-form has the convenient property that $n^\alpha\equiv g^{\alpha\beta}n_\beta = (0,n^a)$, where $n^a=\delta^{ab}n_b$. I will use the  multi-index notation $n^L\equiv n^{i_1}...n^{i_\ell}\equiv n^{i_1...i_\ell}$. Finally, I define the coordinate one-forms $t_\alpha\equiv\partial_\alpha t$ and $x^a_\alpha\equiv \partial_\alpha x^a$.

Since $r\ll 1$ in the buffer region, both the background metric $g$ and the external perturbations $\hmn{E}{\emph{n}}$ can be expanded for small $r$. In particular, the components of the background metric are given by the following standard result:
\begin{align}
g_{tt} &= -(1+ra_in^i)^2-r^2R_{0i0j}n^{ij}+O(r^3),\\
g_{ta} &= -\frac{2}{3}r^2R_{0iaj}n^{ij}+O(r^3), \\
g_{ab} &= \delta_{ab}-\frac{1}{3}r^2R_{aibj}n^{ij}+O(r^3).
\end{align}
Here the components of the Riemann tensor are evaluated on the worldline, and are therefore functions of $t$ (and potentially $\gamma$) only. For later convenience, I define the electric-type tidal field 
\begin{equation}
\etide_{ab}\equiv R_{a0b0},
\end{equation}
and the magnetic-type tidal field
\begin{equation}
\btide_{ab}\equiv \tfrac{1}{2}\epsilon_a{}^{cd}R_{0bcd}.
\end{equation}
Each of these fields is symmetric trace-free (STF) with respect to the Euclidean spatial metric $\delta_{ab}$. Solving the Einstein equation in the buffer region will  determine an analogous expansion for the perturbations.

However, before beginning that calculation, one should note that the coordinate transformation $x^\alpha(t,x^a)$ between Fermi coordinates and the global coordinates is $\e$-dependent, since Fermi coordinates are tethered to an $\e$-dependent worldline. If one were using a regular expansion, then this coordinate transformation would devolve into a background coordinate transformation to a Fermi coordinate system centered on a geodesic worldline, combined with a gauge transformation to account for the $\e$-dependence. But in the present singular expansion, the transformation is purely a background transformation, because the $\e$-dependence in the transformation is reducible to the $\e$-dependence in the fixed worldline.

The transformation hence induces not only new $\e$-dependence into the perturbations $\hmn{E}{\emph{n}}$, but also $\e$-dependence in the background metric $g$. This new $\e$-dependence takes two forms: a functional dependence on $z^\alpha(t)=x^\alpha(t,x^a=0)$, the coordinate form of the worldline written in the global coordinates $x^\alpha$; and a dependence on the acceleration vector $a^\alpha(t)$ on that worldline. For example, the first type of dependence appears in the components of the Riemann tensor in Fermi coordinates, which are related to the components in the global coordinates via the relationship $R_{IJKL}(t)=R_{\alpha\beta\gamma\delta}(z^\mu(t))e^\alpha_I e^\beta_J e^\gamma_K e^\delta_L$. The second type of $\e$-dependence consists of factors of the acceleration $a^\mu(t)$, which has the assumed expansion given in Eq.~\eqref{a expansion}.

Hence, in the buffer region we can opt to work with the quantities $g$ and $h_E$, which are defined with $a$ fixed, or we can opt to reexpand these quantities by substituting into them the expansion of $a$. (In either case, we would still hold fixed the functional dependence on $z^\mu$.) Substituting the expansion of $a$ in Fermi coordinates yields the \emph{buffer-region expansions}
\begin{align}
g_{\mu\nu} & = g_{\mu\nu}\coeff{0}(t,x^a;\gamma)+\order{\e},\label{buffer_expansion g}\\
\hmn{E\alpha\beta}{\emph{n}} &= \hmn{\alpha\beta}{\emph{n}}(t,x^a;\gamma) +\order{\e}\label{buffer_expansion h},
\end{align}
where $g\coeff{0}\equiv g\big|_{a=\an{0}}$ and $\hmn{}{\emph{n}}\equiv\hmn{E}{\emph{n}}\big|_{a=\an{0}}$. In solving the Einstein equation, I will make use of both the original quantities $g$ and $\hmn{E}{\emph{n}}$, and the buffer-region quantities $g\coeff{0}$ and $\hmn{}{\emph{n}}$.

Now, in order to determine a solution to the Einstein equation in the buffer region, I must first determine the general form of an expansion in powers of $r$ for the metric perturbations $\hmn{E}{\emph{n}}$. To accomplish this, I consider the form of the internal metric $g_B+H$. I assume that in the buffer region there exists a smooth coordinate transformation between the local coordinates $(T,R,\Theta^A)$ and the Fermi coordinates $(t,x^a)$ such that $T\sim t$, $R\sim r$, and $\Theta^A\sim\theta^A$. The buffer region corresponds to asymptotic infinity $r\gg\e$ (or $\tilde r\gg1$) in the internal spacetime. So after re-expressing $\tilde r$ as $r/\e$, the internal background metric can be expanded as
\begin{align}
g_{B\alpha\beta}(t,\tilde r,\theta^A) &= \sum_{n\geq0}\left(\frac{\e}{r}\right)^n g\coeff{\emph{n}}_{B\alpha\beta}(t,\theta^A).
\end{align}
There is no a priori reason to exclude negative values of $n$, since $g_B$ is an unknown function of $\tilde r$. However, since the internal and external solutions must be approximations to the same metric, they must agree with one another. And since the external expansion has no negative powers of $\e$, neither has the internal expansion. Furthermore, since $g+h=g_B+H$, we must have $g_B\coeff{0}=g(x^a=0)$, since these are the only terms independent of both $\e$ and $r$. Thus, noting that $g(x^a=0)=\eta$, where $\eta\equiv\text{diag}(-1,1,1,1)$, I can write
\begin{align}
g_{B\alpha\beta}(t,\tilde r,\theta^A) &= \eta_{\alpha\beta}+\frac{\e}{r} g\coeff{1}_{B\alpha\beta}(t,\theta^A) \nonumber\\
&\quad+\left(\frac{\e}{r} \right)^2g\coeff{2}_{B\alpha\beta}(t,\theta^A)+\order{\e^3/r^3},
\end{align}
implying that the internal background spacetime is asymptotically flat.

I assume that the perturbation $H$ can be similarly expanded in powers of $\e$ at fixed $\tilde r$,
\begin{align}
H_{\alpha\beta}(t,\tilde r,\theta^A,\e) &= \e H\coeff{1}_{\alpha\beta}(t,\tilde r,\theta^A;\gamma) \nonumber\\
&\quad +\e^2 H\coeff{2}_{\alpha\beta}(t,\tilde r,\theta^A;\gamma)+\order{\e^3},
\end{align}
and that each coefficient can be expanded in powers of $1/\tilde r=\e/r$ to yield
\begin{align}\label{buffer ansatz}
\e H\coeff{1}_{\alpha\beta}(\tilde r) &= rH\coeff{0,1}_{\alpha\beta}+\e H\coeff{1,0}_{\alpha\beta}+ \frac{\e^2}{r}H\coeff{2,-1}_{\alpha\beta}\nonumber\\&\quad+\order{\e^3/r^2},\\
\e^2 H\coeff{2}_{\alpha\beta}(\tilde r) & = r^2 H\coeff{0,2}_{\alpha\beta} +\e rH\coeff{1,1}_{\alpha\beta}+\e^2 H\coeff{2,0}_{\alpha\beta} \nonumber\\
&\quad+\e^2\ln r H\coeff{2,0,ln}_{\alpha\beta}+\order{\e^3/r},\\
\e^3 H\coeff{3}_{\alpha\beta}(\tilde r) & = \order{\e^3,\e^2r, \e r^2,r^3},
\end{align}
where $H\coeff{\emph{n,m}}$, the coefficient of $\e^n$ and $r^m$, is a function of $t$ and $\theta^A$ (and potentially a functional of $\gamma$). Again, the form of this expansion is constrained by the fact that no negative powers of $\e$ can appear in the buffer region.\footnote{One might think that terms with negative powers of $\e$ could be allowed in the expansion of $g_B$ if they are exactly cancelled by terms in the expansion of $H$, but the differing powers of $r$ in the two expansion makes this impossible.} Note that explicit powers of $r$ appear because $\e\tilde r=r$. Also note that I allow for a logarithmic term at second order in $\e$; this term arises because the retarded time in the internal background includes a logarithmic correction of the form $\e\ln r$ (e.g., $t-r\to t-r^*$ in Schwarzschild coordinates). Since I seek solutions to a wave equation, this correction to the characteristic curves induces a corresponding correction to the first-order perturbations. 

The expansion of $H$ may or may not hold the acceleration fixed. Regardless of this choice, the general form of the expansion remains valid: incorporating the expansion of the acceleration would merely shuffle terms from one coefficient to another. And since the internal metric $g_B+H$ must equal the external metric $g+h$, the general form of the above expansions of the $g_B$ and $H$ completely determines the general form of the external perturbations:
\begin{align}
\hmn{E\alpha\beta}{1} & = \frac{1}{r}\hmn{E\alpha\beta}{1,-1} +\hmn{E\alpha\beta}{1,0} +r\hmn{\alpha\beta}{1,1}+\order{r^2}, \label{h_E1 expansion}\\
\hmn{E\alpha\beta}{2} & = \frac{1}{r^2}\hmn{E\alpha\beta}{2,-2} +\frac{1}{r}\hmn{E\alpha\beta}{2,-1} +\hmn{E\alpha\beta}{2,0}+\ln r\hmn{E\alpha\beta}{2,0,ln}\nonumber\\
&\quad+\order{r},\label{h_E2 expansion}
\end{align}
where each $\hmn{E}{\emph{n,m}}$ depends only on $t$ and $\theta^A$, along with an implicit functional dependence on $\gamma$. If the internal expansion is performed with $a$ held fixed, then the internal and external quantities are related order-by-order: e.g., $\sum_m H\coeff{0,m}=g$, $\hmn{E}{1,-1}=g_B\coeff{1}$, and $\hmn{E}{1,0}=H\coeff{1,0}$. Since I am not concerned with determining the internal perturbations, the only such relationship of interest is $\hmn{E}{\emph{n,-n}}=g_B\coeff{\emph{n}}$. This equality tells us that the most divergent, $r^{-n}$ piece of the $n$th-order perturbation $\hmn{E}{\emph{n}}$ is defined entirely by the $n$th-order piece of the internal background metric $g_B$, which is the metric of the body if it were isolated.

To obtain a general solution to the Einstein equation, I write each $\hmn{E}{\emph{n,m}}$ as an expansion in terms of irreducible symmetric trace-free pieces:
\begin{align}
\hmn{Ett}{{\it n,m}} &= \sum_{\ell\ge0}\A{L}{{\it n,m}}\nhat^L, \\
\hmn{Eta}{{\it n,m}} &= \sum_{\ell\ge0}\B{L}{{\it n,m}}\nhat_a{}^L\nonumber\\
&\quad+\sum_{\ell\ge1}\left[\C{aL-1}{{\it n,m}}\nhat^{L-1} +\epsilon_{ab}{}^c\D{cL-1}{{\it n,m}}\nhat^{bL-1}\right],\\
\hmn{Eab}{{\it n,m}} &= \delta_{ab}\sum_{\ell\ge0}\K{L}{{\it n,m}}\nhat^L+\sum_{\ell\ge0}\E{L}{{\it n,m}}\nhat_{ab}{}^L \nonumber\\
&\quad+\sum_{\ell\ge1}\!\left[\F{L-1\langle a}{{\it n,m}}\nhat^{}_{b\rangle}{}^{L-1} +\epsilon^{cd}{}_{(a}\nhat_{b)c}{}^{L-1}\G{dL-1}{{\it n,m}}\right] \nonumber\\
&\quad+\sum_{\ell\ge2}\!\left[\H{abL-2}{{\it n,m}}\nhat^{L-2}+\epsilon^{cd}{}_{(a}\I{b)dL-2}{{\it n,m}} \nhat_c{}^{L-2}\right].
\end{align}
Here a hat indicates that a tensor is STF with respect to $\delta_{ab}$, angular brackets $\langle\rangle$ indicate the STF combination of enclosed indices, parentheses indicate the symmetric combination of enclosed indices, and all the uppercase script symbols are functions of time (and potentially functionals of $\gamma$) and are STF in all their indices. Each term in this expansion is linearly independent of all the other terms. All the quantities on the right-hand side are flat-space Cartesian tensors; their indices can be raised or lowered with $\delta_{ab}$. Refer to Appendix~\ref{STF tensors} for more details about this expansion.

Now, despite its $\e$-dependence, $g$ is the background metric of the outer expansion, and I will use it to raise and lower indices on $h$. And since the wave equations \eqref{h_E1 eqn} and \eqref{h_E2 eqn} are covariant, they must still hold in the new coordinate system, despite the additional $\e$-dependence. Thus, both equations could be solved for arbitrary acceleration in the buffer region. However, due to the length of the calculations involved, I will instead solve the equations
\begin{align}
E_{\alpha\beta}[\hmn{E}{1}] & = 0, \label{h_1 eqn}\\
E\coeff{0}_{\alpha\beta}[\hmn{}{2}] & = 2\delta^2 R\coeff{0}_{\alpha\beta}[\hmn{}{1}]+\order{\e},\label{h_2 eqn}
\end{align}
where $E\coeff{0}[f]\equiv E[f]\big|_{a=\an{0}}$ and $\delta^2 R\coeff{0}[f]\equiv \delta^2R[f]\big|_{a=\an{0}}$.\footnote{In analogy with the notation used for $L\coeff{\emph{n}}$, $E\coeff{1}[f]$ and $\delta^2 R\coeff{1}[f]$ would be linear in $\an{1}$, $E\coeff{2}[f]$ and $\delta^2 R\coeff{2}[f]$ would be linear in $\an{2}$ and quadratic in $\an{1}$, and so on. For a function $f\sim 1$, $L\coeff{\emph{n}}[f]$, $E\coeff{\emph{n}}[f]$, and $\delta^2 R\coeff{\emph{n}}[f]$ correspond to the coefficients of $\e^n$ in expansions in powers of $\e$.} The first equation is identical to Eq.~\eqref{h_E1 eqn}. The second equation follows directly from substituting Eqs.~\eqref{buffer_expansion g} and \eqref{buffer_expansion h} into Eq.~\eqref{h_E2 eqn}; in the buffer region, it captures the dominant behavior of $\hmn{E}{2}$, represented by the approximation $\hmn{}{2}$, but it does not capture its full dependence on acceleration. If one desired a global second-order solution, one would solve Eq.~\eqref{h_E2 eqn}, but for my purpose, which is to determine the first-order acceleration $\an{1}$, Eq.~\eqref{h_2 eqn} will suffice.

Unlike the wave equations, the gauge conditions \eqref{h_E1 gauge} and \eqref{h_E2 gauge} already incorporate the expansion of the acceleration. As such, they are unmodified by the replacement of the second-order wave equation \eqref{h_E2 eqn} with its approximation \eqref{h_2 eqn}. So we can write
\begin{align}
L\coeff{0}_\mu\big[\hmn{E}{1}\big] &=0, \label{h_1 gauge}\\
L\coeff{1}_\mu\big[\hmn{E}{1}\big] &= -L\coeff{0}_\mu\big[\hmn{}{2}\big], \label{h_2 gauge}
\end{align}
where the first equation is identical to Eq.~\eqref{h_E1 gauge}, and the second to Eq.~\eqref{h_E2 gauge}. (The second identity holds because $L\coeff{0}_\mu\big[\hmn{}{2}\big]=L\coeff{0}_\mu\big[\hmn{E}{2}\big]$, since $\hmn{}{2}$ differs from $\hmn{E}{2}$ by $\an{1}$ and higher acceleration terms, which are set to zero in $L\coeff{0}$.) I remind the reader that while this gauge choice is important for finding the external perturbations globally, any other choice would suffice in the buffer region calculation. For example, one could expand $g$ and $h$ in buffer region expansions that incorporate the expansion of the acceleration, enabling one to solve the full Einstein equation order-by-order in $\e$; it might be difficult to make this mesh with a global expansion in the external spacetime, but it would suffice to determine the acceleration. Alternatively, one could construct a two-timescale expansion in the buffer region, which would mesh with a global two-timescale expansion of the Einstein equation in the external spacetime.

As a final, important point, I assume the partial time-derivative of any term in an expansion is of the same order as the term itself. In what follows, the reader may safely assume that all calculations are lengthy unless noted otherwise.

\subsection{First-order solution in the buffer region}\label{buffer_expansion1}
In principle, solving the first-order Einstein equation in the buffer region is straightforward. One need simply substitute the expansion of $\hmn{E}{1}$, given in Eq.~\eqref{h_E1 expansion},
into the linearized wave equation \eqref{h_1 eqn} and the gauge condition \eqref{h_1 gauge}. Equating powers of $r$ in the resulting expansions then yields a sequence of equations that can be solved for successively higher-order terms in $\hmn{E}{1}$. Solving these equations consists primarily of expressing each quantity in its irreducible STF form, using the decompositions~\eqref{decomposition_1} and \eqref{decomposition_2}; since the terms in this STF decomposition are linearly independent, we can solve each equation term-by-term. This calculation is aided by the fact that $\del{\alpha}=x^a_\alpha\partial_a+\order{r^0}$, so for example, the wave operator $E_{\alpha\beta}$ consists of a flat-space Laplacian $\partial^a\partial_a$ plus corrections of order $1/r$. Appendix B also contains many useful identities, particularly $\partial_\alpha r =n_\alpha$, $n^\alpha\partial_\alpha\nhat^L=0$, and the fact that $\nhat^L$ is an eigenvector of the flat-space Laplacian: \emph{i.e.}, $\partial^a\partial_a\nhat^L =-\frac{\ell(\ell+1)}{r^2}\nhat^L$. Because the calculation consists mostly of simple, albeit lengthy algebra, I will for the most part simply summarize results. 

Of course, the Einstein equation in the buffer region does not completely determine the solution: auxiliary boundary data must also be provided. Since the most singular term, $\hmn{E\alpha\beta}{1,-1}$, is the order-$1/r$ term in the internal background metric $g_B$, it will be fully determined in terms of the mass of the internal spacetime. Some of the subleading terms will also be determined by the mass, while others will remain unknown. The unknowns form the Detweiler-Whiting regular field; they will eventually be expressed in terms of a tail integral in Sec.~\ref{perturbation calculation}.

So, we begin with the the most divergent term in the wave equation: the order-$1/r^3$, flat-space Laplacian term
\begin{equation}
\frac{1}{r}\partial^c\partial_c\hmn{E\alpha\beta}{1,-1} =0.
\end{equation}
The $tt$-component of this equation is
\begin{equation}
0=-\sum_{\ell\ge0}\ell(\ell+1)\A{L}{1,-1}\nhat^L,
\end{equation}
from which we read off that $\A{}{1,-1}$ is arbitrary and $\A{L}{1,-1}$ must vanish for all $\ell\ge1$. The $ta$-component is
\begin{align}
0&=-\sum_{\ell\ge0}(\ell+1)(\ell+2)\B{L}{1,-1}\nhat_a{}^L\nonumber\\
&\quad-\sum_{\ell\ge1}\ell(\ell-1)\C{aL-1}{1,-1}\nhat^{L-1}\nonumber\\
&\quad-\sum_{\ell\ge1}\ell(\ell+1)\epsilon_{abc}\D{cL-1}{1,-1}\nhat_b{}^{L-1},
\end{align}
from which we read off that $\C{a}{1,-1}$ is arbitrary and all other coefficients must vanish. Lastly, the $ab$-component is
\begin{align}
0&=-\delta_{ab}\sum_{\ell\ge0}\ell(\ell+1)\K{L}{1,-1}\nhat^L\nonumber\\
&\quad-\sum_{\ell\ge0}(\ell+2)(\ell+3)\E{L}{1,-1}\nhat_{ab}{}^L\nonumber\\
&\quad-\sum_{\ell\ge1}\ell(\ell+1)\F{L-1\langle a}{1,-1}\nhat^{}_{b\rangle}{}^{L-1} \nonumber\\
&\quad -\sum_{\ell\ge1}(\ell+1)(\ell+2)\epsilon_{cd(a}\nhat_{b)}{}^{cL-1}\G{dL-1}{1,-1} \nonumber\\
&\quad-\sum_{\ell\ge2}(\ell-2)(\ell-1)\H{abL-2}{1,-1}\nhat^{L-2}\nonumber\\
&\quad -\sum_{\ell\ge2}\ell(\ell-1)\epsilon^{}_{cd(a}\I{b)dL-2}{1,-1}\nhat_c{}^{L-2},
\end{align}
from which we read off that $\K{}{1,-1}$ and $\H{ab}{1,-1}$ are arbitrary and all other coefficients must vanish. Thus, we find that the wave equation constrains $\hmn{E}{1,-1}$ to be
\begin{align}
\hmn{E\alpha\beta}{1,-1}& =\A{}{1,-1}t_\alpha t_\beta+ 2\C{a}{1,-1}t_{(\beta}x^a_{\alpha)} \nonumber\\
&\quad+(\delta_{ab}\K{}{1,-1}+\H{ab}{1,-1})x^a_\alpha x^b_\beta.
\end{align}

This is further constrained by the most divergent, $1/r^2$ term in the gauge condition:
\begin{equation}
-\frac{1}{r^2}\hmn{E\alpha c}{1,-1}n^c+\frac{1}{2r^2}n_\alpha\eta^{\mu\nu}\hmn{E\mu\nu}{1,-1}=0.
\end{equation}
From the $t$-component of this equation, we read off $\C{a}{1,-1}=0$; from the $a$-component, $\K{}{1,-1}=\A{}{1,-1}$ and $\H{ab}{1,-1}=0$. Thus, $\hmn{E\alpha\beta}{1,-1}$ depends only on a single function of time, $\A{}{1,-1}$. By the definition of ADM mass, this function (times $\e$) must be twice the mass of the internal background spacetime. Thus, $\hmn{E}{1,-1}$ is fully determined to be
\begin{equation}\label{h1n1}
\hmn{E\alpha\beta}{1,-1}=2m(t)(t_\alpha t_\beta +\delta_{ab}x^a_\alpha x^b_\beta),
\end{equation}
where $m(t)$ is defined to be the mass at time $t$ divided by the initial mass $\e\equiv m_0$. (Alternatively, we could set $\e$ equal to unity at the end of the calculation, in which case $m$ would simply be the mass at time $t$; obviously, the difference between the two approaches is immaterial.)

At the next order, $\hmn{E}{1,0}$, along with the acceleration of the worldline and the time-derivative of the mass, first appears in the Einstein equation. The order-$1/r^2$ term in the wave equation is
\begin{equation}
\partial^c\partial_c\hmn{E\alpha\beta}{1,0}=-\frac{2m}{r^2}a_cn^c(3t_\alpha t_\beta -\delta_{ab}x^a_\alpha x^b_\beta),
\end{equation}
where the terms on the right arise from $\Box$ acting on $\frac{1}{r}\hmn{E}{1,-1}$. This equation constrains $\hmn{E}{1,0}$ to be
\begin{equation}\label{h10}
\begin{split}
\hmn{Ett}{1,0}&=\A{}{1,0}+3ma_cn^c, \\
\hmn{Eta}{1,0}&=\C{a}{1,0}, \\
\hmn{Eab}{1,0}&=\delta_{ab}\left(\K{}{1,0}-ma_cn^c\right)+\H{ab}{1,0}.
\end{split}
\end{equation}
Substituting this result into the order-$1/r$ term in the gauge condition, we find
\begin{equation}
-\frac{4}{r} t_\alpha\partial_t m +\frac{4m}{r}\an{0}_ax^a_\alpha=0.
\end{equation}
Thus, both the leading-order part of the acceleration and the rate of change of the mass of the body vanish:
\begin{equation}
\begin{array}{lcr}
\displaystyle\frac{\partial m}{\partial t}=0\,, && \an{0}_i =0.
\end{array}
\end{equation}

At the next order, $r\hmn{E}{1,1}$, along with squares and derivatives of the acceleration, first appear in the Einstein equation, and the tidal fields of the external background couple to $\frac{1}{r}\hmn{E}{1,-1}$. The order-$1/r$ term in the wave equation becomes
\begin{align}
\left(r\partial^c\partial_c+\frac{2}{r}\right)\hmn{Ett}{1,1} & = -\frac{20m}{3r}\etide_{ij}\nhat^{ij}
-\frac{3m}{r}a_{\langle i}a_{j\rangle}\nhat^{ij} \nonumber \\
&\quad+\frac{8m}{r}a_ia^i, \\
\left(r\partial^c\partial_c+\frac{2}{r}\right)\hmn{Eta}{1,1} & = -\frac{8m}{3r}\epsilon_{aij}\btide^j_k\nhat^{ik}-\frac{4m}{r}\dot a_a,\\
\left(r\partial^c\partial_c+\frac{2}{r}\right)\hmn{Eab}{1,1} & = \frac{20m}{9r}\delta_{ab}\etide_{ij}\nhat^{ij}  -\frac{76m}{9r}\etide_{ab}\nonumber\\
&\quad -\frac{16m}{3r}\etide^i_{\langle a}\nhat_{b\rangle i}
+\frac{8m}{r}a_{\langle a}a_{b\rangle}\nonumber\\
&\quad+\frac{m}{r}\delta_{ab}\!\left(\tfrac{8}{3}a_ia^i\!-3a_{\langle i}a_{j\rangle}\nhat^{ij}\right).
\end{align}
From the $tt$-component, we read off that $\A{i}{1,1}$ is arbitrary, $\A{}{1,1}=4ma_ia^i$, and $\A{ij}{1,1}=\tfrac{5}{3}m\etide_{ij} +\tfrac{3}{4}ma_{\langle i}a_{j\rangle}$; from the $ta$-component, $\B{}{1,1}$, $\C{ij}{1,1}$, and $\D{i}{1,1}$ are arbitrary, $\C{i}{1,1}=-2m\dot a_i$, and $\D{ij}{1,1}=\tfrac{2}{3}m\btide_{ij}$; from the $ab$ component, $\K{i}{1,1}$, $\F{i}{1,1}$, $\H{ijk}{1,1}$, and $\I{ij}{1,1}$ are arbitrary, and $\K{}{1,1}=\tfrac{4}{3}ma_ia^i$, $\K{ij}{1,1}=-\tfrac{5}{9}m\etide_{ij}+\tfrac{3}{4}ma_{\langle i}a_{j\rangle}$, $\F{ij}{1,1}=\tfrac{4}{3}m\etide_{ij}$, and $\H{ij}{1,1}=-\tfrac{38}{9}m\etide_{ij}+4ma_{\langle i}a_{j\rangle}$.

Substituting this into the order-$r^0$ terms in the gauge condition, we find
\begin{align}
0&=(n^i+r\partial^i)\hmn{E\alpha i}{1,1} -\tfrac{1}{2}\eta^{\mu\nu}(n_a-r\partial_a)\hmn{E\mu\nu}{1,1}x^a_\alpha\nonumber\\
&\quad-\partial_t\hmn{E\alpha t}{1,0} -\tfrac{1}{2}\eta^{\mu\nu}\partial_t\hmn{E\mu\nu}{1,0}t_\alpha\nonumber\\
&\quad +\tfrac{4}{3}m\etide_{ij}\nhat^{ij}n_\alpha +\tfrac{2}{3}m\etide_{ai}n^ix^a_\alpha,
\end{align}
where the equation is to be evaluated at $a=\an{0}=0$. From the $t$-component, we read off
\begin{equation}\label{B11}
\B{}{1,1}=\tfrac{1}{6}\partial_t\left(\A{}{1,0}+3\K{}{1,0}\right).
\end{equation}
From the $a$-component,
\begin{equation}\label{F11}
\F{a}{1,1}=\tfrac{3}{10}\left(\K{a}{1,1}-\A{a}{1,1}+\partial_t\C{a}{1,0}\right).
\end{equation}
It is understood that both these equations hold only when evaluated at $a=\an{0}$.

Thus, the order-$r$ component of $\hmn{E}{1}$ is
\begin{equation}\label{h11}
\begin{split}
\hmn{Ett}{1,1}&=4ma_ia^i+\A{i}{1,1}n^i+\tfrac{5}{3}m\etide_{ij}\nhat^{ij} \\&\quad +\tfrac{3}{4}ma_{\langle i}a_{j\rangle}\nhat^{ij}, \\
\hmn{Eta}{1,1}&=\B{}{1,1}n_a-2m\dot a_a+\C{ai}{1,1}n^i+\epsilon_{ai}{}^j\D{j}{1,1}n^i\\
&\quad +\tfrac{2}{3}m\epsilon_{aij}\btide^j_k\nhat^{ik},\\
\hmn{Eab}{1,1}&=\delta_{ab}\big(\tfrac{4}{3}ma_ia^i+\K{i}{1,1}n^i -\tfrac{5}{9}m\etide_{ij}\nhat^{ij}\\
&\quad +\tfrac{3}{4}ma_{\langle i}a_{j\rangle}\nhat^{ij}\big) +\tfrac{4}{3}m\etide^i_{\langle a}\nhat_{b\rangle i}
-\tfrac{38}{9}m\etide_{ab}\\
&\quad+4ma_{\langle a}a_{b\rangle}+\H{abi}{1,1}n^i +\epsilon\indices{_i^j_{(a}}\I{b)j}{1,1}n^i\\
&\quad+\F{\langle a}{1,1}n^{}_{b\rangle}.
\end{split}
\end{equation}
where $\B{}{1,1}$ and $\F{a}{1,1}$ are constrained to satisfy Eqs.~\eqref{B11} and \eqref{F11}.

To summarize the results of this section, we have $\hmn{E\alpha\beta}{1}=\frac{1}{r}\hmn{E\alpha\beta}{1,-1} +\hmn{E\alpha\beta}{1,0} +r\hmn{E\alpha\beta}{1,1}+\order{r^2}$, where $\hmn{E\alpha\beta}{1,-1}$ is given in Eq.~\eqref{h1n1}, $\hmn{E\alpha\beta}{1,0}$ is given in Eq.~\eqref{h10}, and $\hmn{E\alpha\beta}{1,1}$ is given in Eq.~\eqref{h11}. In addition, we have determined that the ADM mass of the internal background spacetime is time-independent, and that the acceleration of the body's worldline vanishes at leading order.

\subsection{Second-order solution in the buffer region}\label{buffer_expansion2}
Though the calculations are much lengthier, solving the second-order Einstein equation in the buffer region is essentially no different than solving the first. I seek to solve the approximate wave equation \eqref{h_2 eqn}, along with the gauge condition \eqref{h_2 gauge}, for the second-order perturbation $\hmn{}{2}\equiv\hmn{E}{2}\big|_{a=\an{0}}$; doing so will also, more importantly, determine the acceleration $\an{1}$. In this calculation, the acceleration is set to $a=\an{0}=0$ everywhere except in the left-hand side of the gauge condition, $L\coeff{1}[\hmn{E}{1}]$, which is linear in $\an{1}$.

Substituting the expansion
\begin{align}
\hmn{\alpha\beta}{2}&=\frac{1}{r^2}\hmn{\alpha\beta}{2,-2} +\frac{1}{r}\hmn{\alpha\beta}{2,-1}+\hmn{\alpha\beta}{2,0} +\ln(r)\hmn{\alpha\beta}{2,0,ln}\nonumber\\
&\quad +\order{\e,r}
\end{align}
and the results for $\hmn{E}{1}$ from the previous section into the wave equation and the gauge condition again yields a sequence of equations that can be solved for coefficients of successively higher-order powers (and logarithms) of $r$. Due to its length, the expansion of the second-order Ricci tensor is given in Appendix~\ref{second-order expansions}. Note that since the approximate wave equation \eqref{h_2 eqn} contains an explicit $\order{\e}$ correction, $\hmn{}{2}$ will be determined only up to $\order{\e}$ corrections. For simplicity, I omit these $\order{\e}$ symbols from the equations in this section; note, however, that these corrections do not effect the gauge condition, as discussed above. 

To begin, the most divergent, order-$1/r^4$ term in the wave equation reads
\begin{align}
\frac{1}{r^4}\left(2+r^2\partial^c\partial_c\right)\hmn{\alpha\beta}{2,-2} & = \frac{4m^2}{r^4}\left(7\nhat_{ab} + \tfrac{4}{3}\delta_{ab}\right)x^a_\alpha x^b_\beta \nonumber\\
&\quad-\frac{4m^2}{r^4}t_\alpha t_\beta,
\end{align}
where the right-hand side is the most divergent part of the second-order Ricci tensor, as given in Eq.~\eqref{ddR0n4}. From the $tt$-component of this equation, we read off $\A{}{2,-2}=-2m^2$, and that $\A{a}{2,-2}$ is arbitrary. From the $ta$-component, $\B{}{2,-2}$, $\C{ab}{2,-2}$, and $\D{c}{2,-2}$ are arbitrary. From the $ab$-component, $\K{}{2,-2}=\tfrac{8}{3}m^2$, $\E{}{2,-2}=-7m^2$, and $\K{a}{2,-2}$, $\F{a}{2,-2}$, $\H{abc}{2,-2}$, and $\I{ab}{2,-2}$ are arbitrary.

The most divergent, order-$1/r^3$ terms in the gauge condition similarly involve only $\hmn{}{2,-2}$; they read
\begin{equation}
\frac{1}{r^3}\left(r\partial^b-2n^b\right)\hmn{\alpha b}{2,-2}-\frac{1}{2r^3}\eta^{\mu\nu}x^a_\alpha\left(r\partial_a -2n_a\right)\!\hmn{\mu\nu}{2,-2}=0.
\end{equation}
After substituting the results from the wave equation, the $t$-component of this equation determines that $\C{ab}{2,-2}=0$. The $a$-component determines that $\H{abc}{2,-2}=0$, $\I{ab}{2,-2}=0$, and
\begin{equation}\label{F2n2}
\F{a}{2,-2}=3\K{a}{2,-2}-3\A{a}{2,-2}.
\end{equation}
Thus, the order-$1/r^2$ part of $\hmn{}{2}$ is given by
\begin{equation}
\begin{split}
\hmn{tt}{2,-2} & = -2m^2+\A{i}{2,-2}n^i, \\
\hmn{ta}{2,-2} & = \B{}{2,-2}n_a+\epsilon_{a}{}^{ij}n_i\D{j}{2,-2},\\
\hmn{ab}{2,-2} & = \delta_{ab}\left(\tfrac{8}{3}m^2+\K{i}{2,-2}n^i\right)\\
&\quad-7m^2\nhat_{ab}+\F{\langle a}{2,-2}n^{}_{b\rangle},
\end{split}
\end{equation}
where $\F{a}{2,-2}$ is given by Eq. \eqref{F2n2}.

The metric perturbation in this form depends on five free functions of time. However, from calculations in flat spacetime, we know that order-$\e^2/r^2$ terms in the metric perturbation can be written in terms of two free functions: a mass dipole and a spin dipole. We transform the perturbation into this ``canonical" form by performing a gauge transformation (\emph{c.f.} Ref.~\cite{STF_2}). The transformation is generated by $\xi_\alpha=-\frac{1}{r}\B{}{2,-2}t_\alpha-\frac{1}{2r}\F{a}{2,-2}x^a_\alpha$, the effect of which is to remove $\B{}{2,-2}$ and $\F{a}{2,-2}$ from the metric.  This transformation is a refinement of the Lorenz gauge. (Effects at higher order in $\e$ and $r$ will be automatically incorporated into the higher-order perturbations.) The condition $\F{a}{2,-2}-3\K{a}{2,-2}+3\A{a}{2,-2}=0$ then becomes $\K{a}{2,-2}=\A{a}{2,-2}$. The remaining two functions are related to the ADM momenta of the internal spacetime: 
\begin{equation}
\begin{array}{lcr}
\A{i}{2,-2} =2M_i\,, && \D{i}{2,-2}=2S_i,
\end{array}
\end{equation}
where $M_i$ is such that $\partial_t M_i$ is proportional to the ADM linear momentum of the internal spacetime, and $S_i$ is the ADM angular momentum. $M_i$ is a mass dipole term; it is what would result from a transformation $x^a\to x^a+M^a/m$ applied to the $1/r$ term in $\hmn{E}{1}$. $S_i$ is a spin dipole term. Thus, the order-$1/r^2$ part of $\hmn{}{2}$ reads
\begin{equation}\label{h2n2}
\begin{split}
\hmn{tt}{2,-2} & = -2m^2+2M_in^i, \\
\hmn{ta}{2,-2} & = 2\epsilon_{aij}n^iS^j,\\
\hmn{ab}{2,-2} & = \delta_{ab}\left(\tfrac{8}{3}m^2+2M_in^i\right)-7m^2\nhat_{ab}.
\end{split}
\end{equation}

At the next order, $1/r^3$, because the acceleration is set to zero, $\hmn{}{2,-2}$ does not contribute to $E\coeff{0}[\hmn{}{2}]$, and $\hmn{}{1,-1}$ does not contribute to $\delta^2R\coeff{0}[\hmn{}{1}]$. The wave equation hence reads
\begin{equation}
\frac{1}{r}\partial^c\partial_c\hmn{\alpha\beta}{2,-1}= \frac{2}{r^3}\ddR{\alpha\beta}{0,-3}{\hmn{}{1}},
\end{equation}
where $\ddR{\alpha\beta}{0,-3}{\hmn{}{1}}$ is given in Eqs.~\eqref{ddR0n3_tt}--\eqref{ddR0n3_ab}. The $tt$-component of this equation implies $r^2\partial^c\partial_c\hmn{tt}{2,-1}=6m\H{ij}{1,0}\nhat^{ij}$, from which we read off that $\A{}{2,-1}$ is arbitrary and $\A{ij}{2,-1}=-m\H{ij}{1,0}$. The $ta$-component implies $r^2\partial^c\partial_c\hmn{ta}{2,-1}=6m\C{i}{1,0}\nhat_a^i$, from which we read off $\B{i}{2,-1}=-m\C{i}{1,0}$ and that $\C{a}{2,-1}$ is arbitrary. The $ab$-component implies
\begin{align}
r^2\partial^c\partial_c\hmn{ab}{2,-1}&=6m\left(\A{}{1,0} +\K{}{1,0}\right)\nhat_{ab}\nonumber\\
&\quad-12m\H{i\langle a}{1,0}\nhat_{b\rangle}{}^{i}\nonumber\\
&\quad+2m\delta_{ab}\H{ij}{1,0}\nhat^{ij},
\end{align}
from which we read off that $\K{}{2,-1}$ is arbitrary, $\K{ij}{2,-1}=-\tfrac{1}{3}m\H{ij}{1,0}$, $\E{}{2,-1}=-m\left(\A{}{1,0}+\K{}{1,0}\right)$, $\F{ab}{2,-1}=2m\H{ab}{1,0}$, and $\H{ab}{2,-1}$ is arbitrary. This restricts $\hmn{}{2,-1}$ to the form
\begin{equation}
\begin{split}
\hmn{tt}{2,-1}&=\A{}{2,-1}-m\H{ij}{1,0}\nhat^{ij}, \\
\hmn{ta}{2,-1}&=-m\C{i}{1,0}\nhat_a^i+\C{a}{2,-1}, \\
\hmn{ab}{2,-1}&=\delta_{ab}\left(\K{}{2,-1} -\tfrac{1}{3}m\H{ij}{1,0}\nhat^{ij}\right)\\
&\quad -m\left(\A{}{1,0}+\K{}{1,0}\right)\nhat_{ab}\\
&\quad +2m\H{i\langle a}{1,0}\nhat^{}_{b\rangle}{}^i+\H{ab}{2,-1}.
\end{split}
\end{equation}

We next substitute $\hmn{}{2,-2}$ and $\hmn{}{2,-1}$ into the order-$1/r^2$ terms in the gauge condition. The $t$-component becomes 
\begin{equation}
\frac{1}{r^2}\left(4m\C{i}{1,0}+12\partial_tM_i+3\C{i}{2,-1}\right)n^i=0,
\end{equation}
from which we read off
\begin{equation}
\C{i}{2,-1}=-4\partial_tM_i-\tfrac{4}{3}m\C{i}{1,0}.
\end{equation}
And the $a$-component becomes
\begin{align}
0&=\frac{1}{r^2}\left(-\tfrac{4}{3}m\A{}{1,0}-\tfrac{4}{3}m\K{}{1,0} -\tfrac{1}{2}\A{}{2,-1}+\tfrac{1}{2}\K{}{2,-1}\right)n_a\nonumber\\
&\quad +\left(\tfrac{2}{3}m\H{ai}{1,0}-\H{ai}{2,-1}\right)n^i -2\epsilon_{ija}n^i\partial_tS^j,
\end{align}
from which we read off
\begin{align}
\A{}{2,-1}&=\K{}{2,-1}-\tfrac{8}{3}m\left(\A{}{1,0}+\K{}{1,0}\right),\\
\H{ij}{2,-1}&=\tfrac{2}{3}m\H{ij}{1,0},
\end{align}
and that the angular momentum of the internal background is constant at leading order:
\begin{equation}
\partial_tS^i=0.
\end{equation}

Thus, the order-$1/r$ term in $\hmn{}{2}$ is given by
\begin{equation}\label{h2n1}
\begin{split}
\hmn{tt}{2,-1}&=\K{}{2,-1}-\tfrac{8}{3}m\left(\A{}{1,0}+\K{}{1,0}\right)\\
&\quad-m\H{ij}{1,0}\nhat^{ij}, \\
\hmn{ta}{2,-1}&=-m\C{i}{1,0}\nhat_a^i-4\partial_tM_i-\tfrac{4}{3}m\C{i}{2,-1}, \\
\hmn{ab}{2,-1}&=\delta_{ab}\left(\K{}{2,-1} -\tfrac{1}{3}m\H{ij}{1,0}\nhat^{ij}\right)\\
&\quad -m\left(\A{}{1,0}+\K{}{1,0}\right)\nhat_{ab}\\
&\quad +2m\H{i\langle a}{1,0}\nhat^{}_{b\rangle}{}^i+\tfrac{2}{3}m\H{ab}{1,0}.
\end{split}
\end{equation}
Note a peculiar feature of this term: the undetermined function $\K{}{2,-1}$ appears in precisely the form of a mass monopole. The value of this function will never be determined (though its time-dependence will be). This ambiguity arises because the mass $m$ that I have defined is the mass of the internal \emph{background} spacetime, which is based on the internal limit process that holds $\e/r$ fixed. A term of the form $\e^2/r$ appears as a perturbation of this background, even when, as in this case, it is part of the mass monopole of the body. This is equivalent to the ambiguity in any expansion in one's choice of small parameter: one could expand in powers of $\e$, or one could expand in powers of $\e+\e^2$, and so on. It is also equivalent to the ambiguity in defining the mass of a non-isolated body; whether the ``mass" of the body is taken to be $m$ or $m+\tfrac{1}{2}\K{}{2,-1}$ is a matter of taste. As we shall discover, the time-dependent part of $\K{}{2,-1}$ is constructed from the tail terms in the first-order metric perturbation. Hence, the ambiguity in the definition of the mass is, at least in part, equivalent to whether or not one chooses to include the free gravitational field induced by the body in what one calls its mass. In any case, I will define the ``correction" to the mass as $\delta m\equiv\tfrac{1}{2}\K{}{2,-1}$.

We next move to the order-$\ln(r)/r^2$ terms in the wave equation, and the order-$\ln(r)/r$ terms in the gauge condition, which read
\begin{align}
\frac{\ln r}{r^2}\partial^c\partial_c\hmn{\alpha\beta}{2,0,ln} &= 0,\\
\ln r \left(\partial^b\hmn{\alpha b}{2,0,ln}-\tfrac{1}{2}\eta^{\mu\nu}x^a_\alpha\partial_a \hmn{\alpha b}{2,0,ln}\right) & = 0.
\end{align}
From this we determine
\begin{align}
\hmn{\alpha\beta}{2,0,ln}& =\A{}{2,0,ln}t_\alpha t_\beta+ 2\C{a}{2,0,ln}t_{(\beta}x^a_{\alpha)} \nonumber\\
&\quad+(\delta_{ab}\K{}{2,0,ln}+\H{ab}{2,0,ln})x^a_\alpha x^b_\beta.
\end{align}

Finally, we arrive at the order-$1/r^2$ terms in the wave equation. At this order, the body's tidal moments become coupled to those of the external background. The equation reads
\begin{equation}\label{2n2 wave equation}
\partial^c\partial_c\hmn{\alpha\beta}{2,0} +\frac{1}{r^2}\!\!\left(\hmn{\alpha\beta}{2,0,ln}\!\! +\tilde{E}_{\alpha\beta}\right)=\frac{2}{r^2}\ddR{\alpha\beta}{0,-2}{\hmn{}{1}},
\end{equation}
where $\tilde E_{\alpha\beta}$ comprises the contributions from $\hmn{}{2,-2}$ and $\hmn{}{2,-1}$, given in Eqs.~\eqref{E_tt}, \eqref{E_ta}, and \eqref{E_ab}. The contribution from the second-order Ricci tensor is given in Eqs.~\eqref{ddR0n2_tt}--\eqref{ddR0n2_ab}.

Foregoing the details, after some algebra we can read off the solution
\begin{align}\label{h20}
\hmn{tt}{2,0} & = \A{}{2,0}+\A{i}{2,0}n^i+\A{ij}{2,0}\nhat^{ij}+\A{ijk}{2,0}\nhat^{ijk}\\
\hmn{ta}{2,0} & = \B{}{2,0}n_a+\B{ij}{2,0}\nhat_a{}^{ij} +\C{a}{2,0}+\C{ai}{2,0}\nhat_a{}^i\nonumber\\
&\quad+\epsilon_a{}^{bc}\left(\D{c}{2,0}n_b+\D{ci}{2,0}\nhat_b{}^i +\D{cij}{2,0}\nhat_b{}^{ij}\right)\\
\hmn{ab}{2,0} & = \delta_{ab}\left(\K{}{2,0}+\K{i}{2,0}n^i +\K{ijk}{2,0}\nhat^{ijk}\right)\nonumber\\
&\quad +\E{i}{2,0}\nhat_{ab}{}^i+\E{ij}{2,0}\nhat_{ab}{}^{ij} +\F{\langle a}{2,0}\nhat_{b\rangle} \nonumber\\
&\quad +\F{i\langle a}{2,0}\nhat_{b\rangle}{}^i+\F{ij\langle a}{2,0}\nhat_{b\rangle}{}^{ij} +\epsilon^{cd}{}_{(a}\nhat_{b)c}{}^i\G{di}{2,0}\nonumber\\
&\quad+\H{ab}{2,0}+\H{abi}{2,0}n^i+\epsilon^{cd}{}_{(a}\I{b)d}{2,0}n_c,
\end{align}
where each of the STF tensors is listed in Table~\ref{h20_tensors}.

In solving Eq.~\eqref{2n2 wave equation}, we also find that the logarithmic term in the expansion becomes uniquely determined:
\begin{equation}\label{h2ln}
\hmn{\alpha\beta}{2,0,ln} = -\tfrac{16}{15}m^2\etide_{ab}x^a_\alpha x^b_\beta.
\end{equation}
This term arises because the sources in the wave equation \eqref{2n2 wave equation} contain a term $\propto\etide_{ab}$, which cannot be equated to any term in $\partial^c\partial_c\hmn{ab}{2,0}$. Thus, the wave equation cannot be satisfied without including a logarithmic term. Recall that the logarithmic term arises at the order we would expect it to: the first-order perturbation alters the null cone of the spacetime, such that, e.g., $t-r\to t-r-2\e m\ln r$, which naturally introduces a correction $\sim \e^2\ln r$ to the order-$\e$ terms in the solution to the wave equation.

\begin{table*}
\caption{Symmetric trace-free tensors appearing in the order-$r^0$ part of the metric perturbation $\hmn{}{2}$ in the buffer region around the body. Each tensor is a function of the proper time $t$ on the worldline $\gamma$, and each is STF with respect to the flat-space metric $\delta_{ij}$.}
\begin{ruledtabular}
\begin{tabular}{l|l}
$\begin{array}{ll}
\A{}{2,0} & \phantom{=} \text{ is arbitrary} \\ 
\A{i}{2,0} &= -\partial^2_tM_i-\tfrac{4}{5}S^j\btide_{ji}+\tfrac{1}{3}M^j\etide_{ji}\\
&\quad-\tfrac{7}{5}m\A{i}{1,1}-\tfrac{3}{5}m\K{i}{1,1} +\tfrac{4}{5}m\partial_t\C{i}{1,0}\\
\A{ij}{2,0} &= -\tfrac{7}{3}m^2\etide_{ij}\\
\A{ijk}{2,0} &= -2S_{\langle i}\btide_{jk\rangle} +\tfrac{5}{3}M_{\langle i} \etide_{jk\rangle} -\tfrac{1}{2}m\H{ijk}{1,1} \\
\B{}{2,0} &= m\partial_t\K{}{1,0} \\
\B{ij}{2,0} &= \tfrac{1}{9}\left(2M^l\btide^k_{(i} -5S^l\etide^k_{(i}\right)\epsilon_{j)kl}-\tfrac{1}{2}m\C{ij}{1,1} \\
\C{i}{2,0} & \phantom{=} \text{ is arbitrary} \\
\C{ij}{2,0} &= 2\left(S^l\etide^k_{(i}-\tfrac{14}{15}M^l\btide^k_{(i}\right)\epsilon_{j)lk}
-m\left(\tfrac{6}{5}\C{ij}{1,1}-\partial_t\H{ij}{1,0}\right)\\
\D{i}{2,0} &= \tfrac{1}{5}\left(6M^j\btide_{ij} -7S^j\etide_{ij}\right)+2m\D{i}{1,1} \\
\D{ij}{2,0} &= \tfrac{10}{3}m^2\btide_{ij} \\
\D{ijk}{2,0} &= \tfrac{1}{3}S_{\langle i}\etide_{jk\rangle} +\tfrac{2}{3}M_{\langle i}\btide_{jk\rangle} \\
\K{}{2,0} &= 2\delta m \\
\vphantom{\K{}{}}
\end{array}$
&
$\begin{array}{ll}
\K{i}{2,0} &= -\partial^2_tM_i-\tfrac{4}{5}S^j\btide_{ij} -\tfrac{5}{9}M^j\etide_{ij}\\
&\quad+\tfrac{13}{15}m\A{i}{1,1}\!+\tfrac{9}{5}m\K{i}{1,1}\! -\tfrac{16}{15}m\partial_t\C{i}{1,0}\\
\K{ijk}{2,0} &= -\tfrac{5}{9}M_{\langle i}\etide_{jk\rangle} +\tfrac{2}{9}S_{\langle i}\btide_{jk\rangle}-\tfrac{1}{6}m\H{ijk}{1,1}\\
\E{i}{2,0} &= \tfrac{2}{15}M^i\etide_{ij}+\tfrac{1}{5}S^j\btide_{ij} +\tfrac{1}{10}m\partial_t\C{i}{1,0}
-\tfrac{9}{20}m\K{i}{1,1} \\
&\quad-\tfrac{11}{20}m\A{i}{1,1}\\
\E{ij}{2,0} &= \tfrac{7}{5}m^2\etide_{ij}\\
\F{i}{2,0} &= \tfrac{184}{75}M^j\etide_{ij}+\tfrac{72}{25}S^j\btide_{ij} +\tfrac{46}{25}m\partial_t\C{i}{1,0}
-\tfrac{28}{25}m\A{i}{1,1}\\
&\quad+\tfrac{18}{25}m\K{i}{1,1} \\
\F{ij}{2,0} &= 4m^2\etide_{ij}\\
\F{ijk}{2,0} &= \tfrac{4}{3}M_{\langle i}\etide_{jk\rangle} -\tfrac{4}{3}S_{\langle i}\btide_{jk\rangle} +m\H{ijk}{1,1}\\
\G{ij}{2,0} &= -\tfrac{4}{9}\epsilon^{}_{lk(i}\etide_{j)}^kM^l\! -\tfrac{2}{9}\epsilon^{}_{lk(i}\btide_{j)}^kS^l\!+\tfrac{1}{2}m\I{ij}{1,1}\\
\H{ij}{2,0} & \phantom{=} \text{ is arbitrary} \\
\H{ijk}{2,0} &= \tfrac{58}{15}M_{\langle i}\etide_{jk\rangle} -\tfrac{28}{15}S_{\langle i}\btide_{jk\rangle}+\tfrac{2}{5}m\H{ijk}{1,1}\\
\I{ij}{2,0} &= -\tfrac{104}{45}\epsilon^{}_{lk(i}\etide_{j)}^kM^l -\tfrac{112}{45}\epsilon^{}_{lk(i}\btide_{j)}^kS^l+\tfrac{8}{5}m\I{ij}{1,1}
\end{array}$
\end{tabular}
\end{ruledtabular}
\label{h20_tensors}
\end{table*}

We now move to the final equation in the buffer region: the order-$1/r$ gauge condition. This condition will determine the acceleration $\an{1}$. At this order, $\hmn{E}{1}$ first contributes to Eq.~\eqref{h_2 gauge}:
\begin{equation}
\gauge{\alpha}{1,-1}{\hmn{E}{1}}=\frac{4m}{r}\an{1}_ax^a_\alpha .
\end{equation}
The contribution from $\hmn{}{2}$ is most easily calculated by making use of Eqs.~\eqref{gauge_help1} and \eqref{gauge_help2}. After some algebra, we find that the $t$-component of the gauge condition reduces to
\begin{align}
0&=-\frac{4}{r}\partial_t\delta m +\frac{4m}{3r}\partial_t\A{}{1,0} +\frac{10m}{3r}\partial_t\K{}{1,0},
\end{align}
and the $a$-component reduces to 
\begin{align}\label{accelerations}
0 &=\frac{4}{r}\partial_t^2 M_a+\frac{4m}{r}\an{1}_a+\frac{4}{r}\etide_{ai}M^i+\frac{4}{r}\btide_{ai}S^i\nonumber\\
&\quad-\frac{2m}{r}\A{a}{1,1}+\frac{4m}{r}\partial_t\C{a}{1,0}.
\end{align}
The reader is reminded that these equations are valid only when evaluated at $a(t)=\an{0}(t)=0$, except in the term $\frac{4m}{r}\an{1}_a$ that arose from $\gauge{\alpha}{1}{\hmn{E}{1}}$. In the following subsection, this will allow me to swap partial derivatives with covariant derivatives on the worldline.

The $t$-component determines the rate of change of the mass correction $\delta m$. It can be immediately integrated to find
\begin{align}\label{mdot}
\delta m(t)&=\delta m(0) +\tfrac{1}{6}m\left[2\A{}{1,0}(t)+5\K{}{1,0}(t)\right]\nonumber\\
&\quad-\tfrac{1}{6}m\left[2\A{}{1,0}(0)+5\K{}{1,0}(0)\right].
\end{align}
If one felt so inclined, one could incorporate $\delta m(0)$ into the leading-order mass $m$. The time-dependent terms correspond to the effective mass created by the gravitational waves emitted by the body.

The $a$-component of the gauge condition determines the acceleration of the worldline. Note the most important feature of Eq.~\eqref{accelerations}, which is that it contains two types of accelerations: $\partial_t^2 M_i$ and $\an{1}_i$. The first type is the second time derivative of the body's mass dipole (or the first derivative of its ADM linear momentum), as measured in a frame centered on the worldline $\gamma$. The second type of acceleration is the covariant acceleration of the worldline relative to the external spacetime. In other words, $\partial_t^2M_i$ corresponds to the acceleration of the body's center of mass relative to the center of the coordinate system, while $a_i$ measures the acceleration of the coordinate system itself. We define the worldline to be that of the body if the mass dipole vanishes for all times, meaning that the body is centered on the worldline for all times. If we start with initial conditions $M_i(0)=0=\partial_t M_i(0)$, then the mass dipole remains zero for all times if and only if the worldline satisfies the equation
\begin{equation}\label{a1}
\an{1}_a=\tfrac{1}{2}\A{a}{1,1}-\partial_t\C{a}{1,0}-\tfrac{1}{m}S_i\btide^i_a.
\end{equation}
This equation of motion contains two types of terms: a Papapetrou spin force, given by $-S_i\btide^i_a$, which arises due to the coupling of the body's spin to the local magnetic-type tidal field of the external spacetime; and a self-force, arising from homogenous terms in the wave equation.

Note that if we had followed the path of Gralla and Wald \cite{Gralla_Wald}, we would have identified $\an{0}$ as the acceleration of the worldline $\gamma\coeff{0}$. This would be the only actual worldline in play; all the corrections to the motion would be vectors defined on it. Hence, when we found $\an{0}=0$, we would have identified the worldline as a geodesic, and there would be no corrections $\an{n}$ for $n>0$. We would then have arrived at the equation of motion
\begin{equation}
\partial^2_tM_a+\etide_{ab}M^b = \tfrac{1}{2}\A{a}{1,1}-\partial_t\C{a}{1,0}-\tfrac{1}{m}S_i\btide^i_a.
\end{equation}
This is precisely the equation of motion derived by Gralla and Wald. It describes the drift of the body away from the reference geodesic $\gamma\coeff{0}$. If the external background is flat, then the mass dipole has a valid meaning as a displacement vector regardless of its magnitude; the second derivative $\partial_t^2M_i$ then provides a perfectly valid definition of the body's acceleration for all times. However, if the external background is curved, then $M_i$ has meaning only if the body is ``close" to the worldline. Thus, $\partial_t^2M_i$ is a meaningful acceleration only for short times, since it will generically grow large as the body drifts away from the reference worldline. On that short timescale of validity, the deviation vector defined by $M^i$ accurately points from $\gamma\coeff{0}$ to a ``corrected" worldline $\gamma$; that worldline, the approximate equation of motion of which is given in Eq.~\eqref{a1}, accurately tracks the motion of the body. After a short time, when the mass dipole grows large and the regular expansion scheme begins to break down, the deviation vector will no longer correctly point to the corrected worldline.

To summarize the results of this section, the second-order perturbation in the buffer region is given by $h\coeff{2}_{\alpha\beta} =\frac{1}{r^2}\hmn{\alpha\beta}{2,-2} +\frac{1}{r}\hmn{\alpha\beta}{2,-1} +\hmn{\alpha\beta}{2,0} +\ln(r)\hmn{\alpha\beta}{2,0,ln} +\order{\e,r}$, where $\hmn{}{2,-2}$ is given in Eq.~\eqref{h2n2}, $\hmn{}{2,-1}$ in Eq.~\eqref{h2n1}, $\hmn{}{2,0}$ in Eq.~\eqref{h20}, and $\hmn{}{2,0,ln}$ in Eq.~\eqref{h2ln}. At order $\e^2/r^2$, the metric is written in terms of the mass and spin dipoles of the internal background metric $g_B$. The mass dipole is set to zero by an appropriate choice of worldline. At leading order in $\e$, the body's spin is constant along the worldline. At order $\e^2/r$, there arises an effective correction to the body's mass, given by Eq.~\eqref{mdot}, and the order-$\e$ term in the expansion of the body's acceleration is given by Eq.~\eqref{a1}.

\subsection{Notes on the force and the field in the buffer region}
The foregoing calculation completes the derivation of the gravitational self-force, in the sense that, given the metric perturbation in the neighbourhood of the body, the self-force is uniquely determined by irreducible pieces of that perturbation. Explicitly, the terms that appear in the self-force are given by
\begin{align}
\A{a}{1,1} &= \frac{3}{4\pi}\int n_a\hmn{tt}{1,1}d\Omega,\\
\C{a}{1,0} &= \hmn{ta}{1,0}.
\end{align}
Making use of the fact that $\hmn{\alpha\beta}{1,-1}$ is a monopole, we can write the acceleration  as
\begin{equation}
\an{1}_i = \lim_{r\to0}\left(\frac{3}{4\pi}\int\frac{n_a}{2r}\hmn{tt}{1}d\Omega -\partial_t\hmn{ta}{1}\right) -\tfrac{1}{m}S_i\btide^i_a.
\end{equation}
This is all that is needed to incorporate the motion of the body into a dynamical system that can be numerically evolved; at each timestep, one simply needs to calculate the field near the worldline and decompose it into irreducible pieces in order to determine the acceleration of the body. (Obviously, such a procedure is vastly more complicated than what I have just implied \cite{regularization1, regularization2, regularization3, regularization4, regularization5, regularization6, regularization7}.) The remaining difficulty is to actually determine the field at each timestep. In the next section, I will write down formal expressions for the metric perturbation, and in particular, I will determine the metric perturbation at the location of the body in terms of a tail integral.

However, before doing so, I will emphasize some important features of the self-force and the field near the body. First, note that the first-order external field $\hmn{E}{1}$ separates into two distinct pieces. There is the singular piece $h^S$, given by
\begin{align}
h^S_{tt} &= \frac{2m}{r}\Big\lbrace 1+\tfrac{3}{2}ra_in^i+2r^2a_ia^i\nonumber\\
&\quad+r^2\left(\tfrac{3}{8}a_{\langle i}a_{j\rangle}+\tfrac{5}{6}\etide_{ij}\right)\nhat^{ij}\Big\rbrace+ \order{r^2} \\
h^S_{ta} &= -2mr\dot a_a +\tfrac{2}{3}m r\epsilon_{aij}\btide^j_k\nhat^{ik}+\order{r^2} \\
h^S_{ab} &= \frac{2m}{r}\Big\lbrace\delta_{ab}\big[1-\tfrac{1}{2}ra_in^i +\tfrac{2}{3}r^2a_ia^i\nonumber\\
&\quad +r^2\left(\tfrac{3}{8}a_{\langle i}a_{j\rangle}-\tfrac{5}{18}\etide_{ij}\right)\nhat^{ij}\big] +2a_{\langle a}a_{b\rangle}\nonumber\\
&\quad -\tfrac{19}{9}r^2\etide_{ab}+\tfrac{2}{3}r^2\etide^i_{\langle a}\nhat_{b\rangle i}\Big\rbrace+\order{r^2}.
\end{align}
This field is a solution to the homogenous wave equation for $r>0$, but it is divergent at $r=0$. It is the generalization of the $1/r$ Newtonian field of the body, as perturbed by the tidal fields of the external spacetime $g$. Following the method used in Sec.~5.3.5 of Ref.~\cite{Eric_review}, one can easily show that this is precisely the Detweiler-Whiting singular field, given by
\begin{equation}
h^S_{\alpha\beta}=4m\int_\gamma \bar G^S_{\alpha\beta\alpha'\beta'}u^{\alpha'}u^{\beta'}dt',
\end{equation}
where $G^S_{\alpha\beta\alpha'\beta'}$ is the singular Green's function (defined in Appendix~\ref{Greens_functions}).

Next, there is the regular field $h^R\equiv \hmn{E}{1}-h^S$, given by
\begin{align}
h^R_{tt} &= \A{}{1,0}+r\A{i}{1,1}n^i+\order{r^2}, \\
h^R_{ta} &= \C{a}{1,0} +r\Big(\B{}{1,1}n_a+\C{ai}{1,1}n^i +\epsilon_{ai}{}^j\D{j}{1,1}n^i\Big)\nonumber\\
&\quad+\order{r^2},\\
h^R_{ab} &= \delta_{ab}\K{}{1,0}+\H{ab}{1,0}+r\Big(\delta_{ab}\K{i}{1,1}n^i+\H{abi}{1,1}n^i \nonumber\\
&\quad+\epsilon\indices{_i^j_{(a}}\I{b)j}{1,1}n^i+\F{\langle a}{1,1}n^{}_{b\rangle}\Big)+\order{r^2}.
\end{align}
This field is a solution to the homogeneous wave equation even at $r=0$. It is a free radiation field in the neighbourhood of the body. And it contains all the free functions in the buffer-region expansion.

Now, the acceleration of the body is given by
\begin{align}
\an{1}_a = \tfrac{1}{2}\partial_ah^R_{tt} -\partial_t h^R_{ta} -\tfrac{1}{m}S_i\btide^i_a,
\end{align}
which we can rewrite as 
\begin{align}
a^\alpha &= -\tfrac{1}{2}\left(g^{\alpha\delta}+u^{\alpha}u^{\delta}\right) \!\left(2h^R_{\delta\beta;\gamma}-h^R_{\beta\gamma;\delta}\right)\!\!\big|_{a=0} u^{\beta}u^{\gamma}\nonumber\\
&\quad +\frac{1}{2m}R^{\alpha}{}_{\beta\gamma\delta}u^\beta S^{\gamma\delta}
\end{align}
where $S^{\gamma\delta}\equiv e_c^\gamma e_d^\delta\epsilon^{cdj}S_j$. In other words, a non-spinning body (for which $S^{\gamma\delta}=0$), moves on a geodesic of a spacetime $g+\e h^R$, where $h^R$ is a free radiation field in the neighbourhood of the body; a local observer would measure the ``background spacetime,'' in which the body is in free fall, to be $g+\e h^R$, rather than $g$. If we performed a transformation into Fermi coordinates in $g+\e h^R$, the metric would contain no acceleration term, and it would take the simple form of a smooth background plus a singular perturbation. These points were first realized by Detweiler and Whiting \cite{Detweiler_Whiting} and since emphasized especially by Detweiler \cite{Detweiler_review}. They are, perhaps, made especially clear in the derivation presented here, which naturally demarcates the singular and regular fields.

As a final note, I remind the reader that the equation of motion would contain an antidamping term \cite{damping, damping2,Quinn_Wald} if I had not assumed that the acceleration possesses an expansion of the form given in Eq.~\eqref{a expansion}.

\section{The metric perturbation}\label{perturbation calculation}

\subsection{Integral formulation of the Einstein equation in the external spacetime}\label{integral_expansion}
\begin{figure}[tb]
\begin{center}
\includegraphics{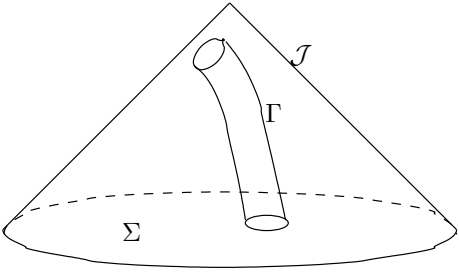}
\end{center}
\caption{The spacetime region $\Omega$ is bounded by the union of the spacelike surface $\Sigma$, the timelike worldtube $\Gamma$, and the null surface $\mathcal{J}$.} 
\label{volume}
\end{figure}

A solution to the self-force problem consists of a pair $(\gamma,h)$. In the previous section, we have determined the equation of motion of $\gamma$; we now require a means of determining the metric perturbation. Specifically, on the external manifold $\man_E$, I seek an approximate solution of Einstein's equation in a vacuum region $\bar\Omega\equiv\Omega\cup\partial\Omega$, where $\Omega$ is a bounded, open subset of $\man_E$. I now specify this region to be the future range of dependence of a surface formed by the union of a worldtube $\Gamma$ and a spatial surface $\Sigma$. This implies that the future boundary of $\Omega$ is a null surface $\mathcal{J}$. Refer to Fig.~\ref{volume} for an illustration. The boundary of the domain is hence $\partial\Omega \equiv \Gamma\cup\mathcal{J}\cup\Sigma$. The worldtube $\Gamma$ is defined by a constant Fermi radial coordinate distance $r=\rad$ from the worldline $\gamma\subset\man_E$. Since the tube is an artificial division of spacetime, and it may be located anywhere in the buffer region, any valid solution cannot depend on $\rad$. The spatial surface $\Sigma$ is chosen to intersect $\Gamma$ at the initial time $t=0$.

In $\Omega$, the Lorenz gauge is imposed on the entire perturbation $h$, splitting the Einstein equation into the weakly nonlinear wave equation
\begin{align}\label{wave_tube}
E_{\alpha\beta}\big[h\big] &=2\delta^2R_{\alpha\beta}\big[h\big]+\order{\e^3}
\end{align}
and the gauge condition $L_{\mu}\big[h\big]=0$. Note that if a solution to the wave equation satisfies the gauge condition on $\partial\Omega$, then the wave equation ensures that the gauge condition is satisfied everywhere. And since I have already determined the equation of motion using the expansion in the buffer region, I will hence not be interested in the gauge condition here.

As discussed in Secs.~\ref{singular_expansion_point_particle} and \ref{outline}, I assume the expansion $h_{\alpha\beta}(x,\e;\gamma)=\sum_n\e^n\hmn{E\alpha\beta}{n}(x;\gamma)$ and arrive at the sequence of wave equations
\begin{align}
E_{\alpha\beta}\big[\hmn{E}{1}\big] & = 0, \label{first_order_wave_tube}\\
E_{\alpha\beta}\big[\hmn{E}{2}\big] & =  2\delta^2R_{\alpha\beta}[\hmn{E}{1}]. \label{second_order_wave_tube}
\end{align}
Following D'Eath \cite{DEath,DEath_paper,Eric_review}, I rewrite the wave equations as integro-differential equations by calculating $E[G^{\text{adv}}]h-E[h]G^{\text{adv}}$ (where $G^{\text{adv}}$ represents the advanced Green's function for $E_{\mu\nu}$), integrating both sides of the resulting equation, making use of Stokes' law, and finally simplifying the result using the reciprocality relation $G^{\text {adv}}_{\alpha'\beta'\alpha\beta}(x',x)= G_{\alpha\beta\alpha'\beta'}(x,x')$. The resulting equations are
\begin{align}
\hmn{E\alpha\beta}{1} & = \frac{1}{4\pi}\oint\limits_{\partial\Omega}\! \Big(G_{\alpha\beta}{}^{\gamma'\delta'}\hmn{E\gamma'\delta';\mu'}{1} -\hmn{E\gamma'\delta'}{1} G_{\alpha\beta}{}^{\gamma'\delta'}{}_{;\mu'}\Big)dS^{\mu'}\!,\label{first_order_Kirchoff}\\
\hmn{E\alpha\beta}{2} & = \frac{1}{4\pi}\oint\limits_{\partial\Omega}\! \Big(G_{\alpha\beta}{}^{\gamma'\delta'}\hmn{E\gamma'\delta';\mu'}{2} -\hmn{E\gamma'\delta'}{2} G_{\alpha\beta}{}^{\gamma'\delta'}{}_{;\mu'}\Big)dS^{\mu'}\nonumber\\
&\quad-\frac{1}{2\pi}\int\limits_\Omega G_{\alpha\beta}{}^{\gamma'\delta'} \delta^2R_{\gamma'\delta'}[\hmn{E}{1}]dV'. \label{second_order_Kirchoff}
\end{align}

Alternatively, we might rewrite Eq.~\eqref{wave_tube} directly:
\begin{align}\label{all_orders_Kirchoff}
h_{\alpha\beta} & = \frac{1}{4\pi}\oint\limits_{\partial\Omega}\Big(G_{\alpha\beta}{}^{\gamma'\delta'} \del{\mu'}h_{\gamma'\delta'}-h_{\gamma'\delta'}\del{\mu'} G\indices{_{\alpha\beta}^{\gamma'\delta'}}\Big) dS^{\mu'}\nonumber\\
&\quad-\frac{1}{2\pi}\int_\Omega G_{\alpha\beta}{}^{\gamma'\delta'}\delta^2R_{\gamma'\delta'}\big[h\big]dV'+O(\e^3).
\end{align}
Note that any solution to Eq.~\eqref{wave_tube} in $\Omega$ will also satisfy this integro-differential equation; however, because of the $\e$-dependence of the true worldline $\gamma$, not every solution to Eq.~\eqref{wave_tube} will admit an expansion satisfying the two equations~\eqref{first_order_wave_tube} and \eqref{second_order_wave_tube} (though a solution to the latter is obviously a solution to the former). In that sense, Eq.~\eqref{all_orders_Kirchoff} is more robust than Eqs.~\eqref{first_order_Kirchoff} and \eqref{second_order_Kirchoff}.

In any case, these integral representations all have several important properties in common. First, the integral over the boundary is, in each case, a homogeneous solution to the wave equation, while the integral over the interior is an inhomogeneous solution.\footnote{The integral over the interior will also contain homogeneous solutions. However, these will be $\rad$-dependent, and they will exactly cancel corresponding $\rad$-dependent terms in the boundary integral.} Second, the integral over the boundary can be split into an integral over the worldtube $\Gamma$ and the spatial surface $\Sigma$; the contribution of the null surface $\mathcal{J}$ vanishes by construction. Also note that $x$ must lie in the interior of $\Omega$; an alternative expression must be derived if $x$ lies on the boundary \cite{Greens_functions}. 

Furthermore, the integral representations avoid any divergence in the second-order solution. Comparing Eq.~\eqref{all_orders_Kirchoff} to the analogous expression for a point particle, given in Eqs.~\eqref{1st_funct} and \eqref{2nd_funct}, we see that the point particle source terms have been replaced by an integral over a worldtube surrounding the small body, as we desired. And the volume integral over the interior of $\Omega$ does not diverge in $\Omega$, as it would in Eq.~\eqref{2nd_funct}, because the region of integration excludes the interior of the worldtube. 

Finally, one should note the essential character of these integrals. They provide a type of Kirchoff representation \cite{Friedlander, Sciama, Eric_review} of a solution to the wave equation \eqref{wave_tube}. However, while the integral representation is satisfied by any solution to the associated wave equation, it does not \emph{provide} a solution. That is, one cannot prescribe arbitrary boundary values on $\Gamma$ and then arrive at a solution. The reason is that the worldtube is a timelike boundary, which means that field data on it can propagate forward in time and interfere with the data at a later time. However, by applying the wave operator $E_{\alpha\beta}$ onto equation \eqref{all_orders_Kirchoff}, we see that the Kirchoff representation of $h$ is guaranteed to satisfy the wave equation at each point $x\in\Omega$. In other words, the problem arises not in satisfying the wave equation in a pointwise sense, but in simultaneously satisfying the boundary conditions. However, since the tube is chosen to lie in the buffer region, these boundary conditions can be supplied by the buffer-region expansion. This can presumably be accomplished in a variety of ways, two of which I will discuss presently. Note that since the buffer-region expansion has been made to satisfy the Lorenz gauge to some order in $\rad$, using it as boundary data will enforce the Lorenz gauge in $\Omega$ to the same order.

Now, recall that in almost all the derivations of the gravitational self-force (excluding those in Refs.~\cite{Fukumoto, Gralla_Wald}), the first-order external perturbation was assumed to be that of a point particle. This was justified to some extent by an argument first made by D'Eath \cite{DEath, DEath_paper} and later used by Rosenthal \cite{Eran_field}. The argument is based on the integral Eq.~\eqref{first_order_Kirchoff} and the asymptotically small size of the worldtube. First, note that the directed area element on the worldtube behaves as $\sim\rad^2(-n^{\mu'})$. Also, in constructing the external solution, we formally assume $r\sim 1$ (since the limit is constructed with fixed coordinate values in mind), which means that we can treat the Green's functions and its derivatives as quantities of order unity. Thus, the dominant term in the worldtube integral is determined by the derivative of the $m/r$ term in $\hmn{E}{1}$; using the result from the buffer-region expansion, this yields
\begin{align}
-\rad^2 n^{\mu'}\del{\mu'}\!\!\left[\frac{2m}{r'}(2u_{\alpha'}u_{\beta'} +g_{\alpha'\beta'})\right]\!\!\bigg|_{r'=\rad} \nonumber\\
=2m(2u_{\alpha'}u_{\beta'}+g_{\alpha'\beta'})+\order{\rad}.
\end{align}
Hence, the boundary integral can be written as
\begin{align}\label{external tube approx}
\hmn{E\alpha\beta}{1} &= \frac{1}{2\pi}\int\limits_{\Gamma} mG_{\alpha\beta}{}^{\gamma'\delta'}(2u_{\alpha'}u_{\beta'} +g_{\alpha'\beta'})dt'd\Omega' \nonumber\\
&\quad+\hmn{\Sigma\alpha\beta}{1}+\order{\rad},
\end{align}
where $\hmn{\Sigma\alpha\beta}{1}$ is the contribution from the initial data surface $\Sigma$. Expanding the Green's function on the worldtube about the worldline $\gamma$, this becomes
\begin{align}\label{external point particle}
\hmn{E\alpha\beta}{1} &= \int\limits_\gamma 2mG_{\alpha\beta\bar\alpha\bar\beta}(2u^{\bar\alpha}u^{\bar\beta} +g^{\bar\alpha\bar\beta})d\bar t \nonumber\\
&\quad+\hmn{\Sigma\alpha\beta}{1}+\order{\rad},
\end{align}
where the barred coordinates correspond to points on the worldline, and $\bar t$ is proper time, running from $\bar t=0$ to $\bar t\sim 1/\e$. Equation \eqref{external point particle} is the solution to the wave equation with a point particle source---except for the corrections of order $\rad$. It can be put in the more usual form of Eq.~\eqref{1st_funct} by using the identity \eqref{Green3}.

In the original derivation presented by D'Eath \cite{DEath,DEath_paper}, $\rad$ was set to zero with no explicit justification. In Rosenthal's later derivations \cite{Eran_field}, this step was justified based on the notion that we are interested in the limit in which the small body shrinks to a point. However, if the size of the body vanishes, then so too does its mass, in which case there is no perturbation at all; and at second order, setting $\rad$ to zero would create a divergent solution. Hence, discarding the order-$\rad$ corrections based on this argument is not justified. We could also argue that the order-$\rad$ terms must be discarded because the external solution cannot depend on the arbitrary radius of the tube. However, this second argument is also specious: One could just as easily express Eq.~\eqref{external tube approx} as an integral over \emph{any} curve in the interior of $\Gamma$, rather than the central curve $\gamma$. But if one did so, then one would introduce mass dipole terms into the metric, and an explicit calculation of the error terms would show that they do \emph{not} vanish. In some sense, this correctly implies that the choice of worldline at leading order is inconsequential, since any choice within the worldtube results only in the introduction of a mass dipole, which is a second-order term, and the self-force will by definition set the resulting mass dipole to zero. However, this resolution becomes murky when we consider that the size of the tube must be left arbitrary to achieve a valid solution, and the mass dipole in the buffer region calculation is precisely order $\e^2$, rather than order $\e\rad$.

Instead, I present here an alternative argument to justify D'Eath's conclusion: Suppose we take our buffer region expansion of $\hmn{E}{1} $ to be valid everywhere in the interior of $\Gamma$ (in $\man_E$), rather than just in the buffer region. This is a meaningful supposition in a distributional sense, since the $1/r$ singularity in $\hmn{E}{1}$ is locally integrable even at $\gamma$. Note that the extension of the buffer-region expansion is not intended to provide an accurate or meaningful approximation in the interior; it is used only as a means of determining the field in the exterior. The reason I can do this is as follows: since the field values in $\Omega$ are entirely determined by the field values on $\Gamma$, using the buffer-region expansion in the entire interior of $\Gamma$ leaves the field values in $\Omega$ unaltered. Now, given the extension of the buffer-region expansion, it follows from Stokes' law that the integral over $\Gamma$ in Eq.~\eqref{first_order_Kirchoff} can be replaced by a volume integral over the interior of the tube, plus two surface integrals over the ``caps" $\mathcal{J}_{cap}$ and $\Sigma_{cap}$, which fill the ``holes" in $\mathcal{J}$ and $\Sigma$, respectively, where they intersect $\Gamma$. Schematically, we can write Stokes' law as $\int_{\text{Int}(\Gamma)}=\int_{\mathcal{J}_{cap}} +\int_{\Sigma_{cap}}-\int_{\Gamma}$, where $\text{Int}(\Gamma)$ is the interior of $\Gamma$; this is valid as a distributional identity in this case.\footnote{Note that the ``interior" here means the region bounded by $\Gamma\cup\Sigma_{cap}\cup\mathcal{J}_{cap}$. $\text{Int}(\Gamma)$ does not refer to the set of interior points in the point-set defined by $\Gamma$.} The minus sign in front of the integral over $\Gamma$ accounts for the fact that the directed surface element in Eq.~\eqref{first_order_Kirchoff} points \emph{into} the tube. Because $\mathcal{J}_{cap}$ does not lie in the past of any point in $\Omega$, it does not contribute to the perturbation at $x\in\Omega$. Hence, we can rewrite Eq.~\eqref{first_order_Kirchoff} as 
\begin{align}
\hmn{E\alpha\beta}{1} &= -\frac{1}{4\pi}\!\!\!\int\limits_{\text{Int}(\Gamma)}\!\!\! \del{\mu'}\Big(G_{\alpha\beta}{}^{\alpha'\beta'}\nabla^{\mu'}\hmn{E\alpha'\beta'}{1} \nonumber\\
&\quad-\hmn{E\alpha'\beta'}{1}\nabla^{\mu'}G_{\alpha\beta}{}^{\alpha'\beta'}\Big)dV' +\hmn{\bar\Sigma\alpha\beta}{1}\nonumber\\
&=-\frac{1}{4\pi}\!\!\!\int\limits_{\text{Int}(\Gamma)}\!\!\! \Big(G_{\alpha\beta}{}^{\alpha'\beta'}E_{\alpha'\beta'}[\hmn{E}{1}] \nonumber\\
&\quad-\hmn{E\alpha'\beta'}{1}E^{\alpha'\beta'}[G_{\alpha\beta}]\Big)dV' +\hmn{\bar\Sigma\alpha\beta}{1},
\end{align}
where $\hmn{\bar\Sigma\alpha\beta}{1}$ is the contribution from the spatial surface $\bar\Sigma\equiv\Sigma\cup\Sigma_{cap}$. Now note that $E^{\alpha'\beta'}[G_{\alpha\beta}]\propto\delta(x,x')$; since $x\notin\text{Int}(\Gamma)$, this term integrates to zero. Next note that $E_{\alpha'\beta'}[\hmn{E}{1}]$ vanishes everywhere except at $\gamma$. This means that the field at $x$ can be written as
\begin{align}
\hmn{E\alpha\beta}{1} &= \frac{-1}{4\pi}\!\lim_{\rad\to 0}\!\!\!\int\limits_{\text{Int}(\Gamma)} \!\!\!\!\!G_{\alpha\beta}{}^{\alpha'\beta'}E_{\alpha'\beta'}[\hmn{E}{1}]dV'+\hmn{\bar\Sigma\alpha\beta}{1}.
\end{align}
Making use of the fact that $E_{\alpha\beta}[\hmn{E}{1}] = \partial^c\partial_c(1/r)\hmn{E\alpha\beta}{1,-1}+\order{r^{-2}}$, along with the identity $\partial^c\partial_c(1/r)=-4\pi\delta^3(x^a)$, where $\delta^3$ is a coordinate delta function in Fermi coordinates, we arrive at the desired result
\begin{equation}
\hmn{E\alpha\beta}{1} = 2m\int_\gamma G_{\alpha\beta\bar\alpha\bar\beta}(2u^{\bar\alpha}u^{\bar\beta} +g^{\bar\alpha\bar\beta})d\bar t+\hmn{\bar\Sigma\alpha\beta}{1}.
\end{equation}
Thus, simply neglecting the $\order{\rad}$ terms in Eq.~\eqref{external point particle} yields the correct result, and in the region $\Omega$, the leading-order perturbation produced by the asymptotically small body is identical to the field produced by a point particle.

Gralla and Wald \cite{Gralla_Wald} have provided an alternative derivation of the same result, using distributional methods to prove that the distributional source for the linearized Einstein equation must be that of a point particle in order for the solution to diverge as $1/r$. One can understand this by considering that the most divergent term in the linearized Einstein tensor is a Laplacian acting on the perturbation, and the Laplacian of $1/r$ is a flat-space delta function; the less divergent corrections are due to the curvature of the background, which distorts the flat-space distribution into a covariant curved-spacetime distribution.

At second-order, the above method can be used to simplify Eq.~\eqref{second_order_Kirchoff} by replacing at least part of the integral over $\Gamma$ with an integral over $\gamma$. I will not pursue this simplification here, however. Instead, I will present an alternative means of determining the metric perturbation. This method is based on a direct calculation of the boundary integral in Eq.~\eqref{all_orders_Kirchoff}. As such, it is somewhat similar in spirit to the Direct Integration of the Relaxed Field Equations (DIRE) used by Will et al. in post-Newtonian theory \cite{DIRE}. While the method used above relied on $E_{\mu\nu}[h]$ being well defined as a distribution, a direct integration of the boundary integral can be performed, in principle, regardless of the behavior of $h$ in the buffer region. Hence, it might be used at any order in perturbation theory.

The method of direct integration proceeds as follows. As noted above, the Kirchoff representation of the solution is guaranteed to satisfy the wave equation at all points in $\Omega$, but it provides a valid solution only if, in addition, it agrees with the data on the boundary $\partial\Omega$. Thus, the Kirchoff representation is guaranteed to be a $C^1$ solution in $\bar\Omega$ if it satisfies the consistency conditions
\begin{equation}
\begin{split}
\lim_{x\to x'}h_{\alpha\beta} & = h_{\alpha'\beta'} \\
\lim_{x\to x'}n^\mu\del{\mu}h_{\alpha\beta} & = n^{\mu'}\del{\mu'}h_{\alpha'\beta'}
\end{split}\quad
\text{for } x'\in\Gamma.
\end{equation}
However, these conditions allow $h$ to contain a term such as $(r-\rad)^2\ln(r-\rad)$; both the term itself and its first derivative vanish in the limit $r\to\rad$, but the second-derivative does not. Since we seek a solution that is smooth and independent of $\rad$, I demand that $h$ satisfy the following, stronger condition: Since the radius $\rad$ of the tube is small, the boundary data $h'$ can be expressed as an expansion in powers of $\rad$ and $\e$---this is the buffer-region expansion. If $x$ is near the worldtube, then $r\sim\rad$, meaning that $h_{\alpha\beta}$ can similarly be expressed as an expansion in powers of $r$ and $\e$. Defining $\expand_s(f(s))$ to be an expansion of $f$ for small $s$, I write the expansion of the boundary values as $\expand_\e(\expand_\rad(h'))$, and I write the expansion of the integral representation of the solution in $\Omega$ as $\expand_\e(\expand_r(h))$. I demand that these expansions are identical:
\begin{equation}\label{consistency}
\expand_\e(\expand_\rad(h'))\big|_{\rad=r}=\expand_\e(\expand_r(h)).
\end{equation}
Hence, by expanding the integral representation of the perturbation near the worldtube and insisting that the result is consistent with the boundary data provided by the buffer-region expansion, all the free functions in the buffer region expansion will be determined. 

Since the equation of motion depends only on first-order terms, for the purposes of this paper I will limit the expansion just described to first order. The expansion is performed only in the buffer region, meaning that it provides an explicit expression for the perturbation only in that region. However, by imposing the consistency condition, the boundary data on $\Gamma$ can be determined to any desired order of accuracy in $\rad$; using this boundary data, the solution in $\Omega$ will then be determined to the same order of accuracy. A similar procedure could be adopted at second order and above. At those orders, the expansion of the boundary integral would yield $\rad$-dependent terms that would be grouped with the volume integral over $\Omega$; this combination would yield an approximation to the inhomogenous part of the solution. The homogenous part of the solution would be dealt with in the same manner as the first-order perturbation.

\subsection{The boundary integral}\label{integral_expansion1}

\begin{figure}[tb]
\begin{center}
\includegraphics{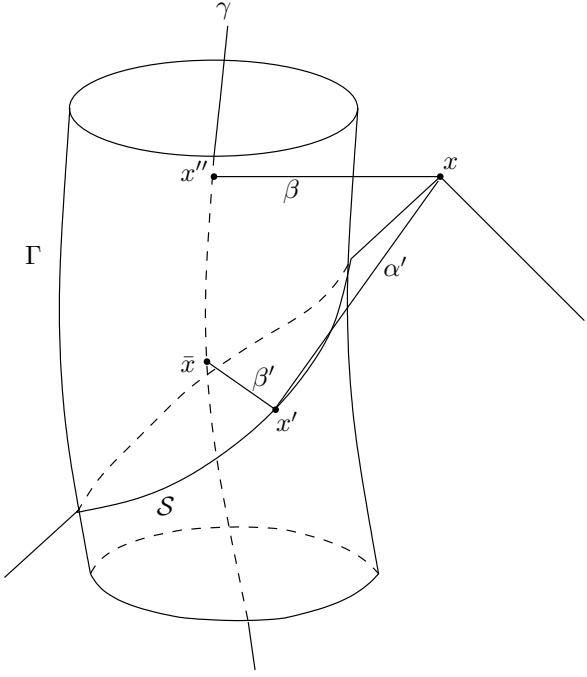}
\end{center}
\caption{The two-dimensional hypersurface $\mathcal{S}$ is defined by the intersection of the worldtube $\Gamma$ with the past light cone of the point $x$. $x$ is linked to a point $x'\in\mathcal{S}$ by a null geodesic $\alpha'$. $x$ and $x'$ are separately linked to points $x''=\gamma(t)$ and $\bar{x}=\gamma(t')$ by spacelike geodesics $\beta$ and $\beta'$, each of which is perpendicular to $\gamma$.} 
\label{tube}
\end{figure}

Since the calculation in this section is intended primarily as a proof of principle, rather than calculating $\hmn{E}{1}$ I will calculate its approximation $\hmn{}{1}$; in other words, I will consistently neglect acceleration terms. Hence I take the boundary data on the tube to be defined by $h=\frac{\e}{r}\hmn{}{1,-1}+\e\hmn{}{1,0}+\e r\hmn{}{1,1}$, and the field outside the tube to be the expansion of 
\begin{align}
h_{\alpha\beta}&=\frac{1}{4\pi}\oint\limits_{\partial\Omega}\! \Big(G_{\alpha\beta}{}^{\gamma'\delta'}\hmn{\gamma'\delta';\mu'}{1}
-\hmn{\gamma'\delta'}{1}G\indices{_{\alpha\beta}^{\gamma'\delta'}_{;\mu'}}\Big) dS^{\mu'}\nonumber\\
&\quad+\order{\e^2}
\end{align}
to order $\e r$. Since the volume integral contributes only $\order{\e^2,r^2}$ terms, it is neglected here.

Two parts of the boundary lie within the causal past of $x$: the spatial hypersurface $\Sigma$ and the worldtube $\Gamma$. The contribution to the field from the data on $\Sigma$ is given by
\begin{equation}\label{Sigma contribution}
\hmn{\Sigma\alpha\beta}{1}=\frac{1}{4\pi}\int\limits_{\Sigma}\! \Big(G_{\alpha\beta}{}^{\gamma'\delta'}\hmn{\gamma'\delta';\mu'}{1}
-\hmn{\gamma'\delta'}{1}G\indices{_{\alpha\beta}^{\gamma'\delta'}_{;\mu'}}\Big) dS^{\mu'},
\end{equation}
where the data $\hmn{\gamma'\delta'}{1}$ is constrained to satisfy the Lorenz gauge and merge smoothly with the buffer region expansion. I assume that $\hmn{\Sigma}{1}$ can be expanded in a regular power series in $r$,
\begin{equation}
\hmn{\Sigma\alpha\beta}{1}=\sum_{m\ge 0}r^m\hmn{\Sigma\alpha\beta}{1,\emph{m}},
\end{equation}
and that each $\hmn{\Sigma\alpha\beta}{1,\emph{m}}$ can be decomposed into irreducible STF pieces. Because this data can only contribute to the homogenous, free functions in the buffer region expansion, we can infer the nonzero pieces of $\hmn{\Sigma\alpha\beta}{1}$ from that expansion:
\begin{align}
\hmn{\Sigma tt}{1,0}&=\A{\Sigma}{1,0}, \\
\hmn{\Sigma ta}{1,0}&=\C{\Sigma a}{1,0}, \\
\hmn{\Sigma ab}{1,0}&=\delta_{ab}\K{\Sigma}{1,0}+\H{\Sigma ab}{1,0},\\
\hmn{\Sigma tt}{1,1}&=\A{\Sigma i}{1,1}n^i, \\
\hmn{\Sigma ta}{1,1}&=\B{\Sigma}{1,1}n_a+\C{\Sigma ai}{1,1}n^i +\epsilon_{ai}{}^j\D{\Sigma j}{1,1}n^i,\\
\hmn{\Sigma ab}{1,1}&=\delta_{ab}\K{\Sigma i}{1,1}n^i+\H{\Sigma abi}{1,1}n^i +\epsilon\indices{_i^j_{(a}}\I{\Sigma b)j}{1,1}n^i \nonumber\\
&\quad+\F{\Sigma\langle a}{1,1}n^{}_{b\rangle}.
\end{align}

Now consider the integration over $\Gamma$. In the buffer region, I define $\lambda(\e)$ such that $r\sim\rad\sim\lambda(\e)$. Since the function $\lambda(\e)$ is arbitrary, except that it must vanish in the limit $\e\to0$, we can use $\e$ and $\lambda$ as independent expansion parameters. The volume element on $\Gamma$ is given by $dS_{\mu'}=-n_{\mu'}N(x')\rad^2dt'd\Omega'$, where $N(x)=1+\tfrac{1}{3}\etide_{cd}(t)x^{cd}+\order{\lambda^3,\e}$, and $t'$, $\rad$, and $\theta'^A$ are Fermi coordinates based at $\gamma$. The boundary data is constructed from 
\begin{equation}
\hmn{\gamma'\delta'}{1}=\frac{1}{\rad}\hmn{\gamma'\delta'}{1,-1} +\hmn{\gamma'\delta'}{1,0} +\rad\hmn{\gamma'\delta'}{1,1} +\order{\lambda^2}
\end{equation}
where $\hmn{\gamma'\delta'}{1,-1}$, $\hmn{\gamma'\delta'}{1,0}$, and $\hmn{\gamma'\delta'}{1,1}$ are obtained by setting the acceleration to zero in Eqs.~\eqref{h1n1}, \eqref{h10}, and \eqref{h11}.

The integral over $\Gamma$ can be divided into two regions: the convex normal neighbourhood $\mathcal{N}$ of $x$---consisting of all the points that are connected to $x$ by unique geodesics---and the complement of the convex normal neighbourhood. In $\mathcal{N}$, the Green's function admits the Hadamard decomposition \cite{Eric_review}
\begin{equation}
G_{\alpha\beta}{}^{\gamma'\delta'}= U_{\alpha\beta}{}^{\gamma'\delta'}\delta_+(\sigma) +V_{\alpha\beta}{}^{\gamma'\delta'}\theta_+(-\sigma),
\end{equation}
where $\sigma(x,x')$ is Synge's world function, which is equal to one-half the squared geodesic distance between $x$ and $x'$. Derivatives of this biscalar will be denoted by, e.g., $\sigma_\mu\equiv\sigma_{;\mu}$. The delta function $\delta_+(\sigma(x,x'))$ has support on the past light cone of $x$, while the Heaviside function $\theta_+(-\sigma(x,x'))$ has support within the past light cone.

Substituting these expressions into the boundary integral, we find that it can be broken into several pieces:
\begin{align}
4\pi h_{\alpha\beta} &= \int\limits_{\Gamma\cap\mathcal{N}}\!\! \Big[\h^{\text{tail}}_{\alpha\beta}\theta_+(-\sigma) +\h^{\text{dir}1}_{\alpha\beta}\delta_+(\sigma)\nonumber\\ &\qquad\quad+\h^{\mathrm{dir}2}_{\alpha\beta}\delta'_+(\sigma)\Big]Ndt'd\Omega'\nonumber\\
&\quad+\!\!\!\!\!\!\!\!\!\int\limits_{\ (\Gamma\setminus\mathcal{N})\cap\past}\!\!\!\!\!\!\!\!\!\h^{\text{tail}}_{\alpha\beta}Ndt' d\Omega' +\hmn{\Sigma\alpha\beta}{1}+\order{\e^2},
\end{align}
where $\past$ is the past of $x$, and $\delta'$ is the derivative of the delta function. Inside the normal neighbourhood, the terms in the integrand are given by
\begin{align}
\h^{\text{tail}}_{\alpha\beta} &= \Big(\hmn{\gamma'\delta'}{1,-1}-\rad\del{n'}\hmn{\gamma'\delta'}{1,-1}\Big) V\indices{_{\alpha\beta}^{\gamma'\delta'}} +\rad\hmn{\gamma'\delta'}{1,-1} \del{n'}V\indices{_{\alpha\beta}^{\gamma'\delta'}}\nonumber\\
&\quad+\order{\lambda^2},\\
\h^{\text{dir}1}_{\alpha\beta} &= \Big(\hmn{\gamma'\delta'}{1,-1}\!\!-\!\rad\del{n'}\hmn{\gamma'\delta'}{1,-1}\!\! -\!\rad^2\del{n'}\hmn{\gamma'\delta'}{1,0}\!\! -\!\rad^2\hmn{\gamma'\delta'}{1,1}\Big) U\indices{_{\alpha\beta}^{\gamma'\delta'}}\nonumber\\
&\quad+\Big(\rad\hmn{\gamma'\delta'}{1,-1} +\rad^2\hmn{\gamma'\delta'}{1,0}\Big)\del{n'} U\indices{_{\alpha\beta}^{\gamma'\delta'}}\nonumber\\
&\quad-\rad \hmn{\gamma'\delta'}{1,-1}V\indices{_{\alpha\beta}^{\gamma'\delta'}}\sigma_{\mu'}n^{\mu'} +\order{\lambda^3},\\
\h^{\text{dir}2}_{\alpha\beta} &= \Big(\rad\hmn{\gamma'\delta'}{1,-1}+\rad^2\hmn{\gamma'\delta'}{1,0} +\rad^3\hmn{\gamma'\delta'}{1,1}\Big)U\indices{_{\alpha\beta}^{\gamma'\delta'}} \sigma_{\mu'}n^{\mu'}\nonumber\\
&\quad+\order{\lambda^5},
\end{align}
where $\del{n'}\equiv n'^\alpha\del{\alpha'}$. The ``direct" and ``tail" titles should be self-explanatory. Outside the normal neighbourhood, the term in the integrand is
\begin{align}
\h^{\text{tail}}_{\alpha\beta} &= \Big(\hmn{\gamma'\delta'}{1,-1}-\rad\del{n'}\hmn{\gamma'\delta'}{1,-1}\Big) G\indices{_{\alpha\beta}^{\gamma'\delta'}} \nonumber\\
&\quad+\rad\hmn{\gamma'\delta'}{1,-1} \del{n'}G\indices{_{\alpha\beta}^{\gamma'\delta'}}+\order{\lambda^2}.
\end{align}

These expressions are completely general; they can be simplified by making use of the fact that $\del{n'}\hmn{\gamma'\delta'}{1,-1} = \order{\lambda^2,\lambda\e} = \del{n'}\hmn{\gamma'\delta'}{1,0}$. Note that $\h^{\text{tail}}_{\alpha\beta}$, $\h^{\text{dir}1}_{\alpha\beta}$, and $\h^{\text{dir}2}_{\alpha\beta}$ are scalars at $x'$ and rank-two tensors at $x$. Also note that we require the final answer to be accurate up to errors of $\order{\lambda^2}$, in order to determine the free functions in $\hmn{}{1,1}$; each of the above expansions is performed to an order sufficient to meet this requirement, given that $\delta_+(\sigma)\sim 1/\lambda^2$ and $\delta'_+(\sigma)\sim 1/\lambda^4$.

It is convenient to adopt $\sigma$ as an integration variable, which can be done using the transformation $dt'=\displaystyle\frac{d\sigma}{\r}$, where $\r\equiv\sigma_{\alpha'}(x,x')u^{\alpha'}$ can be thought of as a measure of the luminosity distance from $x$ to $x'$. Note that the four-velocity at $x'$ is defined as the tangent to a curve of constant $r$ and $\theta^A$: that is, $u^{\alpha'}=\frac{\partial x^{\alpha'}}{\partial t'}\big|_\Gamma$. Since $t'$ is the proper time on the worldline, rather than the proper time on the generators of the worldtube, this four-velocity is not normalized. (This implies that $\r$ is not an affine parameter on the geodesic connecting $x$ to $x'$.)

After performing this change of variables, we eliminate the $\delta'$ term in the boundary integral by using the identity
\begin{equation}
\int\limits_{\Gamma\cap\mathcal{N}}\!\!\!\h^{\text{dir2}}_{\alpha\beta} N\delta'(\sigma)dt'd\Omega' =-\oint\limits_{\mathcal{S}}\frac{1}{\r} \partial_{t'}\bigg(\frac{N}{\r}\h^{\text{dir2}}_{\alpha\beta}\bigg)d\Omega'.
\end{equation}
Here $\mathcal{S}$ is the intersection of the past light cone of $x$ with the worldtube. For simplicity, I assume that the normal neighbourhood of $x$ is large enough for $\mathcal{S}$ to be well defined and for the intersection $\mathcal{S}\cap\mathcal{N}$ to be empty. I also assume that $x$ is late enough in time for $\mathcal{S}$ to be closed, such that it has the topology of a sphere. (If $x$ is not sufficiently late in time, then $\mathcal{S}$ will be ``cut off" where it intersects $\Sigma$.)

We can now express $h$ as
\begin{align}
h_{\alpha\beta}&=\frac{1}{4\pi}\oint\limits_\mathcal{S} \h^{\text{dir}}_{\alpha\beta}Nd\Omega' +\frac{1}{4\pi}\!\!\!\int\limits_{\Gamma\cap\past}\!\!\! \h^{\text{tail}}_{\alpha\beta}Nd\Omega'dt'\nonumber\\ &\quad+\hmn{\Sigma\alpha\beta}{1}+\order{\e^2},
\end{align}
where
\begin{equation}
\h^{\text{dir}}_{\alpha\beta} = \frac{1}{\r}\h^{\text{dir}1}_{\alpha\beta} -\frac{1}{N\r}\partial_{t'}\bigg(\frac{\h^{\text{dir}2}_{\alpha\beta}}{\r}\bigg) +\order{\lambda^2,\e}.
\end{equation}
The metric perturbation has three types of contributions: the ``direct" type arising from data on the light cone; the ``tail" part arising from the interior of the light cone; and the contribution from the initial data surface.

The ``direct" and ``tail" contributions will be calculated explicitly in the following subsections. We begin with the direct contribution.

\subsection{Integral over the past light cone}
Each of the quantities in the bitensor $\h^{\text{dir}}_{\alpha\beta}$ can be expanded in powers of $\lambda$ by first expanding the $x'$-dependence about the point $\bar x=\gamma(t')$ and then expanding the $\bar x$-dependence about the point $x''=\gamma(t)$. See Fig.~\ref{tube} for a depiction of the relationship between these points. The expansions are provided in Appendix~\ref{worldtube expansions}. Most significantly, the distance $\r$ and its time-derivative $\partial_{t'}\r\equiv\dot{\r{}_{\ }}$ are expanded as $\r=\lambda(\r_0+\lambda\r_1+\lambda^2\r_2+...)$ and $\dot{\r{}_{\ }}=\dot{\r_0}+\lambda\dot{\r_1}+\lambda^2\dot{\r_2}+...$, where the leading-order terms are the flat-spacetime values
\begin{align}
\r_0 &= \sqrt{r^2+\rad^2-2r\rad n^an'_a},\\
\dot{\r_0} &= -1.
\end{align}

After making use of these two expansions, and expressing $\hmn{}{1,0}(t')$ and $\hmn{}{1,1}(t')$ in terms of their values at $t$, we can express $\h^{\text{dir}}$ explicitly as
\begin{widetext}
\begin{equation}\label{hdir}
\begin{split}
\h^{\text{dir}}_{\alpha\beta} & =  \frac{1}{\r_0}\left(1-\frac{\r_2}{\r_0}\right)U_{\alpha\beta}{}^{\gamma'\delta'} \hmn{\gamma'\delta'}{1,-1} -\frac{\rad}{\r_0}\Big(U_{\alpha\beta}{}^{\gamma'\delta'}\del{n'} \hmn{\gamma'\delta'}{1,-1} -\del{n'}U_{\alpha\beta}{}^{\gamma'\delta'}\hmn{\gamma'\delta'}{1,-1} +V_{\alpha\beta}{}^{\gamma'\delta'}\hmn{\gamma'\delta'}{1,-1} \sigma_{\mu'}n^{\mu'}\Big)\\
&\quad -\frac{\rad}{\r_0^3}\bigg(1-\dot{\r_2}-\frac{3\r_2}{\r_0}\bigg) U_{\alpha\beta}{}^{\gamma'\delta'}\hmn{\gamma'\delta'}{1,-1} \sigma_{\mu'}n^{\mu'} -\frac{\rad}{\r_0^2}\del{u'}\Big(U_{\alpha\beta}{}^{\gamma'\delta'} \sigma_{\mu'}n^{\mu'} \hmn{\gamma'\delta'}{1,-1}\Big) \\
&\quad -\frac{\rad}{\r_0^3}U_{\alpha\beta}{}^{\gamma'\delta'}e^I_{\gamma'}e^J_{\delta'} \hmn{IJ}{1,0}\!(t)\sigma_{\mu'}n^{\mu'}-\frac{\rad^2}{\r_0}U_{\alpha\beta} {}^{\gamma'\delta'}\hmn{\gamma'\delta'}{1,1}-\frac{\rad^3}{\r_0^3}U_{\alpha\beta} {}^{\gamma'\delta'}e^I_{\gamma'}e^J_{\delta'}\hmn{IJ}{1,1}\!(t) \sigma_{\mu'}n^{\mu'} +\order{\lambda^2,\e}.
\end{split}
\end{equation}
Each term in this expression is further expanded using the results of Appendix~\ref{worldtube expansions}, which details the expansion of the Green's function and $\sigma_{\mu'}(x,x')$. In order to integrate the final, fully expanded expression for $\h^{\text{dir}}$ over the surface $\mathcal{S}$, we make use of the angular integrals displayed in Appendix~\ref{angular integrals}.

The end results of these calculations are as follows: The $\hmn{\alpha\beta}{1,-1}$ terms in $\h^{\text{dir}}$ contribute
\begin{equation}
\begin{split}
\frac{1}{4\pi}\oint\limits_\mathcal{S}\Big(\hmn{\alpha\beta}{1,-1}\text{ terms in \eqref{hdir}}\Big)Nd\Omega' & = 4m\etide_{ab}x^be^a_{(\alpha}e^0_{\beta)}+\frac{2m}{r}\big(1 -\tfrac{1}{6}\etide_{ab}x^{ab}\big)\big(e^0_\alpha e^0_\beta+e_{i\alpha}e^i_\beta\big)-\frac{26m\rad^2}{9r}\etide_{ab}e^a_\alpha e^b_\beta \\
&\quad +\frac{m\rad^4}{15r^3}\Big(-5\etide_{cd}n^{cd}e^0_\alpha e^0_\beta+\tfrac{5}{3}\etide_{cd}n^{cd}e_{i\alpha}e^i_\beta-4\etide_{c\langle a}\nhat_{b\rangle}^c e^a_\alpha e^b_\beta \\
&\quad -4\epsilon_{bdc}\btide^c_a\nhat^{ad}e^0_{(\alpha}e^b_{\beta)}  -\tfrac{4}{3}\etide_{cd}n^{cd}e_{i\alpha}e^i_\beta\Big)+\order{\lambda^2,\e}.
\end{split}
\end{equation}
The $\hmn{\alpha\beta}{1,0}$ terms integrate to zero at the orders of interest. And the $\hmn{\alpha\beta}{1,1}$ terms contribute
\begin{equation}
\begin{split}
\frac{1}{4\pi}\oint\limits_\mathcal{S} \Big(\hmn{\alpha\beta}{1,1} \text{ terms in \eqref{hdir}}\Big)Nd\Omega' & =  \frac{m\rad^4}{3r^3}\etide_{ab}n^{ab}e^0_\alpha e^0_\beta +\frac{4m\rad^4}{15r^3}\epsilon_{abc} \btide^c_d\nhat^{bd}e^0_{(\alpha}e^a_{\beta)} +\frac{38m\rad^2}{9r}\etide_{ab}e^a_\alpha e^b_\beta \\
& +\frac{m\rad^4}{r^3}\Big(-\tfrac{1}{9}\delta_{ab}\etide_{cd}n^{cd} +\tfrac{4}{15}\etide_{c\langle a}\nhat_{b\rangle}^c\Big)e^a_\alpha e^b_\beta +\order{\lambda^2,\e}
\end{split}
\end{equation}
\end{widetext}
Note that the $\hmn{\alpha\beta}{1,1}$ terms are all $\rad$-dependent, and they are necessary to cancel $\rad$-dependent terms arising from $\hmn{\alpha\beta}{1,-1}$. All the actual terms that appear in the buffer region expansion arise from the most singular part of the perturbation, but the regular terms, such as $\hmn{\alpha\beta}{1,1}$, are required on the boundary to ensure the consistency of the solution.

Putting these results together, we arrive at
\begin{align}
\frac{1}{4\pi}\oint\limits_\mathcal{S}\h^{\text{dir}}_{\alpha\beta}Nd\Omega' &= \frac{2m}{r}\big(1-\tfrac{1}{6}\etide_{ab}x^{ab}\big)\big(e^0_\alpha e^0_\beta+e_{i\alpha}e^i_\beta\big)\nonumber\\
&\quad+4m\etide_{ab}x^be^a_{(\alpha}e^0_{\beta)} +\frac{4m\rad^2}{3r}\etide_{ab}e^a_\alpha e^b_\beta\nonumber\\
&\quad+\order{\lambda^2,\e},
\end{align}
where all tensors are evaluated at time $t$. By expressing the tetrad $(e^0_\alpha,e^a_\alpha)$ in terms of the coordinate one-forms $(t_\alpha,x^a_\alpha)$, we can write this result in Fermi coordinates as 
\begin{align}\label{direct terms}
\frac{1}{4\pi}\oint\limits_\mathcal{S}\h^{\text{dir}}_{\alpha\beta}Nd\Omega' &= \frac{2m}{r}\left(t_\alpha t_\beta + \delta_{ab}x^a_\alpha x^b_\beta\right) +\tfrac{5}{3}mr\etide_{ij}\nhat^{ij}t_\alpha t_\beta\nonumber\\
&\quad +4mr\left(\etide_{bi}n^i+\tfrac{1}{3}\epsilon_{bij}\btide^j_k\nhat^{ik}\right) t_{(\alpha}x^b_{\beta)}\nonumber\\
&\quad +\tfrac{1}{9}mr\Big[12\etide_{i\langle a}\nhat_{b\rangle}^i-5\delta_{ab}\etide_{ij}\nhat^{ij} \nonumber\\
&\quad+\left(12\rad^2/r^2-2\right)\etide_{ab}\Big]x^a_\alpha x^b_\beta\nonumber\\
&\quad+\order{\lambda^2,\e}.
\end{align}
Note that the $1/r$ term in this result agrees with the $1/r$ term that was used for boundary data.

We now  proceed to the calculation of the tail terms.

\subsection{Integral over the interior of the past light cone}
The interior of the light cone covers the worldtube from the lower limit $t'=0$, where the worldtube intersects the surface $\Sigma$, to the upper limit defined by the surface $\mathcal{S}$. Because the Fermi time varies over $\mathcal{S}$, I split the integration into two regions: one region from $t'=0$ to $t'=t_{\text{min}}\equiv\displaystyle\min_{\mathcal{S}}\{t'\}$, and another from $t'=t_{\text{min}}$ to $t'=t_{\text{max}}\equiv\displaystyle\max_{\mathcal{S}}\{t'\}$. The integral over the worldtube is then expressed as
\begin{align}
\int\limits_{\Gamma\cap\past}\!\!\!d\Omega'dt' = \int_0^{t_{\text{min}}}\!\!\!\!\!\!\!\!dt'\!\int_0^{4\pi}\!\!\!d\Omega'\!
&+\int_{t_{\text{min}}}^{t_{\text{max}}}\!\!\!\!\!\!\!\!dt' \int_0^{\theta_{\text{max}}(t')}\!\!\!\!\!\!\!\!\!\!\!\!\!\!\!d\theta'\ \int_0^{\phi_{\text{max}}(t',\theta')}\!\!\!\!\!\!\!\!\!\!\!\!\!\!\!\!\!\!\!\!\! d\phi'.\ \ 
\end{align}
In the integral from $t_{\text{min}}$ to $t_{\text{max}}$, the angles $\theta'^A=(\theta',\phi')$ are cut off at some maximum values defined by $\mathcal{S}$.

Because $\h^{\text{tail}}$ is of order $\lambda^0$, and we only seek terms up to order $\lambda$, we can further simplify the integral. For $x'\in\mathcal{S}$, we can write the time difference $t-t'$ as $t-t'=\r_0+\order{\lambda^2}$, where the $\order{\lambda^2}$ error term consists of acceleration and curvature terms (see Eq.~\eqref{Delta t}). I choose $\theta'$ to be the angle between $x'^a$ and $x^a$, such that $t'=t-\sqrt{r^2+\rad^2-2r\rad\cos\theta'}+\order{\lambda^2}$. From this we infer that the maximum and minimum times on $\mathcal{S}$ are given by $t_{\text{max}}=t-(r-\rad)+\order{\lambda^2}$ and $t_{\text{min}}=t-(r+\rad)+\order{\lambda^2}$. The value for $t_{\text{max}}$ corresponds to the time at the point on $\mathcal{S}$ closest to $x$; the value at $t_{\text{min}}$ is the time at the point furthest from $x$. Since the maximum value of $\theta'$ at a given value of $t'$ is determined by the intersection with $\mathcal{S}$, it is given by
\begin{equation}
\cos\theta_{\text{max}}=\frac{r^2+\rad^2-(t-t')^2}{2r\rad} +\order{\lambda}.
\end{equation}
Since $\phi'$ runs from 0 to $2\pi$ everywhere on $\mathcal{S}$, its maximum value is $2\pi$, independent of $t'$ and $\theta'$.

Making use of these approximations, we can expand the first integral, running from $t'=0$ to $t'=t_{\text{min}}$, about the upper limit $t'=t$. This enables us to write the integral over the worldtube as
\begin{align}
\int\limits_{\Gamma\cap\past}\!\!\!d\Omega'dt' & = \int_0^t\!\!\!dt'\!\int_0^{4\pi}\!\!\!d\Omega' -(r+\rad)\!\!\int_0^{4\pi}\!\!\!d\Omega'\Big|_{t'=t}\nonumber\\
&\quad+\int_0^{2\pi}\!\!\!d\phi'\!\int_{t_{\text{min}}}^{t_{\text{max}}}\!\!\!dt' \!\int_{\cos\theta_{\text{max}}}^1\!\!\!\!\!\!\!\!\!d\cos\theta' +\order{\lambda^2}.
\end{align}

With the simplification $\del{n'}\hmn{\gamma'\delta'}{1,-1} = \order{\lambda^2,\lambda\e} = \del{n'}\hmn{\gamma'\delta'}{1,0}$, along with the expansion
\begin{equation}
G\indices{_{\alpha\beta}^{\alpha'\beta'}} = e^{(\alpha'}_{I}e^{\beta')}_{J}\left[G\indices{_{\alpha\beta}^{IJ}}(t') +G\indices{_{\alpha\beta}^{IJ}_{|c}}(t')x'^c+\order{\lambda^2}\right],
\end{equation}
$\h^{\text{tail}}$ can be written as
\begin{align}
\h^{\text{tail}}_{\alpha\beta} &= \hmn{\gamma'\delta'}{1,-1}e_I^{\gamma'}e_J^{\delta'}\left( G\indices{_{\alpha\beta}^{IJ}}(t') +2\rad G\indices{_{\alpha\beta}^{IJ}_{|c}}n'^c\right)\nonumber\\
&\quad+\order{\lambda^2}\\
&=2m(t')\left(\delta^0_I\delta^0_J+\delta_{ij}\delta^i_I\delta^j_J\right)\Big( G\indices{_{\alpha\beta}^{IJ}}(t')\nonumber\\
&\quad +2\rad G\indices{_{\alpha\beta}^{IJ}_{|c}}n'^c\Big)+\order{\lambda^2}.
\end{align}
In addition, in the second and third integral, which lie within the normal neighbourhood of $x$, the Green's function can be replaced with $V\indices{_{\alpha\beta}^{\alpha'\beta'}}$ and we can use the near-coincidence expansion
\begin{equation}
V\indices{_{\alpha\beta}^{\alpha'\beta'}} = e_{(\alpha}^{K}e_{\beta)}^{L}e^{(\alpha'}_{I}e^{\beta')}_{J}R^I{}_K{}^J{}_L(t) +\order{\lambda}
\end{equation}
for $x'$ near $x''=\gamma(t)$.

Substituting these expressions into the integral, and noting that $\int n'^ad\Omega'=0$, yields the result 
\begin{align}
\frac{1}{4\pi}\!\!\!\int\limits_{\Gamma\cap\past}\!\!\! \h^{\text{tail}}_{\alpha\beta}Nd\Omega'dt' & =  \int_0^{t^-}\!\!\!2m\Big(G_{\alpha\beta}{}^{00} +G_{\alpha\beta}{}^{ij}\delta_{ij}\Big)dt'\nonumber\\
&\quad-4m\bigg(r+\frac{\rad^2}{3r}\bigg)e^a_\alpha e^b_\beta \etide_{ab}\nonumber\\
&\quad+\order{\lambda^2}.
\end{align}
Note that the integrals are cut off at $t^-\equiv t-0^+$ to avoid the singular behavior of the Green's function at coincidence. 

Now, since $x$ is near the point $x''$ on the worldline, we can expand the integrand as
\begin{align}
G_{\alpha\beta}{}^{00}+G_{\alpha\beta}{}^{ij}\delta_{ij} &= G_{\alpha\beta\bar\alpha\bar\beta}(2u^{\bar\alpha}u^{\bar\beta} +g^{\bar\alpha\bar\beta})\nonumber\\
&=\left(G_{\alpha''\beta''\bar\alpha\bar\beta} +\del{\gamma''}G_{\alpha''\beta''\bar\alpha\bar\beta} e^{\gamma''}_cx^c\right)\nonumber\\
&\quad\times g^{\alpha''}_{\alpha}g^{\beta''}_{\beta}\left(2u^{\bar\alpha}u^{\bar\beta} +g^{\bar\alpha\bar\beta}\right)
\end{align}
Substituting this into the integral results in the expansion
\begin{align}
\frac{1}{4\pi}\!\int\limits_{\Gamma\cap\past}\!\!\! \h^{\text{tail}}_{\alpha\beta}Nd\Omega'dt' & =  g^{\alpha''}_{\alpha}g^{\beta''}_{\beta}\!\left(\tail_{\Gamma \alpha''\beta''} +\tail_{\Gamma \alpha''\beta''\gamma''}e^{\gamma''}_cx^c\right)\nonumber\\
&\quad-4m\left(r+\frac{\rad^2}{3r}\right)e^a_\alpha e^b_\beta \etide_{ab}\nonumber\\
&\quad+\order{\lambda^2,\e},
\end{align}
where I have defined
\begin{align}
\tail_{\Gamma\alpha''\beta''}&=\int_0^{t^{-}}\!\!\!\! 2mG_{\alpha''\beta''\bar\alpha\bar\beta} \Big(2u^{\bar\alpha}u^{\bar\beta} +g^{\bar\alpha\bar\beta}\Big)d\bar t,\\
\tail_{\Gamma\alpha''\beta''\gamma''}&= \int_0^{t^{-}}\!\!\!\!2m\del{\gamma''} G_{\alpha''\beta''\bar\alpha\bar\beta}\Big(2u^{\bar\alpha}u^{\bar\beta} +g^{\bar\alpha\bar\beta}\Big)d\bar t.
\end{align}
By making use of the identity \eqref{Green3}, we can express these tail terms in their more usual form:
\begin{align}
\tail_{\Gamma\alpha''\beta''}&= \int_0^{t^{-}}\!\!\!\!4m\big(G_{\alpha''\beta''\bar\alpha\bar\beta} \nonumber\\
&\quad-\tfrac{1}{2}g_{\alpha''\beta''} G^{\delta''}{}_{\!\!\delta''\bar\alpha\bar\beta}\big) u^{\bar\alpha}u^{\bar\beta}d\bar t,\\
\tail_{\Gamma\alpha''\beta''\gamma''}&= \int_0^{t^{-}}\!\!\!\!4m\del{\gamma''} \big(G_{\alpha''\beta''\bar\alpha\bar\beta} \nonumber\\ &\quad -\tfrac{1}{2}g_{\alpha''\beta''} G^{\delta''}{}_{\!\!\delta''\bar\alpha\bar\beta}\big) u^{\bar\alpha}u^{\bar\beta}d\bar t.
\end{align}
The complete tail term will consist of the sum of the $\tail_\Gamma$ terms and the $\hmn{\Sigma}{1}$ terms.

In Fermi coordinates, the final result of this section is
\begin{align}\label{tail terms}
\frac{1}{4\pi}\!\!\!\int\limits_{\Gamma\cap\past}\!\!\! \h^{\text{tail}}_{\alpha\beta}Nd\Omega'dt' & =  \Big(\tail_{\Gamma 00}+\tail_{\Gamma 00c}x^c\Big)t_\alpha t_\beta\nonumber\\
&\quad+2\Big(\tail_{\Gamma 0b}+\tail_{\Gamma 0bc}x^c\Big)t_{(\alpha}x^b_{\beta)}\nonumber\\
&\quad+\Big(\tail_{\Gamma ab}+\tail_{\Gamma abc}x^c\Big)x^a_\alpha x^b_\beta \nonumber\\
&\quad-4m\bigg(r+\frac{\rad^2}{3r}\bigg)\etide_{ab}x^a_\alpha x^b_\beta\nonumber\\
&\quad +\order{\lambda^2,\e}.
\end{align}
Note that the $\rad$-dependent term in this equation exactly cancels the $\rad$-dependent term in Eq.~\eqref{direct terms}. In addition, note that this expansion is identical to the one in Sec.~\ref{buffer_expansion} only after explicit factors of the acceleration are set to zero. This means, in effect, that when comparing individual components of our expansion here to those in our previous expansion in the buffer region, we should replace the covariant derivate in the Fermi-coordinate expression for $\tail_{\Gamma\alpha''\beta''\gamma''}$ with a partial derivative.

\subsection{Identification of unknown functions}
We now combine the results of Eqs.~\eqref{Sigma contribution}, \eqref{direct terms}, and \eqref{tail terms} to arrive at an expansion of the form
\begin{equation}
h_{\alpha\beta}=\frac{1}{r}\hmn{\alpha\beta}{1,-1}+\hmn{\alpha\beta}{1,0} +r\hmn{\alpha\beta}{1,1}+\order{\lambda^2,\e},
\end{equation}
which we will identify with the expansion defined by Eqs.~\eqref{h1n1}, \eqref{h10}, and \eqref{h11}. After defining the tail terms 
\begin{align}
\tail_{IJ} &=\tail_{\Gamma IJ}+\hmn{\Sigma IJ}{1,0}, \\ \tail_{IJc}n^c &=\tail_{\Gamma IJc}n^c+\hmn{\Sigma IJ}{1,1},
\end{align}
and decomposing the results into STF pieces, we find
\begin{align}\label{h1n1 tail}
\hmn{\alpha\beta}{1,-1} &= \frac{2m}{r}\left(t_\alpha t_\beta + \delta_{ab}x^a_\alpha x^b_\beta\right),\\
\hmn{\alpha\beta}{1,0} &= \tail_{00}t_\alpha t_\beta +2\tail_{0b}t_{(\alpha}x^b_{\beta)} \nonumber\\
&\quad +\left(\tail_{\av{ab}} +\tfrac{1}{3}\delta_{ab}\delta^{ij}\tail_{ij}\right) x^a_{(\alpha}x^b_{\beta)},\label{h10 tail}
\end{align}
and
\begin{align}\label{h11 tail}
\hmn{tt}{1,1} &= \tfrac{5}{3}m\etide_{ij}\nhat^{ij}+\tail_{00i}n^i, \\
\hmn{ta}{1,1} &= 2m\etide_{ai}n^i+\tfrac{2}{3}m\epsilon_{aij}\btide^j_k\nhat^{ik}+\tail_{0\langle ac\rangle}n^c\nonumber\\
&\quad+\tfrac{1}{3}\tail_{0ij}\delta^{ij}n_a +\tfrac{1}{2}\epsilon_{aci}\epsilon^{ijk}\tail_{0jk}n^c,\\ 
\hmn{ab}{1,1} &= \tfrac{4}{3}m\etide_{i\langle a}\nhat_{b\rangle}^i-\tfrac{5}{9}m\delta_{ab}\etide_{ij}\nhat^{ij} -\tfrac{38}{9}m\etide_{ab}\nonumber\\
&\quad+\mathop{\STF}_{ab}\left[\tfrac{2}{3}\epsilon_{iac} \mathop{\STF}_{ib}\left(\tail_{\langle ij\rangle d}\epsilon_b{}^{jd}\right)n^c+\tfrac{3}{5}\delta^{ij}\tail_{\langle ib\rangle j}n_a\right]\nonumber\\
&\quad+\tfrac{1}{3}\delta_{ab}\delta^{ij}\tail_{ijc}n^c+\tail_{\langle abc\rangle}n^c.
\end{align}
After setting explicit factors of the acceleration to zero, this expansion agrees with Eqs.~\eqref{h1n1}, \eqref{h10}, and \eqref{h11}. By comparing the two sets of equations, we identify all the unknown STF tensors in the buffer region expansion. The results of this identification are listed in Table~\ref{STF wrt tail}. Note that these identifications are modulo the acceleration that appears in the covariant derivative in $\tail_{\Gamma\alpha\beta\gamma}$.

Note that for this solution to agree with the results of the buffer region expansion, it must satisfy the relationships given in Eqs.~\eqref{B11} and \eqref{F11}. In terms of the tail integral, these relationships read 
\begin{align}
\delta^{ij}\tail_{\langle ai\rangle j} &= \tfrac{1}{6}\delta^{ij}\tail_{ija} -\tfrac{1}{2}\tail_{00a} +\partial_t\tail_{0a},\\
\delta^{ij}\tail_{0ij} &=\tfrac{1}{2}\partial_t\Big(\tail_{00} +\delta^{ij}\tail_{ij}\Big),
\end{align}
where it is understood that the equations hold only for $a=0$. By using the Green's functions identities \eqref{Green1}, \eqref{Green2}, and \eqref{Green3}, and neglecting acceleration terms, one can easily show that the tail terms  $\tail_{\Gamma}$ satisfy these relationships. Hence, we must constrain the initial data terms $\hmn{\Sigma}{1}$ to independently satisfy them, which implies 
\begin{align}
\B{\Sigma}{1,1} &= \tfrac{1}{6}\partial_t\left(\A{\Sigma}{1,0} +3\K{\Sigma}{1,0}\right), \\
\F{\Sigma a}{1,1} &= \tfrac{3}{10}\left(\K{\Sigma a}{1,1} -\A{\Sigma a}{1,1} + \partial_t\C{\Sigma a}{1,0}\right).
\end{align}

\begin{table}[tb]
\caption{Symmetric trace-free tensors in the first-order metric perturbation in the buffer region, written in terms of the electric-type tidal field $\etide_{ab}$ and the tail of the perturbation.}  
\begin{ruledtabular}
\begin{tabular}{ll}
$\begin{array}{ll}
\A{}{1,0} &= \tail_{00}\\
\C{a}{1,0} &= \tail_{0a}\\
\K{}{1,0} &= \tfrac{1}{3}\delta^{ab}\tail_{ab}\\
\H{ab}{1,0} &= \tail_{\langle ab\rangle}\\
\A{a}{1,1} &= \tail_{00a}\\
\B{}{1,1} &= \tfrac{1}{3}\tail_{0ij}\delta^{ij}
\end{array}$
& 
$\begin{array}{ll}
\C{ab}{1,1} &= \tail_{0\langle ab\rangle}+2m\etide_{ab} \\
\D{a}{1,1} &= \tfrac{1}{2}\epsilon_a{}^{bc}\tail_{0bc}\\
\K{a}{1,1} &= \frac{1}{3}\delta^{bc}\tail_{bca}\\
\H{abc}{1,1} &= \tail_{\langle abc\rangle}\\
\F{a}{1,1} &= \tfrac{3}{5}\delta^{ij}\tail_{\langle ia\rangle j}\\
\I{ab}{1,1} &= \tfrac{2}{3}\displaystyle{\mathop{\STF}_{ab}} \left(\epsilon_b{}^{ij}\tail_{\langle ai\rangle j}\right)
\end{array}$
\end{tabular}
\end{ruledtabular}
\label{STF wrt tail}
\end{table}  

The reader should take note of two important facts about the metric perturbation derived here. First, as we expected, the expansion displayed above is identical to the expansion of the point particle solution in the neighbourhood of the worldline. Second, and again as we expected, the expansion is completely determined by the most singular, $\e/r$, term in the metric. Although the nonsingular terms are required to maintain consistency at the boundary, one can derive all of them simply by using the $1/r$ term as boundary data.

Now, the principal purpose of the calculation of the boundary integral was to express the equations of motion in terms of the body's past history. The correction to the body's mass, given in Eq.~\eqref{mdot}, can now be written as
\begin{align}\label{mdot_tail}
\delta m(t)=\delta m(0)+\tfrac{1}{3}m\tail_{00} +\tfrac{5}{18}m\delta^{ab}\tail_{ab}.
\end{align}
In covariant form, this is
\begin{equation}
\delta m(t) = \delta m(0)+\tfrac{1}{18}m\left(5g^{\alpha\beta} +11u^{\alpha}u^{\beta}\right)\tail_{\alpha\beta}.
\end{equation}
This is similar, but not identical to, a result found by Mino, Sasaki, and Tanaka~\cite{Mino_Sasaki_Tanaka}. The source of disagreement between the two results is not clear. It is worth noting that both results appear at one order lower than that given by Thorne and Hartle \cite{Thorne_Hartle}, who chose to eliminate the homogeneous field $h^R$.

The leading-order acceleration of the body, given in Eq.~\eqref{a1}, is
\begin{equation}\label{a1_tail}
\an{1}_a=\tfrac{1}{2}\tail_{00a}-\tail_{0a0}-\tfrac{1}{m}S_i\btide^i_a.
\end{equation}
(Here, again, the right-hand side of this equation is to be evaluated at $a=\an{0}=0$.) In covariant form, this result can be written as
\begin{align}
a^{\alpha} &= -\tfrac{1}{2}\e\left(g^{\alpha\delta}+u^{\alpha}u^{\delta}\right) \left(2\tail_{\delta\beta\gamma}-\tail_{\beta\gamma\delta}\right) u^{\beta}u^{\gamma}\nonumber\\
&\quad +\frac{\e}{2m}R^{\alpha}{}_{\beta\gamma\delta}u^\beta S^{\gamma\delta}+\order{\e^2}
\end{align}
where I have again used the definition $S^{\gamma\delta}\equiv e_c^\gamma e_d^\delta\epsilon^{cdj}S_j$. The spin term is the usual Papapetrou spin force. The tail term is the usual MiSaTaQuWa self-force---except that the tail integral is defined as the sum of an integral over the worldline, cut off at $t=0$, and an integral over an initial data surface. Of course, Eqs.~\eqref{mdot_tail} and \eqref{a1_tail} hold only in the Lorenz gauge.

This concludes what might seem to be the most egregiously lengthy derivation of the self-force yet performed. It is hoped, however, that along with the additional length has come additional insight.

\section{Discussion}\label{Conclusion}

\subsection{Summary}
In this paper I have presented a new derivation of the gravitational self-force for a small body. It is based on the familiar technique of using two expansions of the metric: an inner expansion that is more accurate near the body, and an outer expansion that is more accurate far from the body. However, unlike in earlier derivations, I have formulated these expansions in terms of a fixed worldline $\gamma$ defined in the external background spacetime. The self-consistent equation of motion of this worldline then follows directly from solving the Einstein equation. When combined with the first-order metric perturbation, the equation of motion defines a solution to the Einstein equation accurate up to order $\e^2$ errors over times $t\lesssim1/\e$. When combined with the second-order perturbation, it defines a solution accurate up to order $\e^3$ errors on the shorter timescale $\sim 1$.

My approach began with a general analysis of the Einstein equation, up to second order in the body's mass, in a buffer region around the small body. Since the buffer region is assumed to be free of matter, my calculation is valid only for bodies that are sufficiently compact to avoid tidal disruption. An equation of motion for the body's worldline was derived from the condition that the body must possess no mass dipole in coordinates centered on the worldline. From this purely local-in-space analysis, we found an expression for the acceleration in terms of irreducible pieces of a homogeneous solution to the wave equation---the Detweiler-Whiting regular field, which is regular on the worldline. This homogeneous, regular field was not determined by the buffer-region expansion, since it can be determined only by boundary conditions.

A formal expression for the metric perturbation was obtained by casting the Einstein equation in a relaxed form, via the imposition of the Lorenz gauge. This relaxed form can be solved iteratively, with the perturbation at each order given by the sum of (1) an integral over a region outside the body and (2) an integral over an initial data surface and a worldtube surrounding the body. Boundary data on the worldtube are provided by the buffer-region expansion. At first order, it can be shown that the integral representation is identically equal to the perturbation produced by a point particle moving on $\gamma$. At higher orders, because of the increasing singularity of the metric perturbation, only parts of it can be simplified in the same way. Because of this limitation, I introduced a method of direct integration. In this method, the Detweiler-Whiting regular field in the neighbourhood of the body is determined, in terms of initial conditions and the body's past history, by expanding the integral representation in the buffer region and demanding its consistency with the boundary data on the tube.

An essential assumption in this derivation is that the acceleration of the fixed, $\e$-dependent worldline possesses an asymptotic expansion beginning in powers of $\e$. In addition, I made the following assumptions: the exact metric possesses asymptotic expansions of the form given in Sec.~\ref{outline}, there is a smooth coordinate transformation between some internal local coordinates and the external Fermi coordinates in a neighbourhood of the worldtube, the Lorenz gauge condition can be imposed everywhere in the region of interest, and the expansion of the metric perturbation satisfies both the wave equations and (when combined with the expansion of the acceleration) the gauge condition at fixed functional values of the worldline $z^\mu(t)$. While these, especially the last, are strong assumptions, they undoubtedly lead to an eminently useful, systematic approximation scheme. It is worth repeating that while the choice of gauge is not essential in finding an expression for the force in terms of the field in the buffer-region expansion, it \emph{is} essential in my method of determining the field itself. Without making use of the relaxed Einstein equations, no clear method of globally solving the Einstein equation presents itself.

One fruitful avenue of further research might be to explore methods of solving the Einstein equation in alternative gauges but still within the context of a singular expansion that holds $\gamma$ fixed. This might require further thought on the behavior of gauge transformations in such an expansion. However, such details of the formalism are most likely to be made sense of not at the level of the field equations, but at the level of the action. Since partial derivatives do not act directly on the worldline, the dependence of the metric perturbations on the worldline does not appear directly in the field equations. But at the level of the action, using functional derivatives, the role of the worldline becomes transparent. This will be explored in a future publication \cite{perturbation_techniques}.

It is also worth noting that the methods used here would work in many other cases. For example, the direct calculation of the boundary integral can be used to completely determine the force even if the source cannot be represented as a distribution. Also, these methods could be used to derive self-consistent equations of motion for a charged body; the expansion of the acceleration in powers of $\e$ would automatically yield an order-reduced equation of motion, with no runaway solutions (c.f. the recent calculation by Gralla, Harte, and Wald \cite{Gralla_Harte_Wald}).

\subsection{Comparison with alternative methods}
One of the goals of this paper was to construct an approximation scheme that closely mirrors the extremely successful methods of post-Newtonian theory \cite{DIRE, Futamase_review, Blanchet_review, PN_matching, Futamase_particle1, Futamase_particle2, Racine_Flanagan}. As such, many of the methods used here are similar to those used in post-Newtonian expansions. For example, the expansion with a fixed worldline meshes well with the use of the relaxed form of the Einstein equation \cite{relaxed_EFE1,relaxed_EFE2}, which can be solved without specifying the motion of the source, and which is the starting point for post-Minkowski and post-Newtonian expansions. And the use of an inner limit near the body corresponds to the use of the ``strong-field point particle limit" used by Futamase \cite{Futamase_particle1,Futamase_particle2}. In addition, the calculation of the motion of the body in this paper is somewhat similar to the methods used by Futamase and others~\cite{Futamase_review, Futamase_particle1, Futamase_particle2, PN_matching, Racine_Flanagan}, in that it is based on a multipole-expansion of the body's metric in the buffer region. Finally, the direct integration of the relaxed Einstein equation mirrors the approach of Will \emph{et al.}~\cite{DIRE}.

There are, of course, differences between the two cases. In particular, when the finite size of the body is taken into account in post-Newtonian theory, because the background is flat, finding an equation of motion for the mass dipole of the body is equivalent to finding an equation of motion for its worldline. Although this method, or methods similar to it, has also been used in curved spacetimes \cite{Thorne_Hartle, Mino_matching, Fukumoto}, it is somewhat problematic because the mass dipole corresponds to a displacement from the center of a given coordinate system. But in a curved spacetime, such a displacement is meaningful only when it is infinitesimal. Of course, if at a given instant the coordinate system is mass-centered, then the second time-derivative of the mass dipole is equivalent to the acceleration of the worldline; but since there is no unique global time in a curved spacetime, it is more meaningful to speak of a curve about which the body is centered for its entire history, rather than just at a given time.

The more significant goal of this paper was to develop a unified and self-consistent formalism to treat the gravitational self-force problem. Because the problem consists of solving singular perturbation equations, I have emphasized the foundation of the formalism in singular perturbation theory. Because the formalism uses a self-consistent worldline and a finite sized body, it is (potentially) valid on both short and long timescales, and both very near to and far from the small body. As such, it can be used to study (or incorporate studies of) the spacetime near the small body, the long-term motion of the body, and the perturbations produced by it, including the gravitational waves emitted to infinity.

This contrasts with the most recent derivation of the self-force, performed by Gralla and Wald \cite{Gralla_Wald}. In terms of the concrete calculation of the force in the buffer region, my calculation is very similar to theirs, though it differs in many details. (One such difference is that the perturbation I derive satisfies the Lorenz gauge at all orders in $r$ in the local expansion, whereas Gralla and Wald do not impose the Lorenz gauge on the most singular, order-$\e^2/r^2$, term in their calculation.) However, their approach constructs a regular expansion in which both the worldline and the metric perturbation are expanded; they suggest that in order to arrive at a self-consistent set of equations, one must make a ``leap of faith" from the results of their regular expansion. I instead take the stance that the self-consistent equation of motion can, and should, be justified by a more systematic approach; and I have presented one such approach in this paper. From the results of this approach, one can easily derive the results of the regular expansion: simply by expanding the $\e$-dependent worldline, one derives a leading-order metric perturbation sourced by a particle on a geodesic (plus secularly growing corrections); and the usual steps involved in deriving the geodesic-deviation equation leads to an equation of motion for the deviation vector ``connecting" the geodesic to the exact worldline. Contrariwise, one cannot derive the results of the singular expansion from those of the regular expansion.

Other methods have been developed (or suggested) to accomplish the same goals as my own. One such method is the two-timescale expansion suggested by Hinderer and Flanagan \cite{Hinderer_Flanagan}. As discussed in Sec.~\ref{singular_expansion_point_particle}, their method continuously transitions between regular expansion, resulting in a global, uniform-in-time approximation.\footnote{Note that simply patching together a sequence of regular expansions, by shifting to a new geodesic every so often using the deviation vector, would not accomplish this: such a procedure would accumulate a secular error proportional to the number of ``shifts" multiplied by a nonlinear factor depending on the time between ``shifts." And this error would, formally at least, be of the same magnitude as the solution itself.} This method contrasts with the one presented in this paper, in which a single expansion has been constructed by treating the worldline as fixed.

The fundamental difference between the two methods is the following: In the two-timescale method, the Einstein equation, coupled to the equation of motion of the small body, is reduced to a dynamical system that can be evolved in time. The true worldline of the body then emerges from the evolution of this system. In the method presented here, I have instead sought global, formal solutions to the Einstein equation, written in terms of global integrals; to accomplish this, I have treated the worldline of the body as a fixed structure in the external spacetime. The two timescale method is, perhaps, more practical for concrete calculations, while the global solutions presented here are primarily of formal interest. However, the two methods should agree. Note, though, that Hinderer and Flanagan have identified transient resonances in EMRI systems, which lead to half-integer powers of $\e$ in their asymptotic expansions. It is not clear that such effects are correctly accounted for in the method presented in this paper.

\subsection{Prospects for a global solution}
The principal practical goal of solving the self-force problem is to find the waveform emitted from an EMRI. In order to extract the parameters of an EMRI system from its waveform, we must have a model that tracks the wave's phase to within an error of order $\e$ over a time period $1/\e$. This presents several problems.

First amongst these problems is the potential for secular errors. For example, secular errors might arise due to ignoring the slow evolution of the background spacetime. Throughout this paper, I have assumed that the external background metric is $\e$-independent. However, in practice, it might possess a slow time dependence that would account for the backreaction of the perturbations on the background spacetime; for example, in an EMRI, the large black hole's absorption of gravitational waves slowly alters its mass and spin parameters. Any such effect leaves the expression for the self-force unchanged, and it can be easily incorporated into the perturbations presented here \cite{perturbation_techniques}. However, an equation for the slow evolution itself is unknown. Presumably, it can be determined from an averaged version of the Einstein equation, of the form $\av{E_{\mu\nu}[h]} = 2\av{R_{\mu\nu}}+2\av{\delta^2R_{\mu\nu}} + ...$. In an EMRI system, the average of the wave operator will most likely vanish, because the body's orbit is quasi-periodic. The averaged equation will then relate $\delta^2 R$ to the background Ricci tensor $R$, as in the pioneering work of Isaacson \cite{Isaacson}; this corresponds to the effect of quadrupole radiation on the background. In practice, the averaged equation might be solved by using some ansatz for the background metric---e.g., the Kerr metric with slowly varying mass and spin parameters. The feasibility of such a calculation is unclear; the need to perform it will most likely be determined by examining the magnitude of secular growth in a solution that ignores backreaction. See Refs.~\cite{Galley_backreaction, Hinderer_Flanagan} for more information on the backreaction in the self-force problem.

Putting aside the backreaction problem, other secular errors will also arise due to neglected terms in the acceleration and metric perturbation. Although the approach taken in this paper is designed to avoid such errors, a concrete implementation will nevertheless contain them. I have defined the worldline as a fixed curve; proceeding to successively higher orders in perturbation theory yields successively more accurate equations of motion for this curve. However, if we stop at any given order and use any given equation of motion, then the worldline based on that equation of motion will deviate secularly from the true worldline. This in turn implies that the metric perturbation will accumulate secular errors.

Hence, we must have an equation of motion that limits these errors to $\order{\e}$ after a time $1/\e$. If we use the first-order equation of motion, we will be neglecting an acceleration $\sim\e^2$, which will lead to secular errors of order unity after a time $1/\e$. Thus, the second-order self-force is required in order to obtain a sufficiently accurate waveform template.\footnote{Proceeding to second order will also be useful for examining other systems, such as intermediate mass ratio binaries, over shorter timescales.} In order to achieve the correct waveform, we must also obtain the second-order part of the metric perturbation; this can be easily done, at least formally, using the global integral representations outside a worldtube.

A formal expression for the second-order force has already been derived by Rosenthal~\cite{Eran_field, Eran_force}. However, he expresses the second-order force in a very particular gauge in which the first-order self-force vanishes. This is sensible on short time scales, but not on long timescales. Furthermore, it is not a convenient gauge, since it does not provide what we wish it to: a correction to the nonzero leading-order force in the Lorenz gauge.

Thus, we wish to obtain an alternative to Rosenthal's derivation. Based on the methods developed in this paper, there is a clear route to deriving the second-order force. We would construct a buffer-region expansion accurate up to order $\e^3$. Since we would require the order $\e^2 r$ terms in this expansion, in order to determine the acceleration, we would need to increase the order of the expansion in $r$ as well: specifically, we would need terms up to orders $\e^0 r^3$, $\e r^2$, $\e^2 r$, and $\e^3 r^0$. These could be calculated using the methods presented previously.

Unfortunately, such a procedure could be prohibitively difficult. Hence, we might consider a much simpler alternative: the method of matched asymptotic expansions. Using this method, we would need the buffer region expansion to be accurate to order $\e^2 r$---the order at which the second-order acceleration appears in the background metric---meaning that we would need to extend the buffer region expansion by one order in $r$, but not in $\e$. An internal solution is already known to that order \cite{Eric_tidal}, meaning that only the external solution must be found. The equation of motion would then be determined by finding a unique coordinate transformation that makes the external and internal solutions identical in the buffer region. As will be discussed in Ref.~\cite{perturbation_techniques}, this method is somewhat problematic, because the coordinate transformation is \emph{not} unique. However, it should be possible to overcome this problem, and a calculation of the second-order force by this means is entirely feasible.

Even if we can obtain an aproximation with the desired accuracy on the timescale $1/\e$, there remains at least one additional difficulty. The waveform itself is to be calculated at future null infinity, $\mathscr{I}^+$. At first glance, it might seem that we can extend the size of our domain $\Omega$ such that its future null boundary $\mathcal{J}$ is pushed out to $\mathscr{I}^+$ at one end and to the event horizon of the large black hole at the other. However, the size of our domain is intended to be of size $1/\e$. Thus, if we wish to enlarge the domain out to infinity, we must match the solution within it to an outgoing wave solution at its future null boundaries.

\subsection{Conclusion}
Throughout this paper, I have taken the stance that finding a useful approximate solution to the exact Einstein equation, such as that provided by singular perturbation theory, is more important than finding an exact solution to the approximate Einstein equation, such as that provided by regular perturbation theory. In the gravitational self-force problem, a useful approximate solution is one that remains valid on long timescales, self-consistently incorporates the acceleration of the small body, and accounts for its asymptotically small, but finite, size. In this paper, I have developed a formal approximation scheme that promises to satisfy these criteria, and which can be extended to any order in perturbation theory.

However, I have also taken the stance that an approximate solution of the exact Einstein equation must be an approximation to an exact solution if it is to render a meaningful test of General Relativity. As such, I have emphasized how the singular expansions developed in this paper might be related to an exact solution. A far more rigorous, technical, and perhaps altogether unfeasible study would be required to show whether or not the asymptotic solution developed here actually does approximate an exact solution.

Of course, even if the solutions are proven to be asymptotic approximations, they remain purely formal. A practical calculation of the motion of a small body will most likely require a numerical implementation, which will require a formulation of the wave equation, coupled to an equation of motion for the source, that is viable for numerical calculations.

\begin{acknowledgments}
I wish to thank Eric Poisson for providing the initial outline of this work and for many helpful discussions along the way. Thanks are also owed to the participants of the Capra 12 meeting at Indiana State University for indirectly helping me finalize my choice of emphasis in this paper. Finally, thanks to Achim Kemp, Bernie Nickel, and Sam Gralla for helpful comments. This work was supported by the Natural Sciences and Engineering Research Council of Canada.
\end{acknowledgments}

\appendix

\section{Green's Functions}\label{Greens_functions}
I follow the notation and conventions of Ref.~\cite{Eric_review}. The Green's function for the tensor wave operator is defined by the equation
\begin{equation}\label{tensor Green}
\left(g^\rho_\mu g^\sigma_\nu\Box + 2R_\mu{}^\rho{}_\nu{}^\sigma\right) G_{\rho\sigma}{}^{\mu'\nu'}(x,x') = -4\pi g_{(\mu}^{\mu'}g_{\nu)}^{\nu'}\delta(x,x'),
\end{equation}
where $\delta(x,x')=\delta^4(x^\alpha-x^{\alpha'})/\sqrt{|g|}$, and $g_{\mu}^{\mu'}$ is the parallel propagator from $x$ to $x'$. The Green's function for the vector wave operator is defined by
\begin{equation}\label{vector Green}
\left(g^\nu_\mu\Box -R_\mu{}^{\nu}\right)G_\nu{}^{\mu'}(x,x') = -4\pi g_{\mu}^{\mu'}\delta(x,x'),
\end{equation}
and the Green's function for the scalar wave operator is defined by 
\begin{equation}\label{scalar Green}
\left(\Box - \lambda R\right)G(x,x') = -4\pi\delta(x,x').
\end{equation}
All quantities are defined with respect to the background metric $g$.

If the point $x'$ lies in the convex normal neighbourhood of $x$, then the retarded gravitational Green's functions can be written in the Hadamard decomposition
\begin{equation}
G_{\alpha\beta\alpha'\beta'}=U_{\alpha\beta\alpha'\beta'}\delta_+(\sigma) +V_{\alpha\beta\alpha'\beta'}\theta_+(-\sigma),
\end{equation}
where $\delta_+(\sigma)$ is a delta function with support on the past light cone of $x$, $\theta_+(-\sigma)$ is a Heaviside function with support in the interior of the past light cone of $x$, $\sigma$ is Synge's world function, equal to one-half the squared geodesic distance between $x$ and $x'$, and $U_{\alpha\beta\alpha'\beta'}$ and $V_{\alpha\beta\alpha'\beta'}$ are smooth bitensors. The advanced Green's function has an analogous decomposition, with $\delta_+(\sigma)$ and $\theta_+(-\sigma)$ replaced by $\delta_-(\sigma)$ and $\theta_-(-\sigma)$, which, respectively, have support on and within the future lightcone of $x$. The singular Green's function, which satisfies the same defining equation as the retarded and advanced Green's functions but which has support only on and \emph{outside} the past and future light cones, is given by
\begin{equation}
G^S_{\alpha\beta\alpha'\beta'}=\tfrac{1}{2}U_{\alpha\beta\alpha'\beta'}\delta(\sigma) -\tfrac{1}{2}V_{\alpha\beta\alpha'\beta'}\theta(\sigma).
\end{equation}

If $g$ is a vacuum metric, such that $R=0=R_{\mu\nu}$, then one can easily derive the following identities:
\begin{align}
G^{\mu\nu}{}_{\mu'\nu';\nu} & = - G^\mu{}_{(\mu';\nu')}, \label{Green1}\\
G^\mu{}_{\mu';\mu} & = -G_{;\mu'}, \label{Green2}\\
G_{\mu\nu}{}^{\mu'\nu'}g_{\mu'\nu'} & = g_{\mu\nu}G. \label{Green3}
\end{align}
Equation~\eqref{Green1} follows from taking the divergence of Eq.~\eqref{tensor Green} and the covariant derivative of Eq.~\eqref{vector Green}. Equation~\eqref{Green2} follows from taking the divergence of Eq.~\eqref{vector Green} and the covariant derivative of Eq.~\eqref{scalar Green}. Equation~\eqref{Green3} follows from contracting the primed indices in Eq.~\eqref{tensor Green}. In each case, these operations show that the two relevant bitensors satisfy the same differential equation; the equations \eqref{Green1}--\eqref{Green3} hold when the bitensors on the left and right satisfy identical boundary conditions.

Equation \eqref{Green2} appears in Ref.~\cite{DeWitt_Brehme}. To the best of my knowledge, Eqs.~\eqref{Green1} and \eqref{Green3} have not been presented in the literature.


\section{STF tensors}\label{STF tensors}
This appendix briefly reviews the use of STF decompositions and collects several useful formulas. Refer to Ref.~\cite{STF_1,STF_2,STF_3} for thorough reviews. All formulas in this section are either taken directly from Refs.~\cite{STF_1} and \cite{STF_2} or are easily derivable from formulas therein.

Any Cartesian tensor field depending on two angles $\theta^A$ spanning a sphere can be expanded in a unique decomposition in terms of symmetric trace-free tensors. Such a decomposition is equivalent to a decomposition in terms of tensorial harmonics, but it is sometimes more convenient. It begins with the fact that the angular dependence of a Cartesian tensor $T_{S}(\theta^A)$ can be expanded in a series of the form
\begin{equation}\label{nhat_expansion}
T_S(\theta^A)=\sum_{\ell\geq 0}T_{S\langle L\rangle}\nhat^L,
\end{equation}
where $S$ and $L$ denote multi-indices $S=i_1...i_s$ and $L=j_1...j_\ell$, angular brackets denote an STF combination of indices, $n^a$ is a Cartesian unit vector, $n^L\equiv n^{j_1}\ldots n^{j_\ell}$, and $\nhat^L\equiv n^{\langle L\rangle}$. This is entirely equivalent to an expansion in spherical harmonics. Each coefficient $T_{S\langle L\rangle}$ can be found from the formula
\begin{equation}
T_{S\langle L\rangle} = \frac{(2\ell+1)!!}{4\pi\ell!}\int T_S(\theta^A)\nhat_L d\Omega.
\end{equation}
These coefficients can then be decomposed into irreducible STF tensors. For example, for $s=1$, we have
\begin{equation}\label{decomposition_1}
T_{a\langle L\rangle} = \hat T^{(+)}_{aL}+\epsilon^j{}_{a\langle i_\ell}\hat T^{(0)}_{L-1\rangle j}+\delta_{a\langle i_\ell}\hat T^{(-)}_{L-1\rangle},
\end{equation}
where the $\hat T^{(n)}$'s are STF tensors given by
\begin{align}
\hat T^{(+)}_{L+1} & \equiv T_{\langle L+1\rangle}, \\
\hat T^{(0)}_{L} & \equiv \frac{\ell}{\ell+1}T_{pq\langle L-1}\epsilon_{i_\ell\rangle}{}^{pq}, \\
\hat T^{(-)}_{L-1} & \equiv \frac{2\ell-1}{2\ell+1}T^j{}_{jL-1}.
\end{align} 
Similarly, for a symmetric tensor $T_S$ with $s=2$, we have
\begin{align}\label{decomposition_2}
T_{ab\langle L\rangle} & = \hat T^{(+2)}_{abL} + \mathop{\STF}_L\mathop{\STF}_{ab}\Big( \epsilon^p{}_{ai_\ell}\hat T^{(+1)}_{bpL-1} + \delta_{ai_\ell}\hat T^{(0)}_{b L-1} \nonumber\\
&\quad +\delta_{a i_\ell}\epsilon^p{}_{bi_{\ell-1}}\hat T^{(-1)}_{pL-2} +\delta_{ai_\ell}\delta_{bi_{\ell-1}}\hat T^{(-2)}_{L-2}\Big) \nonumber\\
&\quad+\delta_{ab}\hat K_L,
\end{align}
where
\begin{align}
\hat T^{(+2)}_{L+2} & \equiv T_{\langle L+2\rangle}, \\
\hat T^{(+1)}_{L+1} & \equiv \frac{2\ell}{\ell+2}\mathop{\STF}_{L+1}(T_{\langle pi_\ell\rangle qL-1}\epsilon_{i_{\ell+1}}{}^{pq}), \\
\hat T^{(0)}_L & \equiv \frac{6\ell(2\ell-1)}{(\ell+1)(2\ell+3)}\mathop{\STF}_L(T_{\langle ji_\ell\rangle}{}^j{}_{L-1}), \\
\hat T^{(-1)}_{L-1} & \equiv \frac{2(\ell-1)(2\ell-1)}{(\ell+1)(2\ell+1)}\mathop{\STF}_{L-1}(T_{\langle jp\rangle q}{}^j{}_{L-2}\epsilon_{i_{\ell-1}}{}^{pq}), \\
\hat T^{(-2)}_{L-2} & \equiv \frac{2\ell-3}{2\ell+1}T_{\langle jk\rangle}{}^{jk}{}_{L-2} \\
\hat K_L & \equiv \tfrac{1}{3}T^j{}_{jL}.
\end{align}
These decompositions are equivalent to the formulas for addition of angular momenta, $J=S+L$, which results in terms with angular momentum $\ell-s\leq j\leq \ell+s$; the superscript labels $(\pm n)$ in these formulas indicate by how much each term's angular momentum differs from $\ell$.

By substituting Eqs.~\eqref{decomposition_1} and \eqref{decomposition_2} into Eq.~\eqref{nhat_expansion}, we find that a scalar, a Cartesian 3-vector, and the symmetric part of a rank-2 Cartesian 3-tensor can be decomposed as, respectively,
\begin{align}
T(\theta^A) &= \sum_{\ell\ge0}\hat A_L\nhat^L, \\
T_a(\theta^A) &= \sum_{\ell\ge0}\hat B_L\nhat_{aL}\nonumber\\
&\quad+\sum_{\ell\ge1}\left[\hat C_{aL-1}\nhat^{L-1} + \epsilon^i{}_{aj}\hat D_{iL-1}\nhat^{jL-1}\right],\\
T_{(ab)}(\theta^A) &= \delta_{ab}\sum_{\ell\ge0}\hat K_L\nhat^L+\sum_{\ell\ge0}\hat E_L\nhat_{ab}{}^L \nonumber\\
&\quad+\sum_{\ell\ge1}\left[\hat F_{L-1\langle a}\nhat^{}_{b\rangle}{}^{L-1} +\epsilon^{ij}{}_{(a}\nhat_{b)i}{}^{L-1}\hat G_{jL-1}\right] \nonumber\\
&\quad+\sum_{\ell\ge2}\left[\hat H_{abL-2}\nhat^{L-2}+\epsilon^{ij}{}_{(a}\hat I_{b)jL-2}\nhat_{i}{}^{L-2}\right].
\end{align}
Each term in these decompositions is algebraically independent of all the other terms.

We can also reverse a decomposition to ``peel" a fixed index from an STF expression:
\begin{align}
(\ell+1)\mathop{\STF}_{iL}T_{i\langle L\rangle} & = T_{i\langle L\rangle} + \ell\mathop{\STF}_LT_{i_\ell \langle iL-1\rangle} \nonumber\\
&\quad -\frac{2\ell}{2\ell+1}\mathop{\STF}_L T^j{}_{\langle jL-1\rangle}\delta_{i_\ell i}.
\end{align}

In evaluating the action of the wave operator on a decomposed tensor, the following formulas are useful:
\begin{align}
n^c\nhat^L &= \nhat^{cL}+\frac{\ell}{2\ell+1}\delta^{c\langle i_1}\nhat^{i_2...i_\ell\rangle},\\
n_c\nhat^{cL} & = \frac{\ell+1}{2\ell+1}\nhat^L, \\
r\partial_c\nhat_L &= -\ell\nhat_{cL}+\frac{\ell(\ell+1)}{2\ell+1}\delta_{c\langle i_1}\nhat_{i_2...i_\ell\rangle}, \\
\partial^c\partial_c\nhat^L & = -\frac{\ell(\ell+1)}{r^2}\nhat^L, \\
n^c\partial_c\nhat^L &= 0, \\
r\partial_c\nhat^{cL} &= \frac{(\ell+1)(\ell+2)}{(2\ell+1)}\nhat^L.
\end{align}

In evaluating the $t$-component of the Lorenz gauge condition, the following formula is useful for finding the most divergent term (in an expansion in $r$):
\begin{align}\label{gauge_help1}
r\partial^c\hmn{tc}{\emph{n,m}} &= \sum_{\ell\geq 0}\frac{(\ell+1)(\ell+2)}{2\ell+1}\B{L}{\emph{n,m}}\nhat^L \nonumber\\
&\quad -\sum_{\ell\geq 2}(\ell-1)\C{L}{\emph{n,m}}\nhat^L.
\end{align}
And in evaluating the $a$-component, the following formula is useful for the same purpose:
\begin{widetext}
\begin{align}\label{gauge_help2}
r\partial^b\hmn{ab}{\emph{n,m}}-\tfrac{1}{2}r\eta^{\beta\gamma}\partial_a\hmn{\beta\gamma}{\emph{n,m}} &=
\sum_{\ell\geq0}\left[\tfrac{1}{2}\ell(\K{L}{\emph{n,m}}-\A{L}{\emph{n,m}}) +\frac{(\ell+2)(\ell+3)}{2\ell+3}\E{L}{\emph{n,m}} -\tfrac{1}{6}\ell\F{L}{\emph{n,m}}\right]\nhat_a{}^L \nonumber\\
&\quad +\sum_{\ell\geq 1}\left[\frac{\ell(\ell+1)}{2(2\ell+1)}(\A{aL-1}{\emph{n,m}} -\K{aL-1}{\emph{n,m}}) +\frac{(\ell+1)^2(2\ell+3)}{6(2\ell+1)(2\ell-1)}\F{aL-1}{\emph{n,m}} -(\ell-2)\H{aL-1}{\emph{n,m}}\right]\nhat^{L-1} \nonumber\\
&\quad +\sum_{\ell\geq 1}\left[\frac{(\ell+2)^2}{2(2\ell+1)}\G{dL-1}{\emph{n,m}} -\tfrac{1}{2}(\ell-1)\I{dL-1}{\emph{n,m}}\right]\epsilon_{ac}{}^d\nhat^{cL-1}
\end{align}
\end{widetext}
where I have defined $\H{a}{\emph{n,m}}\equiv 0$ and $\I{a}{\emph{n,m}}\equiv 0$.

\subsection{Angular integrals}\label{angular integrals}
The calculation of the boundary integral in Sec.~\ref{integral_expansion1} requires the evaluation of numerous integrals over a cross-section of the worldtube $\Gamma$. This subsection compiles these integrals. Let $x^\alpha=(t,r,\theta,\phi)$ be a point lying outside the worldtube, and let $x^{\alpha'}=(t',\rad, \theta',\phi')$ be a point on the worldtube; here $r$ and $\rad$ denote the Fermi coordinate distances to $x^\alpha$ and $x^{\alpha'}$. Let $n^a$ and $n^{a'}$ be Cartesian unit vectors defined by the angles $(\theta,\phi)$ and $(\theta',\phi')$, respectively. The quantity $\r_0\equiv\sqrt{r^2+\rad^2-2r\rad n_an^{a'}}$ is the leading-order flat-spacetime luminosity distance between $x^\alpha$ and $x^{\alpha'}$. For brevity's sake, I introduce the notation $\av{f}\equiv\frac{1}{4\pi}\int f(\theta',\phi')d\Omega'$ to denote an average over the primed angles, where $d\Omega'\equiv\sin\theta'd\theta'd\phi'$.

I group the integrals according to the power of $\r_0$ that appears in their integrand:
\begin{equation}
\av{\r_0} = r+\frac{\rad^2}{3r},
\end{equation}
\begin{align}
\av{\frac{1}{\r_0}} &= \frac{1}{r},\\
\av{\frac{n'^{a}}{\r_0}} & = \frac{\rad n^a}{3r^2},
\end{align}
\begin{align}
\av{\frac{\nhat'^{ab}}{\r_0}} & = \frac{\rad^2\nhat^{ab}}{5r^3},\\
\av{\frac{\nhat'^{abc}}{\r_0}} & = \frac{\rad^3\nhat^{abc}}{7r^4},
\end{align}
\begin{align}
\av{\frac{1}{\r_0^2}} & = \frac{1}{2r\rad}\ln\left(\frac{r+\rad}{r-\rad}\right) \\
\av{\frac{n'^{a}}{\r_0^2}} & = \frac{1}{2}\left[\frac{r^2+\rad^2}{2r^2\rad^2} \ln\left(\frac{r+\rad}{r-\rad}\right) -\frac{1}{r\rad}\right]n^a \\
\av{\frac{\nhat'^{ab}}{\r_0^2}} & = \frac{3}{8}\frac{r^4+\rad^4+\frac{2}{3}r^2\rad^2}{2r^3\rad^3} \ln\left(\frac{r+\rad}{r-\rad}\right)\nhat^{ab}\nonumber\\
&\quad -\frac{3}{8}\frac{r^2+\rad^2}{r^2\rad^2}\nhat^{ab}
\end{align}
\begin{align}
\av{\frac{1}{\r_0^3}} & = \frac{1}{r(r^2-\rad^2)} \\
\av{\frac{n'^{a}}{\r_0^3}} & = \frac{\rad n^a}{r^2(r^2-\rad^2)} \\
\av{\frac{\nhat'^{ab}}{\r_0^3}} & = \frac{\rad^2\nhat^{ab}}{r^3(r^2-\rad^2)} \\
\av{\frac{\nhat'^{abc}}{\r_0^3}} & = \frac{\rad^3\nhat^{abc}}{r^4(r^2-\rad^2)},
\end{align}
\begin{align}
\av{\frac{1}{\r_0^4}} & = \frac{1}{(r^2-\rad^2)^2}\\
\av{\frac{n'^{a}}{\r_0^4}} & = \frac{r^2+\rad^2}{2r\rad(r^2-\rad^2)^2}n^a\nonumber\\
&\quad-\frac{1}{4r^2\rad^2} \ln\left(\frac{r+\rad}{r-\rad}\right)n^a \\
\av{\frac{\nhat'^{ab}}{\r_0^4}} & = \frac{3}{4}\frac{r^4+\rad^4-\frac{2}{3}r^2\rad^2}{r^2\rad^2(r^2-\rad^2)^2}\nhat^{ab}\nonumber\\
&\quad -\frac{3}{8}\frac{r^2+\rad^2}{r^3\rad^3} \ln\left(\frac{r+\rad}{r-\rad}\right)\nhat^{ab}\\
\av{\frac{\nhat'^{abc}}{\r_0^4}} & = \frac{15}{16}\frac{(r^2+\rad^2)(r^2+\rad^2 -\frac{22}{15}r^2\rad^2)}{r^3\rad^3(r^2-\rad^2)^2}\nhat^{abc}\nonumber\\
&\quad -\frac{15}{32}\frac{r^4+\rad^4+\frac{6}{5}r^2\rad^2}{r^4\rad^4} \ln\left(\frac{r+\rad}{r-\rad}\right)\nhat^{abc}
\end{align}

\begin{align}
\av{\frac{1}{\r_0^5}} & = \frac{3r^2+\rad^2}{3r(r^2-\rad^2)^3} \\
\av{\frac{n'^{a}}{\r_0^5}} & = \frac{\rad(5r^2-\rad^2)n^a}{3r^2(r^2-\rad^2)^3} \\
\av{\frac{\nhat'^{ab}}{\r_0^5}} & = \frac{\rad^2(7r^2-3\rad^2)\nhat^{ab}}{3r^3(r^2-\rad^2)^3} \\
\av{\frac{\nhat'^{abc}}{\r_0^5}} & = \frac{\rad^3(9r^2-5\rad^2)\nhat^{abc}}{3r^4(r^2-\rad^2)^3}
\end{align}


\section{Expansions of bitensors near the worldtube}\label{worldtube expansions}
I present here the expansions of various important bitensors of the form $T(x,x')$, where $x$ is a point in the exterior of the worldtube $\Gamma$, and $x'$ is a nearby point on the worldtube. The expansions are based on standard methods of covariant expansion reviewed in Ref.~\cite{Eric_review}. To keep track of the orders of the expansions, I use the quantity $\lambda\sim r\sim \rad\sim\Delta t$, where $r$ is the radial coordinate at $x$, $\rad$ is the radius of $\Gamma$, and $\Delta t$ is the proper-time difference between $x'$ and $x$. For brevity, I introduce the shorthand notation $x^{ab}\equiv x^a x^b$, $\spb\equiv\sigma(x',\bar x)$, $\sb\equiv\sigma(x,\bar x)$, and $\spp\equiv\sigma(x,x'')$, where $\bar x$ and $x''$ are points on the worldline $\gamma$. The relationship between the various points will eventually be identified with that depicted in Fig.~\ref{tube}.

For completeness, the expansions in this section allow for an arbitrary, non-vacuum background and an arbitrarily accelerating worldline. In the calculations in Sec.~\ref{integral_expansion1}, both the acceleration and the Ricci tensor can be set to zero.

\subsection{General expansions}

Consider a bitensor $T_{\alpha'}(x,x')$. We can expand this in a sequence of steps: First, we expand along a geodesic that connects the point $x'$ on the worldtube to a point $\bar x=\gamma(t')$ on the worldline; this is an expansion in powers of $\rad$. Next, we expand up along the worldline, about a point $x''=\gamma(t)$; this is an expansion in powers of the proper-time difference $\Delta t\equiv t-t'$. These two expansions leave us with bitensors that depend on $x$ and $x''$. The final step in our procedure is a near-coincidence expansion of these bitensors; this last step is an expansion in powers of $r$. This procedure does not rely on any particular relationship between $x'$ and $\bar x$ or between $x$ and $x''$. It becomes a coordinate expansion by fixing these relationships: for example, by connecting $x$ to $x''$ with a geodesic perpendicular to the worldline (and doing likewise for $x'$ and $\bar x$), the covariant expansion becomes an expansion in Fermi coordinates.

We first hold $x$ fixed and expand the $x'$-dependence about $\bar x$:
\begin{equation}\label{xprime_expansion}
T_{\alpha'}(x,x') = g^{\bar\alpha}_{\alpha'}\sum_{k\ge0}\frac{(-1)^k}{k!} T_{\bar\alpha;\bar\gamma_1...\bar\gamma_k}(x,\bar x)\spb^{\bar\gamma_1}\cdots\spb^{\bar\gamma_k}.
\end{equation}
where the reader is reminded that the parallel propagator is given by $g^{\bar\alpha}_{\alpha'}=e^{\bar\alpha}_Ie^I_{\alpha'}$. Next, still holding $x$ fixed, we expand each of the bitensors $T_{\bar\alpha;\bar\gamma_1...\bar\gamma_k}(x,\bar x)$ about $x''$. Since we do not possess a conventient expression for the parallel propagator between $\bar x$ and $x''$, we perform this expansion along the worldline. We do this by expressing $T_{\bar\alpha}$ in terms of its tetrad components, converting to tetrad components via the relationship $T_{\bar\alpha}=T_I(t')e^I_{\bar\alpha}$, and then expanding the tetrad components in powers of the proper time interval $\Delta t$. The time-derivatives along the worldline are evaluated covariantly by reexpressing the tetrad components in terms of the coordinate basis, leading to 
\begin{align}\label{t_expansion}
T_{\bar\alpha;\bar\gamma_1...\bar\gamma_k} & = e^I_{\bar\alpha}e^{J_1}_{\bar\gamma_1}\cdots e^{J_k}_{\bar\gamma_k}\sum_{n\ge0}\frac{(-1)^n}{n!}(\Delta t)^n\times\nonumber\\
&\quad\bigg(\frac{D}{dt''}\bigg)^{\!\!n}\!\!\Big(T_{\alpha'';\gamma''_1...\gamma''_k} e^{\alpha''}_Ie_{J_1}^{\gamma''_1}\cdots e_{J_k}^{\gamma''_k}\Big).
\end{align}
This can be expressed in terms of covariant derivatives of $T$ and combinations of tetrad and acceleration vectors by using the identity
\begin{equation}\label{t_expansion2}
\begin{split}
&\left(\frac{D}{dt''}\right)^{\!\!n}\!\! \left(T_{\alpha'';\gamma''_1...\gamma''_k}e^{\alpha''}_Ie_{J_1}^{\gamma''_1} \cdots e_{J_k}^{\gamma''_k}\right)\\
=&\sum_{i=0}^n{n\choose i}\left(\frac{D}{dt''}\right)^{\!\!n-i}\!\!\! \Big(e^{\alpha''}_Ie_{J_1}^{\gamma''_1}\cdots e_{J_k}^{\gamma''_k}\Big)\times\\
&\sum_{j=0}^iT_{\alpha'';\gamma''_1...\gamma''_k\delta''_1...\delta''_j}A^{\delta''_1...\delta''_j}(i,j),
\end{split}
\end{equation}
where the derivative of the tetrad is given by $\frac{D}{dt}e^{\alpha}_I=(u^\alpha a_\beta-a^\alpha u_\beta)e^\beta_I$. The indexed tensor $A^{\delta''_1...\delta''_j}(i,j)$ is constructed from the four-velocity $u^{\alpha''}$ and its derivatives. Explicitly, 
\begin{align}
A(0,0) & = 1, \\
A^{\delta''_1...\delta''_j}(i,j) & = \frac{D}{dt''}A^{\delta''_1...\delta''_j}(i-1,j) \nonumber\\
&\quad+A^{\delta''_1...\delta''_{j-1}}(i-1,j-1)u^{\delta''_j},
\end{align}
where $0\leq j\leq i$, and $A(i,j)\equiv 0$ for $j<0$ and $j>i$.

Finally, the bitensors $T_{\alpha'';\gamma''_1...\gamma''_k\delta''_1...\delta''_j}(x,x'')$ can be expanded about $x''$ using the near-coincidence expansion
\begin{equation}\label{x_expansion}
\begin{split}
T_{(...)}(x,x'') &= \sum_{m\ge0}\frac{(-1)^m}{m!}t^m_{(...)\mu''_1...\mu''_m}(x'')\times\\
&\quad \spp^{\mu''_1}\cdots\spp^{\mu''_m},
\end{split}
\end{equation}
where $t^m_{(...)\mu''_1...\mu''_m}(x'')$ is defined by the recurrence relation \cite{quasilocal}
\begin{align}
t^0_{(...)}(x'') &= \lim_{x\to x''}T_{(...)}(x,x''),\\
t^m_{(...)\mu''_1...\mu''_m}(x'') & =  \lim_{x\to x''}T_{(...);\mu''_1...\mu''_m}(x,x'')\\
&\quad - \sum_{\ell=0}^{m-1}{m\choose\ell} t^\ell_{\mu''_1...\mu''_\ell;\mu''_{\ell+1}...\mu''_m}(x'')\nonumber.
\end{align}

Substituting Eq. \eqref{x_expansion} into Eq. \eqref{t_expansion2}, Eq. \eqref{t_expansion2} into Eq. \eqref{t_expansion}, and Eq. \eqref{t_expansion} into Eq. \eqref{xprime_expansion}, we arrive at an expression for $T_{\alpha'}(x,x')$ in terms of tensors at $x''$ and the small expansion quantities $\spb^{\bar\alpha}$, $\Delta t$, and $\spp^{\alpha''}$. This procedure is valid in any coordinate system. It can be made into a coordinate expansion in terms of Fermi coordinates $(t,x^a)$ by using the identities $\spb^{\bar\beta}=-e^{\bar\beta}_bx'^b$ and $\spp^{\beta''}=-e^{\beta''}_bx^b$, and identifying $t$ with Fermi time. Analogous identities would generate an expansion in terms of retarded coordinates $(t_{ret},x_{\text{ret}}^a)$ or advanced coordinates $(t_{adv},x_{\text{adv}}^a)$

The generalization of this procedure to tensors of other ranks is obvious.

\subsection{Expansions of $\Delta t$, $\sigma_{\mu'}(x,x')$, and related quantities}

From this point on, I restrict myself to the case in which $x$ and $x'$ are connected by a unique null geodesic. In other words, $x'$ lies on the surface $\mathcal{S}$ in Fig.~\ref{tube}. I begin by expanding $\sigma(x,x')$; since $x'\in\mathcal{S}$, we have $\sigma(x,x')=0$. We can make use of this fact to find $\Delta t$, which is required for all other expansions.

So, following the procedure outlined above, we first expand about $\bar x$:
\begin{align}\label{sigma_expansion}
\sigma(x,x') & = \sb-\sb_{\bar\alpha}\spb^{\bar\alpha} +\tfrac{1}{2}\sb_{\bar\alpha\bar\beta}\spb^{\bar\alpha}\spb^{\bar\beta}
-\tfrac{1}{6}\sb_{\bar\alpha\bar\beta\bar\gamma}\spb^{\bar\alpha} \spb^{\bar\beta}\spb^{\bar\gamma}\nonumber\\
&\quad+\tfrac{1}{24}\sb_{\bar\alpha\bar\beta\bar\gamma\bar\delta} \spb^{\bar\alpha}\spb^{\bar\beta}\spb^{\bar\gamma}\spb^{\bar\delta} +\order{\lambda^5}.
\end{align}
We next expand $\sb_{...}$ about $x''$: for example,
\begin{align}
\sb & = \spp-\spp_{\mu''}u^{\mu''}\Delta t +\tfrac{1}{2}\big(\spp_{\mu''}u^{\mu''}\big)_{\!;\nu''}u^{\nu''}(\Delta t)^2\nonumber\\
&\quad-\tfrac{1}{6} \big(\big(\spp_{\mu''}u^{\mu''}\big)_{\!;\nu''}u^{\nu''}\big)_{\!;\rho''} u^{\rho''}(\Delta t)^3\nonumber\\
&\quad+\tfrac{1}{24} \big(\big(\big(\spp_{\mu''}u^{\mu''}\big)_{\!;\nu''}u^{\nu''}\big)_{\!;\rho''} u^{\rho''}\big)_{\!;\upsilon''}u^{\upsilon''}(\Delta t)^4\nonumber\\
&\quad+\order{\lambda^5}.
\end{align}
Using $\spp=\frac{1}{2}r^2$, $\spp_{\mu''}=-e^a_{\mu''}x_a$, and the standard near-coincidence expansion $\spp_{\mu''\nu''}=g_{\mu''\nu''}-\frac{1}{3}R_{\mu''\gamma''\nu''\delta''} \spp^{\gamma''}\spp^{\delta''}+\mathcal{O}(\lambda^3)$, and dropping terms of order $a^2$, we arrive at the expansion
\begin{align}
\sb & =\tfrac{1}{2}r^2-\tfrac{1}{2}\left(1+a_c(t)x^c +\tfrac{1}{3}R_{0c0d}(t)x^{cd}\right)(\Delta t)^2\nonumber\\
&\quad+\tfrac{1}{6}\left(\dot a_c(t)x^c\right)(\Delta t)^3+\order{\lambda^5}.
\end{align}
The same procedure yields
\begin{align}
\sb_{\bar\alpha} &= -x_ae^a_{\bar\alpha}+\big(e^0_{\bar\alpha}+a_a(t)x^ae^0_{\bar\alpha} +\tfrac{1}{3}R_{0a0b}(t)x^{ab}e^0_{\bar\alpha}\nonumber\\
&\quad+\tfrac{1}{3}R_{ca0b}(t)x^{ab}e^c_{\bar\alpha}\big)\Delta t\nonumber\\
&\quad-\tfrac{1}{2}\left(\tfrac{1}{3}R_{0a0b}(t)x^ae^b_{\bar\alpha} +a_a(t)e^a_{\bar\alpha}\right)(\Delta t)^2\nonumber\\
&\quad+\tfrac{1}{3}\dot a_a(t)e^a_{\bar\alpha}(\Delta t)^3+\order{\lambda^4}\\
\sb_{\bar\alpha\bar\beta} & = g_{\bar\alpha\bar\beta} -\tfrac{1}{3}R_{IaJb}(t)x^{ab}e^I_{\bar\alpha}e^J_{\bar\beta}  \nonumber\\ &\quad+\tfrac{2}{3}R_{0IJb}(t)x^be^I_{(\bar\alpha}e^J_{\bar\beta)}\Delta t\nonumber\\
&\quad -\tfrac{1}{3}R_{a0b0}(t)e^a_{\bar\alpha}e^b_{\bar\beta}(\Delta t)^2+\order{\lambda^3}\\
\sb_{\bar\alpha\bar\beta\bar\gamma} & = -\tfrac{2}{3}R_{KIJa}(t)x^ae^{I}_{(\bar\alpha}e^{J}_{\bar\beta)} e^{K}_{\bar\gamma}\nonumber\\
&\quad -\tfrac{2}{3}R_{KIJ0}(t)e^{I}_{(\bar\alpha} e^{J}_{\bar\beta)}e^{K}_{\bar\gamma}\Delta t+\order{\lambda^2}\\
\sb_{\bar\alpha\bar\beta\bar\gamma\bar\delta} & = \tfrac{2}{3}R_{KIJL}(t)e^I_{(\bar\alpha}e^J_{\bar\beta)} e^K_{\bar\gamma}e^L_{\bar\delta} +\order{\lambda}.
\end{align}

Substituting these expansions into \eqref{sigma_expansion} and setting the result equal to zero, we get
\begin{equation}\label{sigma_equation}
\begin{split}
0 & = \sigma(x,x') \\
 & = \tfrac{1}{2}r^2+\tfrac{1}{2}\rad^2-x_ax'^a-\tfrac{1}{6}R_{acbd}(t)x^{cd}x'^{ab} \\
&\quad +\tfrac{1}{3}R_{0abc}(t)(x'^{ab}x^c+x'^bx^{ac})\Delta t \\
&\quad -\tfrac{1}{2}\Big[1+a_c(t)(x^c+x'^c) \\
&\quad +\tfrac{1}{3}R_{0c0d}(t)(x^{cd}+x^cx'^d+x'^{cd})\Big](\Delta t)^2 \\
&\quad +\tfrac{1}{6}\dot a_c(t)(x^c+2x'^c)(\Delta t)^3+\order{\lambda^5}.
\end{split}
\end{equation}

We next expand $\Delta t$ as 
\begin{equation}
\Delta t = \lambda\left(\Delta t_0+\lambda\Delta t_1+\lambda^2\Delta t_2\right)+\order{\lambda^4}.
\end{equation}
Substituting this into \eqref{sigma_equation} and solving order by order, we find
\begin{equation}\label{Delta t}
\begin{split}
\Delta t_0 & = \r_0 \\
\Delta t_1 & = -\tfrac{1}{2}\r_0a_c(t)(x^c+x'^c)\\
\Delta t_2 & = -\tfrac{1}{6}\r_0^{-1}R_{acbd}(t)x^{cd}x'^{ab}\\
&\quad -\tfrac{1}{6}\r_0R_{0a0b}(t)(x^{ab}+x^ax'^b+x'^{ab})\\
&\quad+\tfrac{1}{3}R_{0abc}(t)(x^{ac}x'^b-x'^{ac}x^b).
\end{split}
\end{equation}
where
\begin{equation}
\r_0=\sqrt{r^2+\rad^2-2r\rad n^an'_a}
\end{equation}
is the flat-spacetime luminosity distance between $x$ and $x'$.

Using this result for $\Delta t$, we can now find an explicit expansion for any bitensor at $x$ and $x'$. In particular, $\r\equiv\sigma_{\mu'}\frac{\partial x^{\mu'}}{\partial t'}$ can be expanded as
\begin{equation}
\r = \lambda(\r_0+\lambda\r_1+\lambda^2\r_2)+\order{\lambda^4},
\end{equation}
where $\r_0$ is given above, and
\begin{equation}
\begin{split}
\r_1  &= \tfrac{1}{2}\r_0a_c(t)(x^c+x'^c),\\
\r_2  &= -\tfrac{1}{6}\r_0^{-1}R_{acbd}(t)x^{cd}x'^{ab}-\r_0^2\dot a_a(t)x'^a,\\
&\quad +\tfrac{1}{6}\r_0R_{0a0b}(t)(x^{ab}+x^ax'^b+x'^{ab}).
\end{split}
\end{equation}
Note that $\r_1=-\Delta t_1$, which is what we would expect in flat spacetime.

The time-derivative of $\r$ can similarly be expanded to find
\begin{equation}
\partial_{t'}\r = \dot{\r_0}+\lambda\dot{\r_1}+\lambda^2\dot{\r_2}+\order{\lambda^3},
\end{equation}
where
\begin{align}
\dot{\r_0} &\equiv -1 \nonumber\\
\dot{\r_1} &\equiv -a_a(t)(x'^a+x^a)\\
\dot{\r_2} &\equiv 2\r_0\dot a_a(t)x'^a-\tfrac{1}{3}R_{0a0b}(t)(x^{ab}+x^ax'^b+x'^{ab})\nonumber
\end{align}

Other useful expansions are
\begin{equation}
\begin{split}
\sigma_{\mu'}(x,x')n^{\mu'} &= \rad-x_an'^a\\
&\quad-\tfrac{1}{2}\r_0a_a(t)n'^a(\r_0-2\r_1)\\
&\quad+\tfrac{1}{3}\dot a_a(t)n'^a\r_0^3\\
&\quad-\tfrac{1}{6}\r_0^2R_{0a0b}(t)(x^a+2x'^a)n'^b\\
&\quad+\tfrac{1}{3}\r_0R_{0acb}(t)(x^{ab}+2x'^ax^b)n'^c\\
&\quad-\tfrac{1}{3}\rad R_{acbd}(t)x^{ab}n'^{cd}+\order{\lambda^4}
\end{split}
\end{equation}
and
\begin{align}
\sigma_{\mu'\nu'}(x,x')n^{\mu'}u^{\nu'} &= \tfrac{1}{3}\r_0R_{0a0b}(t)(x^b-x'^b)n'^a\nonumber\\
&\quad+\tfrac{1}{3}R_{0abc}(t)(x^a-x'^a)x^bn'^c\nonumber\\
&\quad+\order{\lambda^3}.
\end{align}

\subsection{Expansions of Green's function}
I present here the expansion of part of the Green's function for the case in which $x$, $x'$, $\bar x$, and $x''$ lie within one another's convex normal neighbourhood. By following the same procedure and making use of  standard near-coincidence expansions given in Ref.~\cite{Eric_review}, one finds the following expansion for the direct part of the Green's function:
\begin{equation}
\begin{split}
U\indices{_{\alpha\beta}^{\alpha'\beta'}}  & = e^{(\alpha'}_{I}e^{\beta')}_{J}\Big[\U^1_{\alpha\beta}{}^{IJ} +\r_0\U^2_{\alpha\beta}{}^{IJ} +\r_0^2\U^3_{\alpha\beta}{}^{IJ}\\
&\quad +\U^4_{\alpha\beta}{}^{IJ}{}_cx'^c+\r_0\U^5_{\alpha\beta}{}^{IJ}{}_cx'^c\\
&\quad +\U^6_{\alpha\beta}{}^{IJ}{}_{cd}x'^{cd}+\order{\lambda^3}\Big],
\end{split}
\end{equation}
where
\begin{align}
\U^1_{\alpha\beta}{}^{IJ} &= e_{(\alpha}^{I}e_{\beta)}^{J}\left(1+\tfrac{1}{12}R_{kl}(t)x^{kl}\right),\\
\U^2_{\alpha\beta}{}^{IJ} &= e_{(\alpha}^{I}e_{\beta)}^{K}R_K{}^J{}_{0l}(t)x^l +\tfrac{1}{6}e_{(\alpha}^{I}e_{\beta)}^{J}R_{0k}(t)x^k\nonumber\\
&\quad +2e^I_{(\alpha}\left(e^b_{\beta)}\delta^J_0 +e^0_{\beta)}\delta^{Jb}\right)a_b(t),\\
\U^3_{\alpha\beta}{}^{IJ} &= \tfrac{1}{12}e_{(\alpha}^{I}e_{\beta)}^{J}R_{00}(t)\nonumber\\
&\quad -e^I_{(\alpha}\left(e^b_{\beta)}\delta^J_0+e^0_{\beta)}\delta^{Jb}\right)\dot a_b(t),\\
\U^4_{\alpha\beta}{}^{IJ}{}_c &= -e_{(\alpha}^{I}e_{\beta)}^{M}R_M{}^J{}_{cd}(t)x^d\nonumber\\
&\quad -\tfrac{1}{6}e_{(\alpha}^{I}e_{\beta)}^{J}R_{cd}(t)x^d,\\
\U^5_{\alpha\beta}{}^{IJ}{}_c &= -e_{(\alpha}^{I}e_{\beta)}^{M}R_M{}^J{}_{c0}(t) -\tfrac{1}{6}e_{(\alpha}^{I}e_{\beta)}^{J}R_{c0}(t),\\
\U^6_{\alpha\beta}{}^{IJ}{}_{cd} &= \tfrac{1}{12}e_{(\alpha}^{I}e_{\beta)}^{J}R_{cd}(t).
\end{align}
And the expansion of its covariant derivative is given by
\begin{align}
U\indices{_{\alpha\beta}^{\alpha'\beta'}_{;\delta'}} & = e^{(\alpha'}_{I}e^{\beta')}_{J}e^K_{\delta'}\Big[\U^7_{\alpha\beta}{}^{IJ}{}_K +\r_0\U^8_{\alpha\beta}{}^{IJ}{}_K\nonumber\\
&\quad+\U^9_{\alpha\beta}{}^{IJ}{}_{Kc}x'^c+\order{\lambda^2}\Big],
\end{align}
where
\begin{align}
\U^7_{\alpha\beta}{}^{IJ}{}_K  &= -e_{(\alpha}^{I}e_{\beta)}^{L}R_L{}^J{}_{Kc}(t)x^c\nonumber\\
&\quad-\tfrac{1}{6}e_{(\alpha}^{I}e_{\beta)}^{J}R_{Kc}(t)x^c,\\
\U^8_{\alpha\beta}{}^{IJ}{}_K  &= -e_{(\alpha}^{I}e_{\beta)}^{L}R_L{}^J{}_{K0}(t) -\tfrac{1}{6}e_{(\alpha}^{I}e_{\beta)}^{J}R_{K0}(t),\\
\U^9_{\alpha\beta}{}^{IJ}{}_{Kc}  &= e_{(\alpha}^{I}e_{\beta)}^{L}R_L{}^J{}_{Kc}(t) +\tfrac{1}{6}e_{(\alpha}^{I}e_{\beta)}^{J}R_{Kc}(t).
\end{align}

\section{Expansions appearing in the second-order wave equation}\label{second-order expansions}
I present here various expansions used in solving the second-order Einstein equation in Sec.~\ref{buffer_expansion2}.

The part of the Ricci tensor that is quadratic in the perturbation $h$ is written as
\begin{align}
\delta^2R_{\alpha\beta}[h] &=-\tfrac{1}{2}\bar h^{\mu\nu}{}_{;\nu}\left(2h_{\mu(\alpha;\beta)}-h_{\alpha\beta;\mu}\right) +\tfrac{1}{4}h^{\mu\nu}{}_{;\alpha}h_{\mu\nu;\beta}\nonumber\\
&\quad-\tfrac{1}{2}h^{\mu\nu}\left(2h_{\mu(\alpha;\beta)\nu} -h_{\alpha\beta;\mu\nu}-h_{\mu\nu;\alpha\beta}\right)\nonumber\\
&\quad+\tfrac{1}{2}h^{\mu}{}_{\beta}{}^{;\nu}\left(h_{\mu\alpha;\nu} -h_{\nu\alpha;\mu}\right).
\end{align}
I require an expansion of $\delta^2 R\coeff{0}_{\alpha\beta}[\hmn{}{1}]$ in powers of the Fermi radial coordinate $r$, where for a function $f$, $\delta^2 R\coeff{0}_{\alpha\beta}[f]$ consists of $\delta^2 R_{\alpha\beta}[f]$ with the acceleration $a^\mu$ set to zero. Explicitly, I require the coefficients in the expansion
\begin{align}
\delta^2R\coeff{0}_{\alpha\beta}[\hmn{}{1}] &= \frac{1}{r^4}\ddR{\alpha\beta}{0,-4}{\hmn{}{1}}\!\! +\frac{1}{r^3}\ddR{\alpha\beta}{0,-3}{\hmn{}{1}}\ \ \nonumber\\
&\quad+\frac{1}{r^2}\ddR{\alpha\beta}{0,-2}{\hmn{}{1}} +\order{1/r},
\end{align}
where the second superscript index in parentheses denotes the power of $r$. Making use of the expansion of $\hmn{}{1}$, obtained by setting the acceleration to zero in the results for $\hmn{E}{1}$ found in Sec.~\ref{buffer_expansion1}, one finds
\begin{align}
\ddR{\alpha\beta}{0,-4}{\hmn{}{1}} &= 2m^2\left(7\nhat_{ab}+\tfrac{4}{3}\delta_{ab}\right)x^a_\alpha x^b_\beta \nonumber\\
&\quad -2m^2t_\alpha t_\beta,\label{ddR0n4}
\end{align}
and
\begin{align}
\ddR{tt}{0,-3}{\hmn{}{1}} &= 3m\H{ij}{1,0}\nhat^{ij},\label{ddR0n3_tt}\\
\ddR{ta}{0,-3}{\hmn{}{1}} &= 3m\C{i}{1,0}\nhat_{a}^i,\\
\ddR{ab}{0,-3}{\hmn{}{1}} &= 3m\big(\A{}{1,0}+\K{}{1,0}\big)\nhat_{ab}\nonumber\\
&\quad-6m\H{i\langle a}{1,0}\nhat_{b\rangle}^i\nonumber\\
&\quad+m\delta_{ab}\H{ij}{1,0}\nhat^{ij},\label{ddR0n3_ab}
\end{align}
and
\begin{widetext}
\begin{align}
\ddR{tt}{0,-2}{\hmn{}{1}} &= -\tfrac{20}{3}m^2\etide_{ij}\nhat^{ij}+3m\H{ijk}{1,1}\nhat^{ijk}
+\tfrac{7}{5}m\A{i}{1,1}n^i+\tfrac{3}{5}m\K{i}{1,1}n^i-\tfrac{4}{5}m\partial_t\C{i}{1,0}n^i,\label{ddR0n2_tt}\\
\ddR{ta}{0,-2}{\hmn{}{1}}&= -m\partial_t\K{}{1,0}n_a+3m\C{ij}{1,1}\nhat_{a}{}^{ij}
+m\Big(\tfrac{6}{5}\C{ai}{1,1}-\partial_t\H{ai}{1,0}\Big)n^i+2m\epsilon_a{}^{ij}\D{i}{1,1}n_j\nonumber\\ &\quad+\tfrac{4}{3}m^2\epsilon_{aik}\btide^k_j\nhat^{ij},\\
\ddR{ab}{0,-2}{\hmn{}{1}} &= \delta_{ab}\left[m\left(\tfrac{16}{15}\partial_t\C{i}{1,0} -\tfrac{13}{15}\A{i}{1,1} -\tfrac{9}{5}\K{i}{1,1}\right)n^i-\tfrac{50}{9}m^2\etide_{ij}\nhat^{ij}+m\H{ijk}{1,1}\nhat^{ijk}\right] -\tfrac{14}{3}m^2\etide_{ij}\nhat_{ab}{}^{ij}\nonumber\\
&\quad+m\left(\tfrac{33}{10}\A{i}{1,1}+\tfrac{27}{10}\K{i}{1,1} -\tfrac{3}{5}\partial_t\C{i}{1,0}\right)\nhat_{ab}{}^i+m\left(\tfrac{28}{25}\A{\langle a}{1,1}-\tfrac{18}{25}\K{\langle a}{1,1}-\tfrac{46}{25}\partial_t\C{\langle a}{1,0}\right)\nhat^{}_{b\rangle}\nonumber\\
&\quad-\tfrac{8}{3}m^2\etide_{i\langle a}\nhat_{b\rangle}{}^i-6m\H{ij\langle a}{1,1}\nhat^{}_{b\rangle}{}^{ij}+3m\epsilon_{ij(a}\nhat_{b)}{}^{jk}\I{ik}{1,1} +\tfrac{2}{45}m^2\etide_{ab}\nonumber\\
&\quad-\tfrac{2}{5}m\H{abi}{1,1}n^i +\tfrac{8}{5}m\epsilon^i{}_{j(a}^{}\I{b)i}{1,1}n^j.\label{ddR0n2_ab}
\end{align}

I require an analogous expansion of $E\coeff{0}_{\alpha\beta}\left[\frac{1}{r^2}\hmn{}{2,-2} +\frac{1}{r}\hmn{}{2,-1}\right]$, where $E\coeff{0}_{\alpha\beta}[f]$ is defined for any $f$ by setting the acceleration to zero in $E_{\alpha\beta}[f]$. The coefficients of the $1/r^4$ and $1/r^3$ terms in this expansion can be found in Sec.~\ref{buffer_expansion2}; the coefficient of $1/r^2$ will be given here. For compactness, I define  this coefficient to be $\tilde E_{\alpha\beta}$. The $tt$-component of this quantity is given by 
\begin{align}
\tilde E_{tt} &= 2\partial^2_tM_in^i+\tfrac{8}{5}S^j\btide_{ij}n^i-\tfrac{2}{3}M^j\etide_{ij}n^i +\tfrac{82}{3}m^2\etide_{ij}\nhat^{ij} + 24S_{\langle i}\btide_{jk\rangle}\nhat^{ijk}-20M_{\langle i}\etide_{jk\rangle}\nhat^{ijk}.\label{E_tt}
\end{align}
The $ta$-component is given by
\begin{align}
\tilde E_{ta} &= \tfrac{44}{15}\epsilon_{aij}M^k\btide^j_kn^i -\tfrac{2}{15}\left(11S^i\etide^j_k+18M^i\btide^j\right)\epsilon_{ija}n^k +\tfrac{2}{15}\left(41S^j\etide^k_a -10M^j\btide^k_a\right)\epsilon_{ijk}n^i\nonumber\\
&\quad +4\epsilon_{aij}\left(S^j\etide_{kl} +2M_k\btide^j_l\right)\nhat^{ikl}+4\epsilon^{}_{ij\langle k}\etide_{l\rangle}^jS^i\nhat_a{}^{kl} +\tfrac{68}{3}m^2\epsilon_{aij}\btide^j_k\nhat^{ik}.
\end{align}
This can be decomposed into irreducible STF pieces via the identities
\begin{align}
\epsilon_{aij}S^i\etide^j_k & = S^i\etide^j_{(k}\epsilon_{a)ij}+\tfrac{1}{2}\epsilon_{akj}S^i\etide_i^j\\
\epsilon_{aj\langle i}\etide_{kl\rangle}S^j&=\mathop{\STF}_{ikl}\!\left[\epsilon^j{}_{al}S_{\langle i}\etide_{jk\rangle} -\tfrac{2}{3}\delta_{al}S^p\etide^j_{(i}\epsilon^{}_{k)jp}\right]\\
\epsilon_{aj\langle i}M_l\btide_{k\rangle}{}^j&=\mathop{\STF}_{ikl}\!\left[\epsilon^j{}_{al}M_{\langle i}\btide_{jk\rangle}\! +\!\tfrac{1}{3}\delta_{al}M^p\btide^j_{(i}\epsilon^{}_{k)jp}\right],
\end{align}
which follow from Eqs.~\eqref{decomposition_1} and \eqref{decomposition_2}, and which lead to
\begin{align}
\tilde E_{ta} &=\tfrac{2}{5}\epsilon_{aij}\Big(6M^k\btide^j_k-7S^k\etide^j_k\Big)n^i+\tfrac{4}{3}\left(2M^l\btide^k_{(i} -5S^l\etide^k_{(i}\right)\epsilon^{}_{j)kl}\nhat_a{}^{ij} +\left(4S^j\etide^k_{(a} -\tfrac{56}{15}M^j\btide^k_{(a}\right)\epsilon_{i)jk}n^i\nonumber\\
&\quad + 4\epsilon_{ai}{}^l\left(S_{\langle j}\etide_{kl\rangle}+2M_{\langle j}\btide_{kl\rangle}\right)\nhat^{ijk}+\tfrac{68}{3}m^2\epsilon_{aij}\btide^j_k\nhat^{ik}.\label{E_ta}
\end{align}
The $ab$-component is given by
\begin{align}
\tilde E_{ab} &= \tfrac{56}{3}m^2\etide_{ij}\nhat_{ab}{}^{ij} +\tfrac{52}{45}m^2\etide_{ab}-\delta_{ab}\left[\left(2\partial^2_tM_i+\tfrac{8}{5}S^j\btide_{ij} +\tfrac{10}{9}M^j\etide_{ij}\right)n^i+\left(\tfrac{20}{3}M_{\langle i}\etide_{jk\rangle} -\tfrac{8}{3}S_{\langle i}\btide_{jk\rangle}\right)\nhat^{ijk}+\tfrac{100}{9}m^2\etide_{ij}\nhat^{ij}\right] \nonumber\\
&\quad +\tfrac{8}{15}M_{\langle a}\etide_{b\rangle i}n^i+\tfrac{8}{15}M^i\etide_{i\langle a}n_{b\rangle}+\tfrac{56}{3}m^2\etide_{i\langle a}\nhat^{}_{b\rangle}{}^i+16M_i\etide_{j\langle a}\nhat_{b\rangle}{}^{ij}-\tfrac{32}{5}S_{\langle a}\btide_{b\rangle i}n^i+\tfrac{4}{15}\left(10S_i\btide_{ab}+27M_i\etide_{ab}\right)n^i\nonumber\\
&\quad+\tfrac{16}{3}S^i\btide_{i\langle a}n_{b\rangle}-8\epsilon_{ij\langle a}\epsilon_{b\rangle kl}S^j\btide^l_m\nhat^{ikm}+\tfrac{16}{15}\epsilon_{ij\langle a}\epsilon_{b\rangle kl}S^j\btide^{il}n^k.
\end{align}
Again, this can be decomposed, using the identities
\begin{align}
S_{\langle a}\btide_{b\rangle i} &=S_{\langle a}\btide_{bi\rangle} +\mathop{\STF}_{ab}\tfrac{1}{3}\epsilon_{ai}{}^j\epsilon_{kl(b} \btide_{j)}{}^lS^k +\tfrac{1}{10}\delta_{i\langle a}\btide_{b\rangle j}S^j,\\
S_i\btide_{ab} &=S_{\langle a}\btide_{bi\rangle} -\mathop{\STF}_{ab}\tfrac{2}{3}\epsilon_{ai}{}^j\epsilon_{kl(b} \btide_{j)}{}^lS^k +\tfrac{3}{5}\delta_{i\langle a}\btide_{b\rangle j}S^j,\\
\epsilon_{ij\langle a}\epsilon_{b\rangle kl}S^j\btide^{il} & = \mathop{\STF}_{ab}\epsilon_{akj}S^l\btide^i_{(j}\epsilon^{}_{b)il} -\tfrac{1}{2}\delta_{k\langle a}\btide_{b\rangle i}S^i,\\
\mathop{\STF}_{ikm}\epsilon_{ij\langle a}\epsilon_{b\rangle kl}S^j\btide^l_m & = \mathop{\STF}_{ikm}\mathop{\STF}_{ab}\left(2\delta_{ai}S_{\langle b}\btide_{km\rangle}+\tfrac{1}{3}\delta_{ai}\epsilon^l{}_{bk}S^j \btide^p_{(l}\epsilon^{}_{m)jp}-\tfrac{3}{10}\delta_{ai}\delta_{bk}\btide_{mj}S^j\right),
\end{align}
which lead to
\begin{align}
\tilde E_{ab} & = -2\delta_{ab}\left[\left(\partial^2_tM_i+\tfrac{4}{5}S^j\btide_{ij} +\tfrac{5}{9}M^j\etide_{ij}\right)n^i +\tfrac{50}{9}m^2\etide_{ij}\nhat^{ij}+\left(\tfrac{10}{3}M_{\langle i}\etide_{jk\rangle} -\tfrac{4}{3}S_{\langle i}\btide_{jk\rangle} \right)\nhat^{ijk}\right]\nonumber\\ &\quad+\tfrac{1}{5}\left(8M^j\etide_{ij} +12S^j\btide_{ij}\right)\nhat_{ab}{}^i +\tfrac{56}{3}m^2\etide_{ij}\nhat_{ab}{}^{ij} +\tfrac{4}{75}\left(92M^j\etide_{j\langle a} +108S^j\btide_{j\langle a}\right)n_{b\rangle}^{} +\tfrac{56}{3}m^2\etide_{i\langle a}\nhat_{b\rangle}{}^i \nonumber\\
&\quad+16\mathop{\STF}_{aij}\left(M_i\etide_{j\langle a}-S_i\btide_{j\langle a}\right)\nhat^{}_{b\rangle}{}^{ij} -\tfrac{8}{3}\epsilon^{pq}{}_{\langle j}\left(2\etide_{k\rangle p}M_q+\btide_{k\rangle p}S_q\right)\epsilon^k{}_{i(a}\nhat_{b)}{}^{ij} +\tfrac{16}{15}m^2\etide_{ab} \nonumber\\
&\quad+\tfrac{4}{15}\left(29M_{\langle a}\etide_{bi\rangle}-14S_{\langle a}\btide_{bi\rangle}\right)n^i-\tfrac{16}{45}\mathop{\STF}_{ab}\epsilon_{ai}{}^jn^i\epsilon^{pq}{}_{(b} \left(13\etide_{j)q}M_p+14\btide_{j)q}S_p\right).\label{E_ab}
\end{align}
\end{widetext}

\bibliography{self_force}

\begin{thebibliography}{81}
\expandafter\ifx\csname natexlab\endcsname\relax\def\natexlab#1{#1}\fi
\expandafter\ifx\csname bibnamefont\endcsname\relax
  \def\bibnamefont#1{#1}\fi
\expandafter\ifx\csname bibfnamefont\endcsname\relax
  \def\bibfnamefont#1{#1}\fi
\expandafter\ifx\csname citenamefont\endcsname\relax
  \def\citenamefont#1{#1}\fi
\expandafter\ifx\csname url\endcsname\relax
  \def\url#1{\texttt{#1}}\fi
\expandafter\ifx\csname urlprefix\endcsname\relax\def\urlprefix{URL }\fi
\providecommand{\bibinfo}[2]{#2}
\providecommand{\eprint}[2][]{\url{#2}}

\bibitem[{\citenamefont{Einstein et~al.}(1938)\citenamefont{Einstein, Infeld,
  and Hoffmann}}]{Einstein}
\bibinfo{author}{\bibfnamefont{A.}~\bibnamefont{Einstein}},
  \bibinfo{author}{\bibfnamefont{L.}~\bibnamefont{Infeld}}, \bibnamefont{and}
  \bibinfo{author}{\bibfnamefont{B.}~\bibnamefont{Hoffmann}},
  \bibinfo{journal}{Ann. Math.} \textbf{\bibinfo{volume}{39}},
  \bibinfo{pages}{65} (\bibinfo{year}{1938}).

\bibitem[{\citenamefont{Einstein and Infeld}(1949)}]{Einstein2}
\bibinfo{author}{\bibfnamefont{A.}~\bibnamefont{Einstein}} \bibnamefont{and}
  \bibinfo{author}{\bibfnamefont{L.}~\bibnamefont{Infeld}},
  \bibinfo{journal}{Can. J. Math.} \textbf{\bibinfo{volume}{1}},
  \bibinfo{pages}{209} (\bibinfo{year}{1949}).

\bibitem[{\citenamefont{Futamase and Itoh}(2007)}]{Futamase_review}
\bibinfo{author}{\bibfnamefont{T.}~\bibnamefont{Futamase}} \bibnamefont{and}
  \bibinfo{author}{\bibfnamefont{Y.}~\bibnamefont{Itoh}},
  \bibinfo{journal}{Living Rev. Relativity} \textbf{\bibinfo{volume}{10}},
  \bibinfo{pages}{2} (\bibinfo{year}{2007}),
  \eprint{http://www.livingreviews.org/lrr-2007-2}.

\bibitem[{\citenamefont{Blanchet}(2006)}]{Blanchet_review}
\bibinfo{author}{\bibfnamefont{L.}~\bibnamefont{Blanchet}},
  \bibinfo{journal}{Living Rev. Relativity} \textbf{\bibinfo{volume}{9}},
  \bibinfo{pages}{4} (\bibinfo{year}{2006}),
  \eprint{http://www.livingreviews.org/lrr-2006-4}.

\bibitem[{\citenamefont{Infeld and Schild}(1949)}]{Infeld}
\bibinfo{author}{\bibfnamefont{L.}~\bibnamefont{Infeld}} \bibnamefont{and}
  \bibinfo{author}{\bibfnamefont{A.}~\bibnamefont{Schild}},
  \bibinfo{journal}{Rev. Mod. Phys.} \textbf{\bibinfo{volume}{21}},
  \bibinfo{pages}{408} (\bibinfo{year}{1949}).

\bibitem[{\citenamefont{D'Eath}(1996)}]{DEath}
\bibinfo{author}{\bibfnamefont{P.}~\bibnamefont{D'Eath}},
  \emph{\bibinfo{title}{Black holes: gravitational interactions}}
  (\bibinfo{publisher}{Oxford University Press}, \bibinfo{address}{New York},
  \bibinfo{year}{1996}).

\bibitem[{\citenamefont{Kates}(1980{\natexlab{a}})}]{Kates_motion}
\bibinfo{author}{\bibfnamefont{R.~E.} \bibnamefont{Kates}},
  \bibinfo{journal}{Phys. Rev. D} \textbf{\bibinfo{volume}{22}},
  \bibinfo{pages}{1853} (\bibinfo{year}{1980}{\natexlab{a}}).

\bibitem[{\citenamefont{Geroch and Jang}(1975)}]{Geroch_particle1}
\bibinfo{author}{\bibfnamefont{R.}~\bibnamefont{Geroch}} \bibnamefont{and}
  \bibinfo{author}{\bibfnamefont{P.}~\bibnamefont{Jang}}, \bibinfo{journal}{J.
  Math. Phys.} \textbf{\bibinfo{volume}{16}}, \bibinfo{pages}{65}
  (\bibinfo{year}{1975}).

\bibitem[{\citenamefont{Ehlers and Geroch}(2004)}]{Geroch_particle2}
\bibinfo{author}{\bibfnamefont{J.}~\bibnamefont{Ehlers}} \bibnamefont{and}
  \bibinfo{author}{\bibfnamefont{R.}~\bibnamefont{Geroch}},
  \bibinfo{journal}{Ann. Phys.} \textbf{\bibinfo{volume}{309}},
  \bibinfo{pages}{232} (\bibinfo{year}{2004}).

\bibitem[{\citenamefont{D'Eath}(1975)}]{DEath_paper}
\bibinfo{author}{\bibfnamefont{P.}~\bibnamefont{D'Eath}},
  \bibinfo{journal}{Phys. Rev. D} \textbf{\bibinfo{volume}{11}},
  \bibinfo{pages}{1387} (\bibinfo{year}{1975}).

\bibitem[{LIG()}]{LIGO}
\bibinfo{note}{The LIGO website is located at http://www.ligo.caltech.edu}.

\bibitem[{LIS()}]{LISA}
\bibinfo{note}{The LISA website is located at http://lisa.jpl.nasa.gov}.

\bibitem[{\citenamefont{Hughes}(2009)}]{binary_review}
\bibinfo{author}{\bibfnamefont{S.~A.} \bibnamefont{Hughes}}
  (\bibinfo{year}{2009}), \eprint{arXiv:astro-ph/0903.4877}.

\bibitem[{\citenamefont{Baumgarte and Shapiro}(2003)}]{numerical_review}
\bibinfo{author}{\bibfnamefont{T.~W.} \bibnamefont{Baumgarte}}
  \bibnamefont{and} \bibinfo{author}{\bibfnamefont{S.~L.}
  \bibnamefont{Shapiro}}, \bibinfo{journal}{Phys. Rept.}
  \textbf{\bibinfo{volume}{376}}, \bibinfo{pages}{41} (\bibinfo{year}{2003}),
  \eprint{arXiv:gr-qc/0211028}.

\bibitem[{\citenamefont{Pretorius}(2007)}]{Pretorius_review}
\bibinfo{author}{\bibfnamefont{F.}~\bibnamefont{Pretorius}}
  (\bibinfo{year}{2007}), \eprint{arXiv:gr-qc/0710.1338}.

\bibitem[{\citenamefont{Drasco}(2006)}]{Drasco_review}
\bibinfo{author}{\bibfnamefont{S.}~\bibnamefont{Drasco}},
  \bibinfo{journal}{Class. Quant. Grav.} \textbf{\bibinfo{volume}{23}},
  \bibinfo{pages}{S769} (\bibinfo{year}{2006}), \eprint{arXiv:gr-qc/0604115}.

\bibitem[{\citenamefont{Amaro-Seoane et~al.}(2007)}]{EMRI_review}
\bibinfo{author}{\bibfnamefont{P.}~\bibnamefont{Amaro-Seoane}}
  \bibnamefont{et~al.}, \bibinfo{journal}{Class. Quant. Grav.}
  \textbf{\bibinfo{volume}{24}}, \bibinfo{pages}{R113} (\bibinfo{year}{2007}),
  \eprint{arXiv:astro-ph/0703495}.

\bibitem[{\citenamefont{Mino et~al.}(1997{\natexlab{a}})\citenamefont{Mino,
  Sasaki, and Tanaka}}]{Mino_Sasaki_Tanaka}
\bibinfo{author}{\bibfnamefont{Y.}~\bibnamefont{Mino}},
  \bibinfo{author}{\bibfnamefont{M.}~\bibnamefont{Sasaki}}, \bibnamefont{and}
  \bibinfo{author}{\bibfnamefont{T.}~\bibnamefont{Tanaka}},
  \bibinfo{journal}{Phys. Rev. D} \textbf{\bibinfo{volume}{55}},
  \bibinfo{pages}{3457} (\bibinfo{year}{1997}{\natexlab{a}}),
  \bibinfo{note}{arXiv:gr-qc/9606018}.

\bibitem[{\citenamefont{Quinn and Wald}(1997)}]{Quinn_Wald}
\bibinfo{author}{\bibfnamefont{T.}~\bibnamefont{Quinn}} \bibnamefont{and}
  \bibinfo{author}{\bibfnamefont{R.}~\bibnamefont{Wald}},
  \bibinfo{journal}{Phys. Rev. D} \textbf{\bibinfo{volume}{56}},
  \bibinfo{pages}{3381} (\bibinfo{year}{1997}), \eprint{arXiv:gr-qc/9610053}.

\bibitem[{\citenamefont{Gralla and Wald}(2008)}]{Gralla_Wald}
\bibinfo{author}{\bibfnamefont{S.~E.} \bibnamefont{Gralla}} \bibnamefont{and}
  \bibinfo{author}{\bibfnamefont{R.~M.} \bibnamefont{Wald}},
  \bibinfo{journal}{Class. Quant. Grav.} \textbf{\bibinfo{volume}{25}},
  \bibinfo{pages}{205009} (\bibinfo{year}{2008}),
  \eprint{arXiv:gr-qc/0806.3293}.

\bibitem[{\citenamefont{Rosenthal}(2006{\natexlab{a}})}]{Eran_force}
\bibinfo{author}{\bibfnamefont{E.}~\bibnamefont{Rosenthal}},
  \bibinfo{journal}{Phys. Rev. D} \textbf{\bibinfo{volume}{74}},
  \bibinfo{pages}{084018} (\bibinfo{year}{2006}{\natexlab{a}}),
  \eprint{arXiv:gr-qc/0609069}.

\bibitem[{\citenamefont{Detweiler and Whiting}(2003)}]{Detweiler_Whiting}
\bibinfo{author}{\bibfnamefont{S.}~\bibnamefont{Detweiler}} \bibnamefont{and}
  \bibinfo{author}{\bibfnamefont{B.}~\bibnamefont{Whiting}},
  \bibinfo{journal}{Phys. Rev. D.} \textbf{\bibinfo{volume}{67}},
  \bibinfo{pages}{024025} (\bibinfo{year}{2003}), \eprint{arXiv:gr-qc/0202086}.

\bibitem[{\citenamefont{Galley and Hu}(2009{\natexlab{a}})}]{Galley_Hu}
\bibinfo{author}{\bibfnamefont{C.~R.} \bibnamefont{Galley}} \bibnamefont{and}
  \bibinfo{author}{\bibfnamefont{B.~L.} \bibnamefont{Hu}},
  \bibinfo{journal}{Phys. Rev. D} \textbf{\bibinfo{volume}{79}},
  \bibinfo{pages}{064002} (\bibinfo{year}{2009}{\natexlab{a}}),
  \eprint{arXiv:gr-qc/0801.0900}.

\bibitem[{\citenamefont{Fukumoto et~al.}(2006)\citenamefont{Fukumoto, Futamase,
  and Itoh}}]{Fukumoto}
\bibinfo{author}{\bibfnamefont{T.}~\bibnamefont{Fukumoto}},
  \bibinfo{author}{\bibfnamefont{T.}~\bibnamefont{Futamase}}, \bibnamefont{and}
  \bibinfo{author}{\bibfnamefont{Y.}~\bibnamefont{Itoh}},
  \bibinfo{journal}{Prog. Theor. Phys.} \textbf{\bibinfo{volume}{116}},
  \bibinfo{pages}{423} (\bibinfo{year}{2006}), \eprint{arXiv:gr-qc/0606114}.

\bibitem[{\citenamefont{Mino et~al.}(1997{\natexlab{b}})\citenamefont{Mino,
  Sasaki, and Tanaka}}]{Mino_matching}
\bibinfo{author}{\bibfnamefont{Y.}~\bibnamefont{Mino}},
  \bibinfo{author}{\bibfnamefont{M.}~\bibnamefont{Sasaki}}, \bibnamefont{and}
  \bibinfo{author}{\bibfnamefont{T.}~\bibnamefont{Tanaka}},
  \bibinfo{journal}{Prog. Theor. Phys. Suppl.} \textbf{\bibinfo{volume}{128}},
  \bibinfo{pages}{373} (\bibinfo{year}{1997}{\natexlab{b}}),
  \eprint{arXiv:gr-qc/9712056}.

\bibitem[{\citenamefont{Gal'tsov et~al.}(2007)\citenamefont{Gal'tsov, Spirin,
  and Staub}}]{Gal'tsov}
\bibinfo{author}{\bibfnamefont{D.}~\bibnamefont{Gal'tsov}},
  \bibinfo{author}{\bibfnamefont{P.}~\bibnamefont{Spirin}}, \bibnamefont{and}
  \bibinfo{author}{\bibfnamefont{S.}~\bibnamefont{Staub}}
  (\bibinfo{year}{2007}), \eprint{arXiv:gr-qc/0701004}.

\bibitem[{\citenamefont{Galley et~al.}(2006)\citenamefont{Galley, Hu, and
  Lin}}]{Galley_QFT}
\bibinfo{author}{\bibfnamefont{C.~R.} \bibnamefont{Galley}},
  \bibinfo{author}{\bibfnamefont{B.~L.} \bibnamefont{Hu}}, \bibnamefont{and}
  \bibinfo{author}{\bibfnamefont{S.-Y.} \bibnamefont{Lin}},
  \bibinfo{journal}{Phys. Rev. D} \textbf{\bibinfo{volume}{74}},
  \bibinfo{pages}{024017} (\bibinfo{year}{2006}), \eprint{arXiv:gr-qc/0603099}.

\bibitem[{\citenamefont{Detweiler}(2005)}]{Detweiler_review}
\bibinfo{author}{\bibfnamefont{S.}~\bibnamefont{Detweiler}},
  \bibinfo{journal}{Class. Quant. Grav.} \textbf{\bibinfo{volume}{22}},
  \bibinfo{pages}{S681} (\bibinfo{year}{2005}), \eprint{arXiv:gr-qc/0501004}.

\bibitem[{\citenamefont{Poisson}(2004)}]{Eric_review}
\bibinfo{author}{\bibfnamefont{E.}~\bibnamefont{Poisson}},
  \bibinfo{journal}{Living Rev. Relativity} \textbf{\bibinfo{volume}{7}},
  \bibinfo{pages}{6} (\bibinfo{year}{2004}),
  \eprint{http://www.livingreviews.org/lrr-2004-6}.

\bibitem[{\citenamefont{Geroch and Traschen}(1987)}]{linear_distributions}
\bibinfo{author}{\bibfnamefont{R.}~\bibnamefont{Geroch}} \bibnamefont{and}
  \bibinfo{author}{\bibfnamefont{J.}~\bibnamefont{Traschen}},
  \bibinfo{journal}{Phys. Rev. D} \textbf{\bibinfo{volume}{36}},
  \bibinfo{pages}{1017} (\bibinfo{year}{1987}).

\bibitem[{\citenamefont{Steinbauer and
  Vickers}(2006)}]{nonlinear_distributions}
\bibinfo{author}{\bibfnamefont{R.}~\bibnamefont{Steinbauer}} \bibnamefont{and}
  \bibinfo{author}{\bibfnamefont{J.~A.} \bibnamefont{Vickers}},
  \bibinfo{journal}{Class. Quant. Grav.} \textbf{\bibinfo{volume}{23}},
  \bibinfo{pages}{R91} (\bibinfo{year}{2006}), \eprint{arXiv:gr-qc/0603078}.

\bibitem[{\citenamefont{Mino}(2006)}]{Mino_expansion2}
\bibinfo{author}{\bibfnamefont{Y.}~\bibnamefont{Mino}}, \bibinfo{journal}{Prog.
  Theor. Phys.} \textbf{\bibinfo{volume}{115}}, \bibinfo{pages}{43}
  (\bibinfo{year}{2006}), \eprint{arXiv:gr-qc/0601019}.

\bibitem[{\citenamefont{Mino and Price}(2008)}]{Mino_Price}
\bibinfo{author}{\bibfnamefont{Y.}~\bibnamefont{Mino}} \bibnamefont{and}
  \bibinfo{author}{\bibfnamefont{R.}~\bibnamefont{Price}},
  \bibinfo{journal}{Phys. Rev. D} \textbf{\bibinfo{volume}{77}},
  \bibinfo{pages}{064001} (\bibinfo{year}{2008}),
  \eprint{arXiv:gr-qc/0801.0179}.

\bibitem[{\citenamefont{Mino}(2005)}]{Mino_expansion1}
\bibinfo{author}{\bibfnamefont{Y.}~\bibnamefont{Mino}}, \bibinfo{journal}{Prog.
  Theor. Phys.} \textbf{\bibinfo{volume}{113}}, \bibinfo{pages}{733}
  (\bibinfo{year}{2005}), \eprint{arXiv:gr-qc/0506003}.

\bibitem[{\citenamefont{Hinderer and Flanagan}(2008)}]{Hinderer_Flanagan}
\bibinfo{author}{\bibfnamefont{T.}~\bibnamefont{Hinderer}} \bibnamefont{and}
  \bibinfo{author}{\bibfnamefont{E.~E.} \bibnamefont{Flanagan}},
  \bibinfo{journal}{Phys. Rev. D} \textbf{\bibinfo{volume}{78}},
  \bibinfo{pages}{064028} (\bibinfo{year}{2008}),
  \eprint{arXiv:gr-qc/0805.3337}.

\bibitem[{\citenamefont{Dirac}(1938)}]{Dirac}
\bibinfo{author}{\bibfnamefont{P.}~\bibnamefont{Dirac}},
  \bibinfo{journal}{Proc. R. Soc. London, Ser. A}
  \textbf{\bibinfo{volume}{167}}, \bibinfo{pages}{148} (\bibinfo{year}{1938}).

\bibitem[{\citenamefont{DeWitt and Brehme}(1960)}]{DeWitt_Brehme}
\bibinfo{author}{\bibfnamefont{B.}~\bibnamefont{DeWitt}} \bibnamefont{and}
  \bibinfo{author}{\bibfnamefont{R.}~\bibnamefont{Brehme}},
  \bibinfo{journal}{Ann. Phys. (N.Y.)} \textbf{\bibinfo{volume}{9}},
  \bibinfo{pages}{220} (\bibinfo{year}{1960}).

\bibitem[{\citenamefont{Holmes}(1995)}]{Holmes}
\bibinfo{author}{\bibfnamefont{M.~H.} \bibnamefont{Holmes}},
  \emph{\bibinfo{title}{Introduction to perturbation methods}}
  (\bibinfo{publisher}{Springer-Verlag}, \bibinfo{address}{New York},
  \bibinfo{year}{1995}).

\bibitem[{\citenamefont{Verhulst}(2005)}]{Verhulst}
\bibinfo{author}{\bibfnamefont{F.}~\bibnamefont{Verhulst}},
  \emph{\bibinfo{title}{Methods and applications of singular perturbations}}
  (\bibinfo{publisher}{Springer}, \bibinfo{address}{New York},
  \bibinfo{year}{2005}).

\bibitem[{\citenamefont{Lagerstrom}(1988)}]{Lagerstrom}
\bibinfo{author}{\bibfnamefont{P.~A.} \bibnamefont{Lagerstrom}},
  \emph{\bibinfo{title}{Matched asymptotic expansions}}
  (\bibinfo{publisher}{Springer-Verlag}, \bibinfo{address}{New York},
  \bibinfo{year}{1988}).

\bibitem[{\citenamefont{Kevorkian and Cole}(1996)}]{Kevorkian_Cole}
\bibinfo{author}{\bibfnamefont{J.}~\bibnamefont{Kevorkian}} \bibnamefont{and}
  \bibinfo{author}{\bibfnamefont{J.~D.} \bibnamefont{Cole}},
  \emph{\bibinfo{title}{Multiple scale and singular perturbation methods}}
  (\bibinfo{publisher}{Springer}, \bibinfo{address}{New York},
  \bibinfo{year}{1996}).

\bibitem[{\citenamefont{Eckhaus}(1979)}]{Eckhaus}
\bibinfo{author}{\bibfnamefont{W.}~\bibnamefont{Eckhaus}},
  \emph{\bibinfo{title}{Asymptotic Analysis of Singular Perturbations}}
  (\bibinfo{publisher}{Elsevier North-Holland}, \bibinfo{address}{New York},
  \bibinfo{year}{1979}).

\bibitem[{\citenamefont{Pound}()}]{perturbation_techniques}
\bibinfo{author}{\bibfnamefont{A.}~\bibnamefont{Pound}}, \bibinfo{note}{in
  preparation}.

\bibitem[{\citenamefont{Kates}(1981)}]{Kates_structure}
\bibinfo{author}{\bibfnamefont{R.}~\bibnamefont{Kates}}, \bibinfo{journal}{Ann.
  Phys. (N.Y.)} \textbf{\bibinfo{volume}{132}}, \bibinfo{pages}{1}
  (\bibinfo{year}{1981}).

\bibitem[{\citenamefont{Ehlers et~al.}(1976)\citenamefont{Ehlers, Rosenblum,
  Goldberg, and Havas}}]{relaxed_EFE1}
\bibinfo{author}{\bibfnamefont{J.}~\bibnamefont{Ehlers}},
  \bibinfo{author}{\bibfnamefont{A.}~\bibnamefont{Rosenblum}},
  \bibinfo{author}{\bibfnamefont{J.}~\bibnamefont{Goldberg}}, \bibnamefont{and}
  \bibinfo{author}{\bibfnamefont{P.}~\bibnamefont{Havas}},
  \bibinfo{journal}{Astrophys. J. Lett.} \textbf{\bibinfo{volume}{208}},
  \bibinfo{pages}{L77} (\bibinfo{year}{1976}).

\bibitem[{\citenamefont{Walker and Will}(1980)}]{relaxed_EFE2}
\bibinfo{author}{\bibfnamefont{M.}~\bibnamefont{Walker}} \bibnamefont{and}
  \bibinfo{author}{\bibfnamefont{C.}~\bibnamefont{Will}},
  \bibinfo{journal}{Astrophys. J.} \textbf{\bibinfo{volume}{242}},
  \bibinfo{pages}{L129} (\bibinfo{year}{1980}).

\bibitem[{\citenamefont{Havas}(1957)}]{damping}
\bibinfo{author}{\bibfnamefont{P.}~\bibnamefont{Havas}},
  \bibinfo{journal}{Phys. Rev.} \textbf{\bibinfo{volume}{108}},
  \bibinfo{pages}{1351} (\bibinfo{year}{1957}).

\bibitem[{\citenamefont{Havas and Goldberg}(1962)}]{damping2}
\bibinfo{author}{\bibfnamefont{P.}~\bibnamefont{Havas}} \bibnamefont{and}
  \bibinfo{author}{\bibfnamefont{J.~N.} \bibnamefont{Goldberg}},
  \bibinfo{journal}{Phys. Rev.} \textbf{\bibinfo{volume}{128}},
  \bibinfo{pages}{398} (\bibinfo{year}{1962}).

\bibitem[{\citenamefont{Anderson et~al.}(2005)\citenamefont{Anderson, Flanagan,
  and Ottewill}}]{quasilocal}
\bibinfo{author}{\bibfnamefont{W.~G.} \bibnamefont{Anderson}},
  \bibinfo{author}{\bibfnamefont{E.~E.} \bibnamefont{Flanagan}},
  \bibnamefont{and} \bibinfo{author}{\bibfnamefont{A.~C.}
  \bibnamefont{Ottewill}}, \bibinfo{journal}{Phys. Rev. D}
  \textbf{\bibinfo{volume}{71}}, \bibinfo{pages}{024036}
  (\bibinfo{year}{2005}), \eprint{arXiv:gr-qc/0412009}.

\bibitem[{\citenamefont{Casals et~al.}(2009)\citenamefont{Casals, Dolan,
  Ottewill, and Wardell}}]{quasilocal2}
\bibinfo{author}{\bibfnamefont{M.}~\bibnamefont{Casals}},
  \bibinfo{author}{\bibfnamefont{S.~R.} \bibnamefont{Dolan}},
  \bibinfo{author}{\bibfnamefont{A.~C.} \bibnamefont{Ottewill}},
  \bibnamefont{and} \bibinfo{author}{\bibfnamefont{B.}~\bibnamefont{Wardell}}
  (\bibinfo{year}{2009}), \eprint{arXiv:gr-qc/0903.0395}.

\bibitem[{\citenamefont{Anderson and Wiseman}(2005)}]{quasilocal3}
\bibinfo{author}{\bibfnamefont{W.~G.} \bibnamefont{Anderson}} \bibnamefont{and}
  \bibinfo{author}{\bibfnamefont{A.~G.} \bibnamefont{Wiseman}},
  \bibinfo{journal}{Class. Quant. Grav.} \textbf{\bibinfo{volume}{22}},
  \bibinfo{pages}{S783} (\bibinfo{year}{2005}), \eprint{arXiv:gr-qc/0506136}.

\bibitem[{\citenamefont{Hughes}(2000)}]{Hughes_adiabatic}
\bibinfo{author}{\bibfnamefont{S.~A.} \bibnamefont{Hughes}},
  \bibinfo{journal}{Phys. Rev.} \textbf{\bibinfo{volume}{D61}},
  \bibinfo{pages}{084004} (\bibinfo{year}{2000}), \eprint{arXiv:gr-qc/9910091}.

\bibitem[{\citenamefont{Harte}(2008)}]{Harte}
\bibinfo{author}{\bibfnamefont{A.~I.} \bibnamefont{Harte}},
  \bibinfo{journal}{Class. Quant. Grav.} \textbf{\bibinfo{volume}{25}},
  \bibinfo{pages}{235020} (\bibinfo{year}{2008}),
  \eprint{arXiv:gr-qc/0807.1150}.

\bibitem[{\citenamefont{Dixon}(1974)}]{Dixon}
\bibinfo{author}{\bibfnamefont{W.~G.} \bibnamefont{Dixon}},
  \bibinfo{journal}{Phil. Trans. R. Soc. Lond. A}
  \textbf{\bibinfo{volume}{277}}, \bibinfo{pages}{59} (\bibinfo{year}{1974}).

\bibitem[{\citenamefont{Futamase}(1985)}]{Futamase_particle1}
\bibinfo{author}{\bibfnamefont{T.}~\bibnamefont{Futamase}},
  \bibinfo{journal}{Phys. Rev. D} \textbf{\bibinfo{volume}{32}},
  \bibinfo{pages}{2566} (\bibinfo{year}{1985}).

\bibitem[{\citenamefont{Futamase}(1987)}]{Futamase_particle2}
\bibinfo{author}{\bibfnamefont{T.}~\bibnamefont{Futamase}},
  \bibinfo{journal}{Phys. Rev. D} \textbf{\bibinfo{volume}{36}},
  \bibinfo{pages}{321} (\bibinfo{year}{1987}).

\bibitem[{\citenamefont{Manasse}(1963)}]{Manasse}
\bibinfo{author}{\bibfnamefont{F.~K.} \bibnamefont{Manasse}},
  \bibinfo{journal}{J. Math. Phys.} \textbf{\bibinfo{volume}{4}},
  \bibinfo{pages}{746} (\bibinfo{year}{1963}).

\bibitem[{\citenamefont{Poisson}(2005)}]{Eric_tidal}
\bibinfo{author}{\bibfnamefont{E.}~\bibnamefont{Poisson}},
  \bibinfo{journal}{Phys. Rev. Lett.} \textbf{\bibinfo{volume}{94}},
  \bibinfo{pages}{161103} (\bibinfo{year}{2005}).

\bibitem[{\citenamefont{Thorne and Hartle}(1985)}]{Thorne_Hartle}
\bibinfo{author}{\bibfnamefont{K.~S.} \bibnamefont{Thorne}} \bibnamefont{and}
  \bibinfo{author}{\bibfnamefont{J.~B.} \bibnamefont{Hartle}},
  \bibinfo{journal}{Phys. Rev. D} \textbf{\bibinfo{volume}{31}},
  \bibinfo{pages}{1815} (\bibinfo{year}{1985}).

\bibitem[{\citenamefont{Kates}(1980{\natexlab{b}})}]{Kates_Lorenz_force}
\bibinfo{author}{\bibfnamefont{R.~E.} \bibnamefont{Kates}},
  \bibinfo{journal}{Phys. Rev. D} \textbf{\bibinfo{volume}{22}},
  \bibinfo{pages}{1879} (\bibinfo{year}{1980}{\natexlab{b}}).

\bibitem[{\citenamefont{Kates}(1980{\natexlab{c}})}]{Kates_PN}
\bibinfo{author}{\bibfnamefont{R.~E.} \bibnamefont{Kates}},
  \bibinfo{journal}{Phys. Rev. D} \textbf{\bibinfo{volume}{22}},
  \bibinfo{pages}{1871} (\bibinfo{year}{1980}{\natexlab{c}}).

\bibitem[{\citenamefont{Taylor and Poisson}(2008)}]{PN_matching}
\bibinfo{author}{\bibfnamefont{S.}~\bibnamefont{Taylor}} \bibnamefont{and}
  \bibinfo{author}{\bibfnamefont{E.}~\bibnamefont{Poisson}},
  \bibinfo{journal}{Phys. Rev. D} \textbf{\bibinfo{volume}{78}},
  \bibinfo{pages}{084016} (\bibinfo{year}{2008}),
  \eprint{arXiv:gr-qc/0806.3052}.

\bibitem[{\citenamefont{Rosenthal}(2006{\natexlab{b}})}]{Eran_field}
\bibinfo{author}{\bibfnamefont{E.}~\bibnamefont{Rosenthal}},
  \bibinfo{journal}{Phys. Rev. D} \textbf{\bibinfo{volume}{73}},
  \bibinfo{pages}{044034} (\bibinfo{year}{2006}{\natexlab{b}}),
  \eprint{arXiv:gr-qc/0602066}.

\bibitem[{\citenamefont{Racine and Flanagan}(2005)}]{Racine_Flanagan}
\bibinfo{author}{\bibfnamefont{E.}~\bibnamefont{Racine}} \bibnamefont{and}
  \bibinfo{author}{\bibfnamefont{E.~E.} \bibnamefont{Flanagan}},
  \bibinfo{journal}{Phys. Rev. D} \textbf{\bibinfo{volume}{71}},
  \bibinfo{pages}{044010} (\bibinfo{year}{2005}), \eprint{arXiv:gr-qc/0404101}.

\bibitem[{\citenamefont{Sciama et~al.}(1969)\citenamefont{Sciama, Waylen, and
  Gilman}}]{Sciama}
\bibinfo{author}{\bibfnamefont{D.}~\bibnamefont{Sciama}},
  \bibinfo{author}{\bibfnamefont{P.}~\bibnamefont{Waylen}}, \bibnamefont{and}
  \bibinfo{author}{\bibfnamefont{R.~C.} \bibnamefont{Gilman}},
  \bibinfo{journal}{Phys. Rev.} \textbf{\bibinfo{volume}{187}},
  \bibinfo{pages}{1762} (\bibinfo{year}{1969}).

\bibitem[{\citenamefont{Damour and Iyer}(1991)}]{STF_2}
\bibinfo{author}{\bibfnamefont{T.}~\bibnamefont{Damour}} \bibnamefont{and}
  \bibinfo{author}{\bibfnamefont{B.}~\bibnamefont{Iyer}},
  \bibinfo{journal}{Phys. Rev. D} \textbf{\bibinfo{volume}{43}},
  \bibinfo{pages}{3259} (\bibinfo{year}{1991}).

\bibitem[{\citenamefont{Vega and Detweiler}(2008)}]{regularization1}
\bibinfo{author}{\bibfnamefont{I.}~\bibnamefont{Vega}} \bibnamefont{and}
  \bibinfo{author}{\bibfnamefont{S.}~\bibnamefont{Detweiler}},
  \bibinfo{journal}{Phys. Rev. D} \textbf{\bibinfo{volume}{77}},
  \bibinfo{pages}{084008} (\bibinfo{year}{2008}),
  \eprint{arXiv:gr-qc/0712.4405}.

\bibitem[{\citenamefont{Barack et~al.}(2007)\citenamefont{Barack, Golbourn, and
  Sago}}]{regularization2}
\bibinfo{author}{\bibfnamefont{L.}~\bibnamefont{Barack}},
  \bibinfo{author}{\bibfnamefont{D.~A.} \bibnamefont{Golbourn}},
  \bibnamefont{and} \bibinfo{author}{\bibfnamefont{N.}~\bibnamefont{Sago}},
  \bibinfo{journal}{Phys. Rev. D} \textbf{\bibinfo{volume}{76}},
  \bibinfo{pages}{124036} (\bibinfo{year}{2007}),
  \eprint{arXiv:gr-qc/0709.4588}.

\bibitem[{\citenamefont{Haas and Poisson}(2006)}]{regularization3}
\bibinfo{author}{\bibfnamefont{R.}~\bibnamefont{Haas}} \bibnamefont{and}
  \bibinfo{author}{\bibfnamefont{E.}~\bibnamefont{Poisson}},
  \bibinfo{journal}{Phys. Rev. D} \textbf{\bibinfo{volume}{74}},
  \bibinfo{pages}{044009} (\bibinfo{year}{2006}), \eprint{arXiv:gr-qc/0605077}.

\bibitem[{\citenamefont{Haas}(2007)}]{regularization4}
\bibinfo{author}{\bibfnamefont{R.}~\bibnamefont{Haas}}, \bibinfo{journal}{Phys.
  Rev. D} \textbf{\bibinfo{volume}{75}}, \bibinfo{pages}{124011}
  (\bibinfo{year}{2007}), \eprint{arXiv:gr-qc/0704.0797}.

\bibitem[{\citenamefont{Barack and Ori}(2003)}]{regularization5}
\bibinfo{author}{\bibfnamefont{L.}~\bibnamefont{Barack}} \bibnamefont{and}
  \bibinfo{author}{\bibfnamefont{A.}~\bibnamefont{Ori}},
  \bibinfo{journal}{Phys. Rev. D} \textbf{\bibinfo{volume}{67}},
  \bibinfo{pages}{024029} (\bibinfo{year}{2003}), \eprint{arXiv:gr-qc/0209072}.

\bibitem[{\citenamefont{Barack et~al.}(2002)\citenamefont{Barack, Mino, Nakano,
  Ori, and Sasaki}}]{regularization6}
\bibinfo{author}{\bibfnamefont{L.}~\bibnamefont{Barack}},
  \bibinfo{author}{\bibfnamefont{Y.}~\bibnamefont{Mino}},
  \bibinfo{author}{\bibfnamefont{H.}~\bibnamefont{Nakano}},
  \bibinfo{author}{\bibfnamefont{A.}~\bibnamefont{Ori}}, \bibnamefont{and}
  \bibinfo{author}{\bibfnamefont{M.}~\bibnamefont{Sasaki}},
  \bibinfo{journal}{Phys. Rev. Lett.} \textbf{\bibinfo{volume}{88}},
  \bibinfo{pages}{091101} (\bibinfo{year}{2002}), \eprint{arXiv:gr-qc/0111001}.

\bibitem[{\citenamefont{Mino et~al.}(2003)\citenamefont{Mino, Nakano, and
  Sasaki}}]{regularization7}
\bibinfo{author}{\bibfnamefont{Y.}~\bibnamefont{Mino}},
  \bibinfo{author}{\bibfnamefont{H.}~\bibnamefont{Nakano}}, \bibnamefont{and}
  \bibinfo{author}{\bibfnamefont{M.}~\bibnamefont{Sasaki}},
  \bibinfo{journal}{Prog. Theor. Phys.} \textbf{\bibinfo{volume}{108}},
  \bibinfo{pages}{1039} (\bibinfo{year}{2003}), \eprint{arXiv:gr-qc/0111074}.

\bibitem[{\citenamefont{Roach}(1982)}]{Greens_functions}
\bibinfo{author}{\bibfnamefont{G.~F.} \bibnamefont{Roach}},
  \emph{\bibinfo{title}{Green's Functions}} (\bibinfo{publisher}{Cambridge
  University Press}, \bibinfo{address}{Cambridge}, \bibinfo{year}{1982}).

\bibitem[{\citenamefont{Friedlander}(1975)}]{Friedlander}
\bibinfo{author}{\bibfnamefont{F.~G.} \bibnamefont{Friedlander}},
  \emph{\bibinfo{title}{The Wave Equation on a Curved Space-Time}}
  (\bibinfo{publisher}{Cambridge University Press},
  \bibinfo{address}{Cambridge}, \bibinfo{year}{1975}).

\bibitem[{\citenamefont{Pati and Will}(2000)}]{DIRE}
\bibinfo{author}{\bibfnamefont{M.}~\bibnamefont{Pati}} \bibnamefont{and}
  \bibinfo{author}{\bibfnamefont{C.}~\bibnamefont{Will}},
  \bibinfo{journal}{Phys. Rev. D} \textbf{\bibinfo{volume}{62}},
  \bibinfo{pages}{1} (\bibinfo{year}{2000}).

\bibitem[{\citenamefont{Gralla et~al.}(2009)\citenamefont{Gralla, Harte, and
  Wald}}]{Gralla_Harte_Wald}
\bibinfo{author}{\bibfnamefont{S.~E.} \bibnamefont{Gralla}},
  \bibinfo{author}{\bibfnamefont{A.~I.} \bibnamefont{Harte}}, \bibnamefont{and}
  \bibinfo{author}{\bibfnamefont{R.~M.} \bibnamefont{Wald}}
  (\bibinfo{year}{2009}), \eprint{arXiv:gr-qc/0905.2391}.

\bibitem[{\citenamefont{Isaacson}(1968)}]{Isaacson}
\bibinfo{author}{\bibfnamefont{R.~A.} \bibnamefont{Isaacson}},
  \bibinfo{journal}{Phys. Rev.} \textbf{\bibinfo{volume}{166}},
  \bibinfo{pages}{1272} (\bibinfo{year}{1968}).

\bibitem[{\citenamefont{Galley and
  Hu}(2009{\natexlab{b}})}]{Galley_backreaction}
\bibinfo{author}{\bibfnamefont{C.~R.} \bibnamefont{Galley}} \bibnamefont{and}
  \bibinfo{author}{\bibfnamefont{B.-L.} \bibnamefont{Hu}}
  (\bibinfo{year}{2009}{\natexlab{b}}), \eprint{arXiv:gr-qc/0906.0968}.

\bibitem[{\citenamefont{Damour and Blanchet}(1986)}]{STF_1}
\bibinfo{author}{\bibfnamefont{T.}~\bibnamefont{Damour}} \bibnamefont{and}
  \bibinfo{author}{\bibfnamefont{L.}~\bibnamefont{Blanchet}},
  \bibinfo{journal}{Phil. Trans. R. Soc. Lond. A}
  \textbf{\bibinfo{volume}{320}}, \bibinfo{pages}{379} (\bibinfo{year}{1986}).

\bibitem[{\citenamefont{Thorne}(1980)}]{STF_3}
\bibinfo{author}{\bibfnamefont{K.~S.} \bibnamefont{Thorne}},
  \bibinfo{journal}{Rev. Mod. Phys.} \textbf{\bibinfo{volume}{52}},
  \bibinfo{pages}{299} (\bibinfo{year}{1980}).

\end{thebibliography}

\end{document}